\documentclass[traditabstract]{aa}
\usepackage{epsfig}
\usepackage{float}
\restylefloat{table}

\newcommand{\ms}{\mbox{m\,s$^{-1}$} }

\begin{document}

\title{The $Spitzer$ search for the transits of HARPS low-mass planets - II. Null results for 19 
planets\thanks{The photometric and radial velocity time series used in this work are only available in electronic form at the CDS via anonymous ftp to  cdsarc.u-strasbg.fr (130.79.128.5) or via http://cdsweb.u-strasbg.fr/cgi-bin/qcat?J/A+A/}}
\author{M.~Gillon$^{1}$,  B.-O.~Demory$^{2,3}$,  C.~Lovis$^4$, D.~Deming$^5$,  D. Ehrenreich$^4$,  G. Lo Curto$^6$, M. Mayor$^4$, F.~Pepe$^4$, D.~Queloz$^{3, 4}$,  S.~Seager$^7$, D.~S\'egransan$^4$, S.~Udry$^4$} 

\offprints{michael.gillon@ulg.ac.be}
\institute{
$^1$ Space sciences, Technologies and Astrophysics Research (STAR) Institute, Universit\'e de Li\`ege,  All\'ee du 6 Ao\^ut 17,  Bat.  B5C, 4000 Li\`ege, Belgium \\
$^2$ University of Bern, Center for Space and Habitability, Sidlerstrasse 5, CH-3012, Bern, Switzerland\\
$^3$ Cavendish Laboratory, J.~J. Thomson Avenue, Cambridge CB3 0HE, UK\\
$^4$ Observatoire de Gen\`eve, Universit\'e de Gen\`eve, 51 Chemin des Maillettes, 1290 Sauverny, Switzerland\\
$^5$ Department of Astronomy, University of Maryland, College Park, MD 20742-2421, USA\\
$^6$ European Southern Observatory, Karl-Schwarzschild-Str. 2, D-85478 Garching bei M\"unchen, Germany\\
$^7$ Department of Earth, Atmospheric and Planetary Sciences, Department of Physics, Massachusetts Institute of 
Technology, 77 Massachusetts Ave., Cambridge, MA 02139, USA
}

\date{Received date / accepted date}
\authorrunning{M. Gillon et al.}
\titlerunning{Results of the {\it Warm Spitzer} transit search}
\abstract{Short-period super-Earths and Neptunes  are now known to be very frequent around solar-type 
stars. Improving our understanding of these mysterious planets requires the detection of a significant sample of 
objects suitable for detailed characterization.  Searching for the transits of the low-mass planets
 detected by Doppler surveys is a straightforward  way to achieve this goal. Indeed, Doppler
  surveys target  the most nearby main-sequence stars, they regularly detect close-in low-mass planets with significant 
  transit probability, and their radial velocity data constrain strongly the ephemeris of possible transits.
In this context, we initiated in 2010 an ambitious {\it Spitzer} multi-Cycle transit search project that targeted 
25 low-mass planets detected by radial velocity,  focusing mainly on the shortest-period planets detected 
by the HARPS spectrograph.  We report here  null results for 19 targets of the project. For 16 planets out of 
19, a transiting configuration is strongly disfavored or firmly rejected by our data
for most planetary compositions. We derive  a posterior probability of 83\% that none of the probed 19 planets transits  (for a
 prior probability of 22\%), which still leaves a significant probability of 17\% that at least one of them does transit. 
 Globally, our {\it Spitzer} project  revealed or confirmed transits for three of its 25 
targeted planets, and discarded or disfavored the transiting nature of 20 of them.   
Our light curves demonstrate for {\it Warm Spitzer} excellent photometric precisions:
  for 14 targets out of 19, we were able to reach standard deviations that were better than 50ppm per 30 min intervals. 
  Combined with its Earth-trailing orbit, which makes it capable of pointing any star in the sky and to monitor it continuously for days, 
  this work confirms {\it Spitzer} as  an optimal instrument to detect sub-mmag-deep transits on the bright nearby stars
targeted by Doppler surveys.

 }

\maketitle

\section{Introduction}

Starting from 2004 (Butler et al. 2004, Santos et al. 2004), exoplanet search projects have been 
detecting  planets of a few to $\sim$20 Earth masses at an ever-increasing rate, 
revealing them to  be very frequent around solar-like stars (e.g. Howard et al. 2012) where they 
tend to form compact multiple systems (Rowe et al. 2014). Based on their mass (or minimal mass 
for planets detected by radial velocity - RV), these objects are loosely classified as super-Earths 
($M_p \le 10 M_\oplus$) and Neptunes ($M_p > 10 M_\oplus$). This division is based on the 
theoretical limit for the runaway accretion of H/He by a protoplanet, $\sim 10 M_\oplus$ (Rafikov 
2006), and thus implicitly assumes that Neptunes are predominantly ice giants and that most 
super-Earths are massive terrestrial planets. However, the growing sample of transiting low-mass 
exoplanets with precise mass and radius measurements exhibit a wide diversity of densities that 
reveals a very heterogeneous population, making simplistic inferences hazardous  when based on the 
comparison with solar system planets. A better understanding of this ubiquitous class of planets 
requires the thorough characterization of a significant sample, not only the precise 
measurements  of their physical dimensions but also the exploration of their atmospheric 
composition to alleviate the strong degeneracies of composition models in this mass range 
(e.g. Seager et al. 2007, Valencia et al. 2013).

Among the known transiting low-mass planets, GJ\,436\,b (Butler et al. 2004, Gillon et al. 2007) 
and GJ\,1214\,b (Charbonneau et al. 2009) are the most thoroughly characterized planets in
the Neptune and super-Earth mass ranges, respectively, thanks to the small size ($\sim$ 0.45 and
 0.2 $R_\odot$) and proximity (about a dozen of pc) of their M-dwarf host stars. First constraints on their 
 atmospheric properties have indeed been obtained by several programs (e.g. Kreidberg et al. 2014, Knutson 
et al. 2014a, Ehrenreich et al. 2015). Extending this kind of detailed studies to Neptunes and 
super-Earths orbiting solar-type hosts requires the detection of such planets in transit in front of extremely bright
and nearby stars. A straightforward method to achieve this goal is to search for the 
 transits of the low-mass planets detected by RV surveys. Indeed, these surveys target  the most 
 nearby  main-sequence stars, and they have now  detected enough short-period low-mass planets  
 to make it highly probable that a handful of them transit their parent stars, as demonstrated by the  
previous detections by the MOST space telescope of the transits of the super-Earths 55\,Cnc\,e (Winn et al. 2011)\footnote{
Our {\it Spitzer} program independently revealed the transiting nature of 55\,Cnc\,e (Demory et al. 2011).
and  HD\,97658\,b  (Dragomir  et al. 2013)}. 
Thanks to the brightness of their host stars, the  atmospheric 
  characterization of these two planets has already started (e.g. Demory et al. 2012,  Knutson et al. 2014b).
  More recently, the same approach enabled us to reveal with  {\it Spitzer} the transiting configuration of the rocky planet
   HD\,219134\,b (Motalebi et al. 2015) which, at 6.5pc, is  the nearest known transiting exoplanet.
    
 Searching for the transits of RV low-mass planets is one of the main objectives of  the future European 
 space mission CHEOPS (Broeg et al. 2013). However, CHEOPS is not due  to launch before the end of 2017. 
 To set it on its path,  back in 2010 we set up an ambitious project using  {\it Spitzer}/IRAC (Fazio et al. 2004) 
 to search for the transits of the RV low-mass planets that have the highest geometric transit probabilities, 
 focusing mainly on the shortest-period planets detected by the HARPS spectrograph (Mayor et al. 2003). Our 
 {\it Spitzer} transit search was composed of a cryogenic program targeting HD\,40307\,b (ID 495, 27.5hr), and  
  three so-called {\it Warm} (i.e. non cryogenic)  programs (ID 60027, 90072, and 11180; 100hr, 300hr, and 9.5hr) that targeted 
 24 other RV low-mass planets. Its published results  have so far been the non-detection of the transits of 
 HD\,40307\,b (Gillon et al. 2010, hereafter G10) and GJ\,3634\,b (Bonfils et al. 2011), the detection and 
 confirmation of the transits of 55\,Cnc\,e (Demory et al. 2011, Gillon et al. 2012a), the confirmation of 
 the transiting nature of HD\,97658\,b (Van Grootel et al. 2014), and the detection of a transit of 
 HD\,219134\,b (Motalebi et al. 2015). Results for another planet will be presented in a forthcoming
 paper (S\'egransan et al., in prep.) We report here null results for the 19 other targets of the project. 
 
 We first present our targets and our determination of their transit  ephemeris. 
 In Sect. 3, we present our $Spitzer$ data and their reduction.  Section 4 describes our data analysis. Its
 main results are presented in Sect. 5. Finally,  we discuss our global results and give our conclusions in Sect. 6. 

\section{Targets and transit ephemeris determination}

Table 1 and Tables A.1 to A.4 list the 19 targets of this work. For each target, we performed an analysis of the available 
RVs, including new measurements for some of them gathered by HARPS\footnote{Most HARPS measurements used in this work 
are available on the ESO/HARPS online archive at http://archive.eso.org/wdb/wdb/eso/repro/form}, to derive 
the most accurate transit ephemeris. This analysis was done with the adaptative Markov Chain Monte-Carlo 
(MCMC) algorithm described in G10 (see also Gillon et al. 2012a, 2012b, 2014).  Our nominal model was based 
on a star and one or several planets on Keplerian orbits around their common center of mass. For some cases, 
we added a linear or quadratic trend to the model, based on the minimization of the Bayesian information criterion 
(BIC, Schwarz 1974) to elect our  final model. We checked that planet-planet interactions had negligible influence 
on our solutions, using, for this  purpose, the {\it Systemic Console} software (Meschiari et al. 2009). We also 
used this software to perform an  initial optimization of parameters  to check that no solution existed with a higher 
likelihood than the published one. During this initial stage
 of the analysis, the RV errors   were assumed to be measurement uncertainties.  Once this initial stage 
 was completed, we measured the  quadrature difference between the rms of the residuals and the mean error, 
 and this ''jitter noise'' (Wright  2005) was quadratically summed to the measurement uncertainties. We  then 
 performed a MCMC analysis to  probe the posterior  probability distribution functions (PDF) of the model and 
 physical parameters

Each MCMC analysis was composed of five Markov chains of 10$^5$ steps, the first 20\% of each being 
considered as its burn-in phase and discarded. For each run, the convergence of the five Markov chains was 
checked using the statistical test of Gelman \& Rubin (1992). The resulting posterior PDFs for the transit time 
and period were then used to schedule the {\it Spitzer} observations, with monitoring the $2-\sigma$ transit 
window as the goal, i.e. keeping the probability to miss a transit below 5\%. 

Even if the resulting posterior PDFs for the orbital eccentricity of most target planets were compatible with zero,
we did not assume a purely circular orbit for them to ensure the reliability of our derived transit ephemeris. Indeed, 
several examples of low-mass exoplanets with significantly eccentric few-days orbits  (e.g. GJ\,436\,b, HAT-P-11\,b, 
HD215497\,b) remind us of the risks of systematically assuming circular orbits for close-in exoplanets based on tidal 
circularization arguments. An established theory for tidal dissipation mechanisms is still a long-term goal, and reaching it 
relies mostly on gathering new observational constraints. 

Once the {\it Spitzer} observations had been performed for a planet, we made a second MCMC analysis of the most up-to-date 
RV dataset, this time with the fitted transit time corresponding to the epoch covered by the {\it Spitzer} run. The purpose of this
analysis was to derive the most accurate orbital parameters, transit ephemeris, and minimal masses for the planets
under consideration. Indeed, for most of them HARPS gathered a significant number of RV measurements between the  scheduling of the {\it Spitzer}  observations and the final analysis of the {\it Spitzer} images. 

Tables A.1 to A.4 present the results obtained for each targeted planet from these MCMC analyses of the most up-to-date RV dataset. In addition to some basic parameters for the host stars, 
these tables give the origin of the RVs used as input in our MCMC analysis for each target, and provide the most 
relevant results of our MCMC analysis: the transit and occultation ephemeris, the minimal mass, the expected minimum 
transit depth corresponding to a pure iron planet (Seager et al. 2007), the equilibrium dayside  temperature, the orbital 
parameters, and the expected duration for a central transit. The radius of the star was  derived from the luminosity and 
effective temperature, taking bolometric corrections from Flower (1996). 
 
 Tables A.1 to A.4 also present the median value and the 1-$\sigma$ errors for the prior transit  probability. It was computed 
 at each step of the MCMC with the following formula:
\begin{equation}
P(tr)  =   \bigg(\frac{R_\ast}{a}\bigg) \bigg( \frac{1+e\sin\omega}{1-e^2}\bigg) \textrm{,} 
\end{equation} where $R_\ast$ is the stellar radius, $a$ is the orbital semi-major axis, $e$ is the orbital eccentricity 
and $\omega$ is the argument of periastron. This probability estimate does not take into account that planets 
are more likely to be discovered by  RV if  their orbit is significantly inclined (e.g. Wisniewski et al. 2012). By 
performing Bayesian simulations that assume different prior PDFs for the planetary masses,
 Stevens \& Gaudi (2013) have shown that this bias increases the transit probability of short-period low-mass RV planets
like the ones considered here by only $\sim$20\% in average. Furthermore, its actual estimation for a given planet depends 
strongly on the assumed prior PDFs for the planetary masses. We have thus neglected it in the context of this work.

\section{{\it Warm  Spitzer} photometry}

Tables A.5 to A.11 provide a summary of the {\it Spitzer} observations. Since all our targets are very bright ($K$ between
 2.96 and 6.90), all of them were observed in subarray mode (32x32 pixels windowing of the detector), the extremely
  fast Fowler sampling ($\sim$0.01s), which maximizes the duty cycle and signal-to-noise ratio (S/N). No dithering pattern 
  was applied to the telescope (continuous staring). For each target, the exposure time was selected to maximize the 
S/N while staying in the linear regime of the detector, basing on the  {\it Warm Spitzer} flux density estimator 
tool\footnote{http://ssc.spitzer.caltech.edu/warmmission/propkit/pet/starpet} and on the instructions of the 
{\it Warm Spitzer} Observer Manual\footnote{http://ssc.spitzer.caltech.edu/warmmission/propkit/som}.

The observations of program 60027 (Cycle 6) and 90072 (Cycle 9) were performed between 2009 Dec 14 and 2010 
Sep 11, and between 2012 Dec 03 and 2014 May 14, respectively. For several of the Cycle 9 targets, we benefitted 
from the  newly introduced PCRS peak-up mode (Ingalls et al. 2012). This mode provides enhanced accuracy in the 
position of the target on the detector,  to a significant decrease of the so-called `pixel phase effect' that is the most important 
source of correlated noise in high-S/N staring mode observation with IRAC InSb arrays (e.g. Knutson et al. 2008). 
For HD\,1461\,b, we supplemented our data with the IRAC 4.5$\mu$m observations presented by 
Kammer et al. (2014), as described in Sect. 5.2. 

On a practical level, each observation run was divided in one or several science astronomical observational requests (AOR) 
of 12hr at most, preceded by a short (20-30 min) AOR to allow the pointing of the telescope and the instrument to stabilize.
The IDs of all AORs are given for each target in Tables A.5 to A.11. These tables also give the version of the 
{\it Spitzer} pipeline used to calibrate the corresponding images, the resulting files being called basic calibrated 
data (BCD) in the {\it Spitzer} nomenclature. Each subarray mode BCD is composed of a cube of 64 subarray
 images of 32$\times$32 pixels (pixel scale  = 1.2 arc second).

The following reduction strategy was used for all the {\it Spitzer} AOR. We first converted fluxes from the 
{\it Spitzer} units of specific intensity (MJy/sr) to photon counts, then aperture photometry was performed on
 each subarray image with the {\tt IRAF/DAOPHOT}\footnote{IRAF is distributed by the National Optical 
 Astronomy Observatory, which is operated by the Association of Universities for Research in Astronomy, Inc., 
 under cooperative agreement with the National Science Foundation.} software (Stetson, 1987). For each AOR, we 
 tested different aperture radii and background annuli, and selected the combination minimizing the 
white and red noises in the residuals of a short data fitting analysis. The center and width of the point-spread
 functions (PSF) were measured by fitting a 2D-Gaussian profile on each image. The $x-y$ distribution
of the measurements was then studied, and measurements that had a visually discrepant position relative to 
 the bulk of the data were then discarded. For each block of 64 subarray images, we then discarded the discrepant 
 values for the measurements of flux, background, $x$- and $y$-positions using a 10-$\sigma$ median clipping for 
 the four parameters, and the resulting values were averaged, the photometric errors being taken as the errors
on the average flux measurements. Finally, a 50-$\sigma$ slipping median clipping was used on the resulting 
 light curves to discard outliers (owing to, for example, cosmic hits). 
 
\section{Global {\it Warm Spitzer} + RV data analysis}

We analyzed the {\it Spitzer} photometric time-series supplemented by the RVs with our MCMC code. 
For each target, our model for the RVs was the same as the one presented in Sect. 2. The assumed photometric 
model consisted of the eclipse model of Mandel \& Agol (2002) to represent the possible eclipses of the probed 
planets, multiplied for each  light curve by a baseline model that aimed to represent the other astrophysical and instrumental 
effects at the source of photometric variations. We assumed a quadratic limb-darkening law for the stars. For each light curve 
that corresponded to a specific AOR, we based  the selection of the baseline model on the minimization of the BIC. Tables A.5 
to A.11 present the baseline function elected for each AOR.

Following Gillon et al. (2014), the instrumental models included three types of low-order polynomials. The first one 
had as variables the $x$- and $y$-positions of the center of the PSF to represent the so-called pixel phase effect of the IRAC
 InSb arrays (e.g. Knutson et al. 2008). The second one had  the PSF widths in the $x$- and/or the 
 $y$-direction as variables, its inclusion in the baseline model strongly increasing the quality of the fit for all AORs (the so-called PSF breathing effect, see also Lanotte et al. 2014). The third  function was a polynomial of the logarithm of time to represent a sharp increase of the detector response at the start of some AORs (the so-called ramp effect, Knutson et al. 2008). To improve the quality of the modeling of the
  pixel phase effect, especially the fitting of its highest frequency components,for most AORs  we supplemented the 
  $x$- and $y$-polynomial with the bi-linearly-interpolated sub-pixel sensitivity (BLISS) mapping method (Stevenson 
  et al. 2012). The sampling of the position space was selected so that at least five measurements fall within the 
  same sub-pixel. See Gillon et al. (2014) for more details.  

The jump parameters of the MCMC, i.e. the parameters randomly perturbed at each step of the Markov Chains, were as follows:
 \begin{itemize}
\item The stellar mass $M_\ast$, radius $R_\ast$, effective temperature $T_{eff}$, and metallicity [Fe/H]. For 
these four parameters, normal prior PDFs  were assumed based on the values given in Tables A.1-4.
\item For the potential transiting planet, the parameter  $b' = a \cos{i}/R_\ast$, where $a$ is the orbital
 semi-major axis and $i$ is the orbital inclination. $b'$ would correspond to the transit impact parameter in the case of a circular orbit. 
 The step was rejected if $b' > a/R_\ast$. For the other planets of the system,  $b'$ was fixed to 0.
 \item The parameter $K_2 = K  \sqrt{1-e^2}   \textrm{ }  P^{1/3}$ for all planets of the system, $K$ being the RV orbital 
semi-amplitude, $e$ the orbital eccentricity, and $P$ the orbital period.
 \item The orbital period $P$ of each planet.
 \item For each planet, the two parameters $\sqrt{e} \cos{\omega}$ and $\sqrt{e} \sin{\omega}$, $e$ being the 
orbital eccentricity and $\omega$ being the argument of periastron.
 \item The planet/star area ratio $dF = (R_p/R_\ast)^2$ for the potential transiting planet.  At each step of the MCMC, 
 the planetary radius  corresponding to a pure iron composition was computed under the formalism given by Seager et al. 
 (2007), and if the planetary radius derived from $dF$ and $R_\ast$ was smaller, the step was rejected. A similar rejection 
 was done for $R_p > 11 R_\oplus$, an  implausibly large size for the low-mass planets considered here. The 
goal of these prior constraints on $R_p$ was to avoid  fitting extremely shallow transits and 
ultra-grazing transits of unrealistically big planets in the correlated noise of the light curves to ensure an 
unbiased posterior transit probability. For the other planets of the multiple systems, $dF$ was fixed to 0. 
In all cases, we checked that a transit of another planet of these systems was not expected to occur
during the {\it Spitzer} observations. 
\item The time of inferior conjunction $T_0$ for all planets of the system. For the potential transiting 
planets that we considered, $T_0$ corresponds approximatively to the mid-time of the transit searched for by {\it Spitzer}. 
\end{itemize}

The limb-darkening of the star was modeled by a quadratic law (Claret 2000). For both {\it Warm Spitzer}
 bandpasses (3.6 and 4.5 $\mu$m), values for the two quadratic limb-darkening coefficients $u_1$ and
  $u_2$ were drawn at each step of the MCMC from normal distributions whose expectations and 
  standard deviations were drawn from the tables of Claret \& Bloemen (2011) for the corresponding bandpasses 
  and for the stellar atmospheric parameters given in Tables A.1-4.

Five chains of 100000 steps were performed for each analysis, their convergences being checked using 
the statistical test of Gelman and Rubin (1992). They followed a preliminary chain of 100000 steps, 
which was performed to estimate the need to rescale the photometric errors. For each light curve, the standard deviation of 
the residuals was then compared to the mean photometric errors, and the resulting ratios $\beta_w$ were stored.  
$\beta_w$ represents the under-  or overestimation of the white noise of each measurement. On its side, the red 
noise present in the light curve (i.e. the inability of our model to represent perfectly the data) was taken into 
account as described in G10, i.e. a scaling factor $\beta_r$ was determined from the standard 
deviations of the binned and unbinned residuals for different binning intervals ranging from 5 to 120 minutes, the 
largest values being kept as $\beta_r$. At the end, the error bars  were multiplied by the correction factor $CF = 
\beta_r \times \beta_w$. The derived values for $\beta_r $ and  $\beta_w$ are given for each light curves in Tables A.5-11.

\section{Results}

For each planet searched for transit, Table 1 presents the derived posterior full transit probability $P(f,D)$,  i.e. the probability that the planet undergoes full transits given the {\it Spitzer} data. Bayes theorem shows that 
\begin{equation}
P(f,D) = \frac{P(f) P(D,f)}{P(f) P(D,f) + P(g) P(D,g) +  P(n) P(D,n)}
\end{equation} $P(f)$, $P(g)$, and $P(n)$ are the prior (geometric) probabilities of full, grazing and no transit, respectively ($P(g)$ is close to zero, and $P(n) \sim 1 - P(t)$).  $P(D,f)$, $P(D,g)$, and $P(D,n)$ are the probabilities (likelihoods) to have the observed data given the three mutually exclusive hypotheses. All the terms of the right-hand side of Eq. 2 are  probed by the MCMC analysis, resulting in accurate estimates of $P(f,D)$. 

Figs. 1-4, 6-14, and 16-20 show the resulting detrended {\it Spitzer} light curves and the derived posterior PDFs for the inferior conjunction. Below, we provide relevant details for our 19 targets.

\subsection{BD-061339}

BD-061339 (a.k.a. GJ\,221) is a V=9.7 late-K dwarf around which two planets were detected by HARPS (Lo Curto et al. 2013),  a super-Earth on a $\sim$3.9d orbit and a planet of $\sim 50 M_\oplus$ minimal mass on a $\sim$126d orbit.
Our model selection process for the RVs (HARPS + PFS) favored a model with a slope in addition to these two planets (Table A.1) . The derived value for this slope is $-0.73 \pm 0.15$ \ms per year,  which could correspond to a giant planet in outer orbit or to the imprint of a stellar magnetic cycle. 

As can be seen in Fig. 1, the high-precision of the {\it Spitzer} photometry (42 ppm per 30 min time bin) discards any transit 
for BD-061339\,b during our observations,  even for an unrealistic pure iron composition. However, the edge of the right wing
 of the PDF for $T_0$ was not explored by our observations, so a late transit is still possible. The resulting posterior full 
 transit probability is 0.53\% (Table 1), high enough to justify a future exploration of the late part of the transit window.

\begin{figure}
\label{fig:1}
\centering                     
\includegraphics[width=9cm]{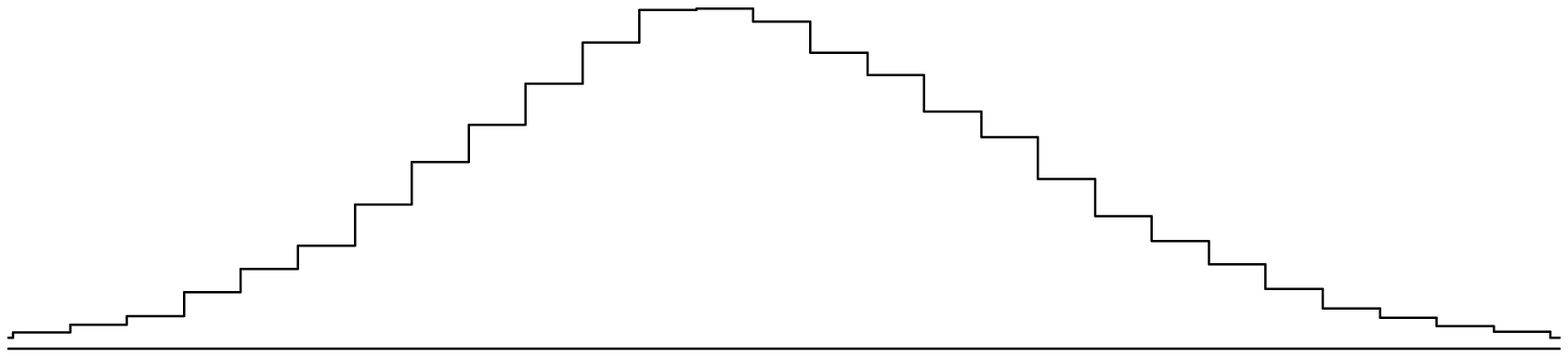}
\includegraphics[width=9cm]{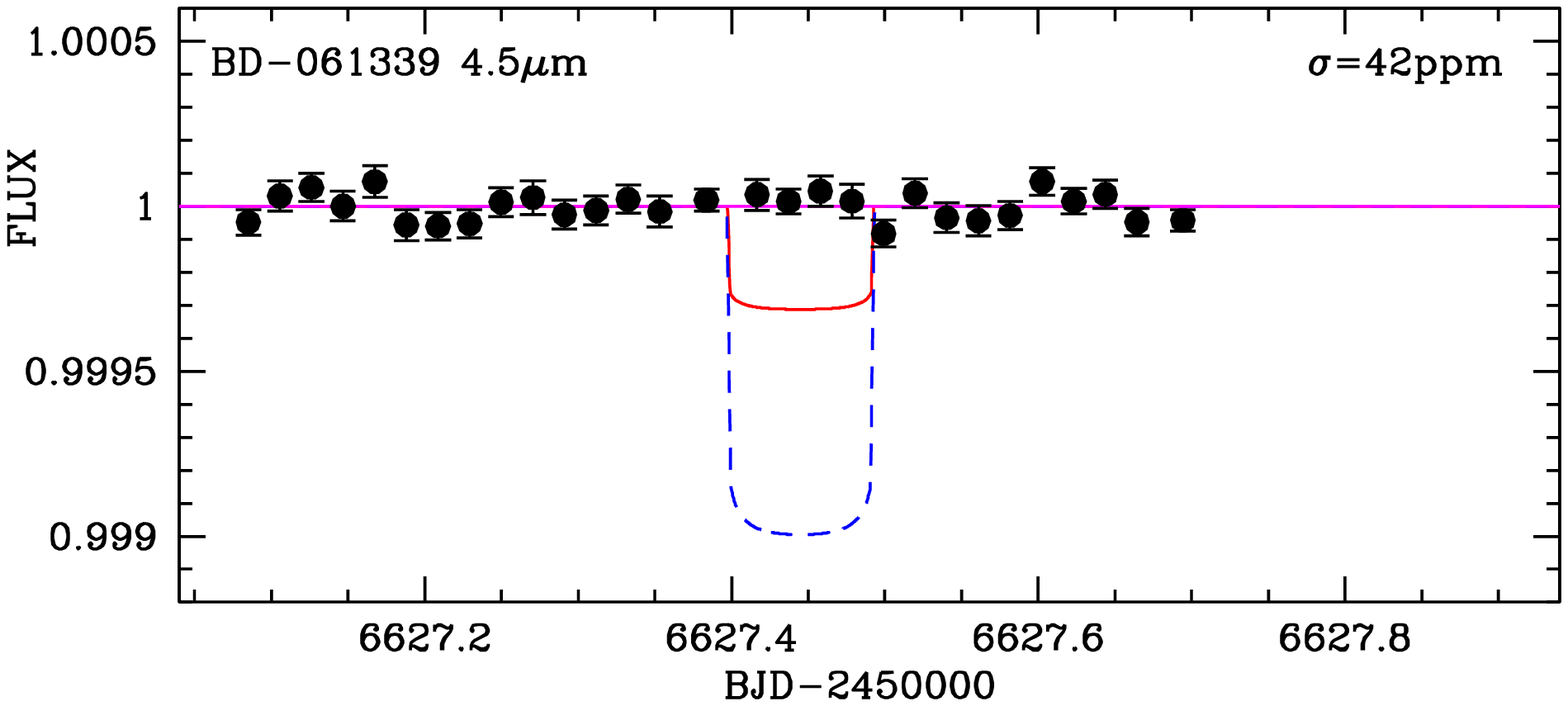}
\caption{{\it Warm Spitzer} photometry for BD-061339 after correction for the instrumental effects, normalization, and 
binning per 30 min. Models for a central transit of BD-061339\,b are shown, assuming a pure iron (red solid line) and pure 
water ice (blue dashed line) composition. The highest posterior probability model, which assumes no transit, is also shown (purple line). The posterior PDF for the transit timing derived from the RV analysis is shown above the figure.}
\end{figure}

\subsection{HD\,1461}

HD\,1461 is a V=6.5 solar-type star known to host a super-Earth on a 5.77\,d orbit (Rivera et al. 2010). Recently,
a second super-Earth on a 13.5\,d orbit was announced by D\'iaz  et al. (2015).
In our analysis of the HARPS + Keck RVs for this star, the convergence of the MCMC  appeared to be significantly improved 
by removing the 20 first Keck measurements for which the exposure time and the resulting precision were
 significantly lower than for the rest of the data. Our analysis favored a three planet  model (5.77, 13.5, and 377 
 days period) in addition to a second-order time polynomial, suggesting the presence of a fourth low-frequency signal, 
 and an activity model consisting of a sum of second-order polynomials in the  cross-correlation function (CCF, Baranne et al. 1996,
  Queloz et al. 2001) parameters (Table A.1): width, contrast, and bisector. We inferred that the 377d period, close to the duration of a year, is caused by a systematic effect (the stitching) recently revealed to affect HARPS data (Dumusque et al. 2015). The data used by D\'iaz  et al. (2015) were corrected for this effect. We made several tests that showed that the  inclusion or not of this 377d Doppler signal in the RV model did not affect the results for the closest-in  planet, including its transit ephemeris,  so we kept it in our final analysis. 
 
 A search for a transit of HD\,1461\,b with {\it Spitzer} was  presented by Kammer et al. in 2014 (program 80220). 
 We first reduced their data and used them with the HARPS+Keck RVs as input for a global MCMC analysis. The
 resulting  posterior full transit probability was 0.5\%, suggesting that  a small but significant fraction of  the transit window 
 was not covered by these {\it Spitzer} observations. As can be seen in Fig. 2, a transit that had ended just before the 
 {\it Spitzer} observations remained possible.  This possibility is amplified by the ramp effect that affected the first hour 
 of the {\it Spitzer} data, resulting in an increase of brightness that could be degenerated with a transit egress. 
 We thus complemented these {\it Spitzer} archive data with new {\it Spitzer} observations covering the first part of the 
 transit window. We performed a global analysis of all {\it Spitzer} + RVs that made  a  transiting configuration 
 for HD\,1461\,b very unlikely (see Fig. 2), the resulting posterior full transit probability now being of only 0.14\% (Table 1). 
 
\begin{figure}
\label{fig:2}
\centering                     
\includegraphics[width=9cm]{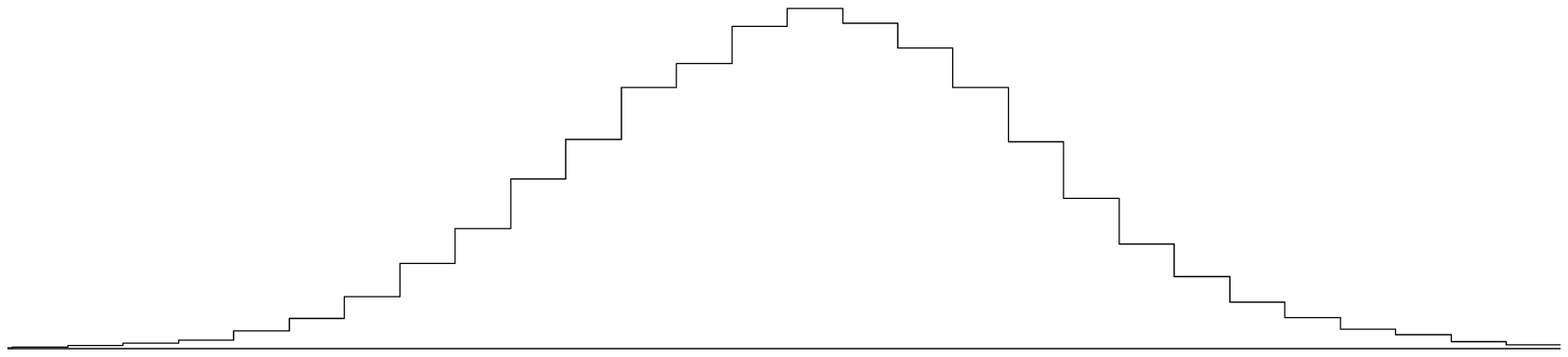}
\includegraphics[width=9cm]{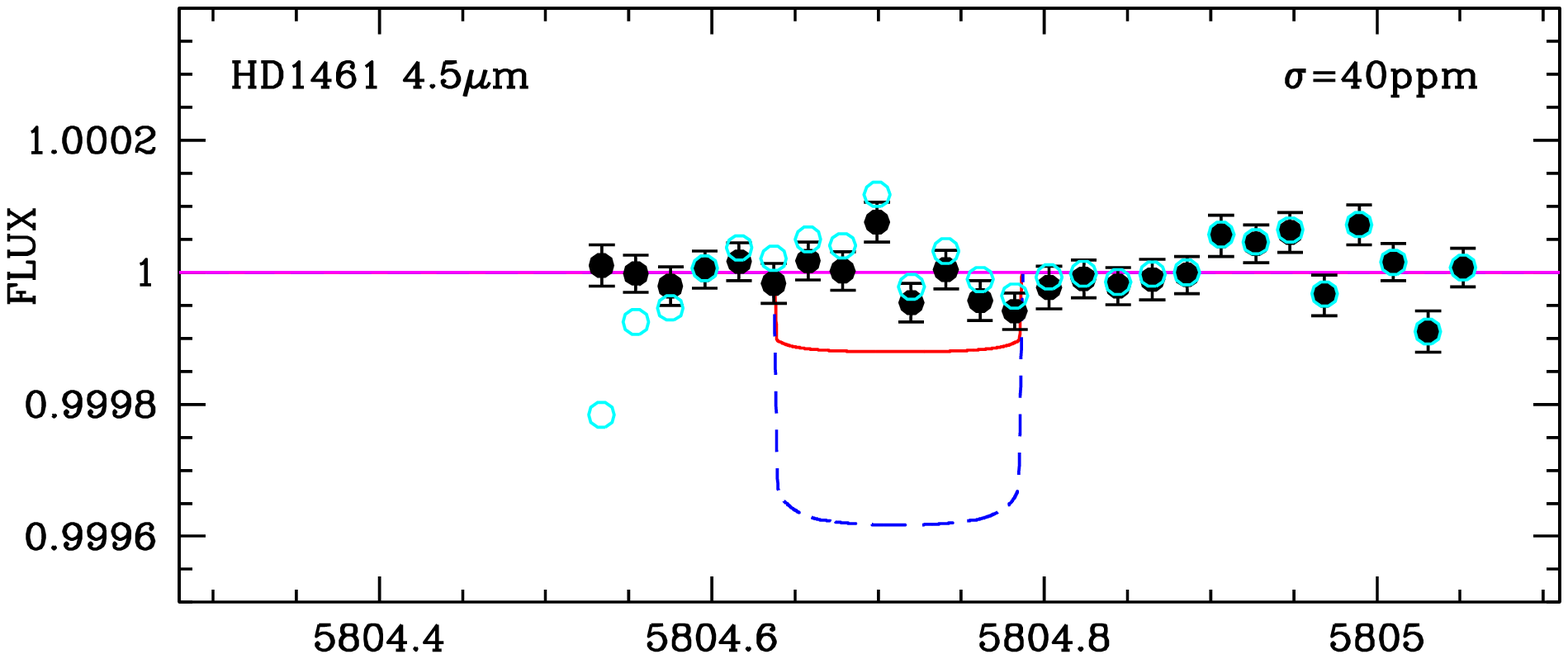}
\includegraphics[width=9cm]{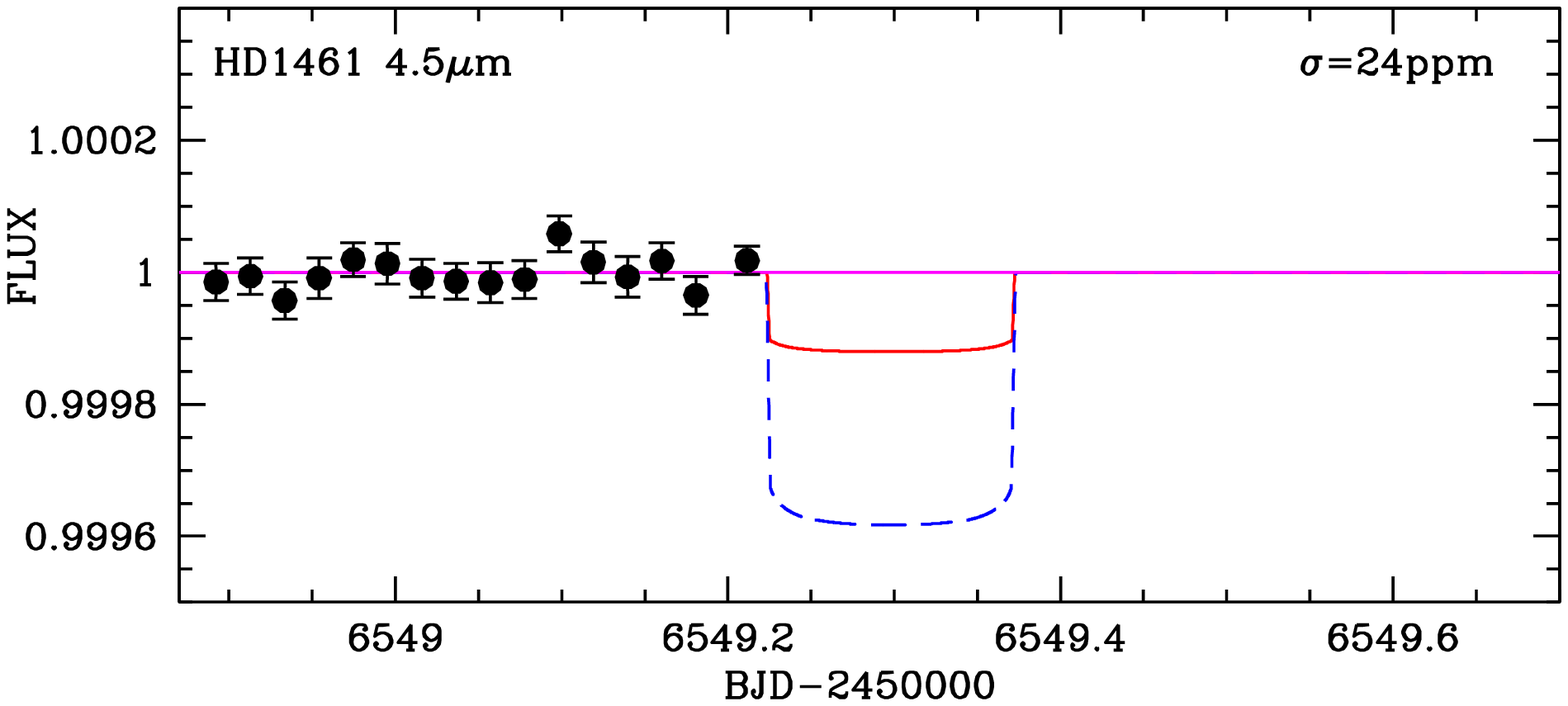}
\caption{Same as Fig. 1 for HD\,1461\,b. Upper panel corresponds to archive data (program 80220, Kammer et al. 2014), and the lower panel corresponds to the new data gathered in our program 90072. The upper panel also shows the light curve derived without correction of the ramp effect (open pale green circles).  }
\end{figure} 

\subsection{HD\,10180}

HD\,10180 is a V=7.3 solar-type star hosting a particularly interesting system of six planets of relatively low-masses (Lovis et al. 2011). At first, we did not consider searching for the transit of the 1.18d-period Earth-mass planet candidate HD\,10180\,b  presented by Lovis et al. (2011)  because its Doppler detection was not firmly secured when we planned our {\it Spitzer} Cycle 6 observations of the star. Furthermore,  we estimated that its  transit would, in any case, be too shallow (a few dozens of ppm, at most 100 ppm) to be firmly detected with {\it Spitzer}. We thus focused on the planet HD\,10180\,c ($M\sin{i} = 13 M_\oplus$, $P$ = 5.76~d), for which the Doppler signal was clearly detected in the HARPS data, and for which the expected transit depth ($> 150$ ppm) was large enough to ensure a sure detection with {\it Spitzer}. 
enabled us to discard a transit of the planet (Fig. 3), the resulting posterior full transit probability being of 0.14\% (Table 1). Still, we noticed a shallow structure in the detrended photometry that occurs at a time consistent with  a transit of HD\,10180\,b (Fig. 4), the best-fit transit depth of $\sim$90 ppm translating into a radius of 1.2 $R_\oplus$ consistent with a planetary mass of $1 M_\oplus$.  Based on this result, we decided to observe in our Cycle 9 program 90072 two more transit windows of HD\,10180\,b, this time at 4.5 $\mu$m. No transit structure was clearly detected in the resulting light curves (Fig. 4, bottom),  and a short MCMC analysis of the data led us to conclude that these new data do not  increase the significance of the 2010 tentative detection. We thus conclude that the low-amplitude  structure in our Cycle 6 {\it Spitzer} light curve is probably just correlated noise of instrumental origin that is not perfectly represented by our baseline model   and that can easily be modeled by a shallow enough transit profile. For {\it Spitzer}, the average amplitude of these correlated noise structures is of a few dozen of ppm (see Sec. 6), making the firm detection of a unique transit shallower than $\sim$100ppm  impossible  - similarly to the one expected for HD\,10180\,b (see discussion in Sec. 6 and our estimated detection thresholds in Table 1). This so-called red noise limit can  be surpassed, but only by gathering more observations of the transit window, as we did here (see also the case of HD\,40307\,b in Gillon et al. 2010).

\begin{figure}
\label{fig:3}
\centering                     
\includegraphics[width=9cm]{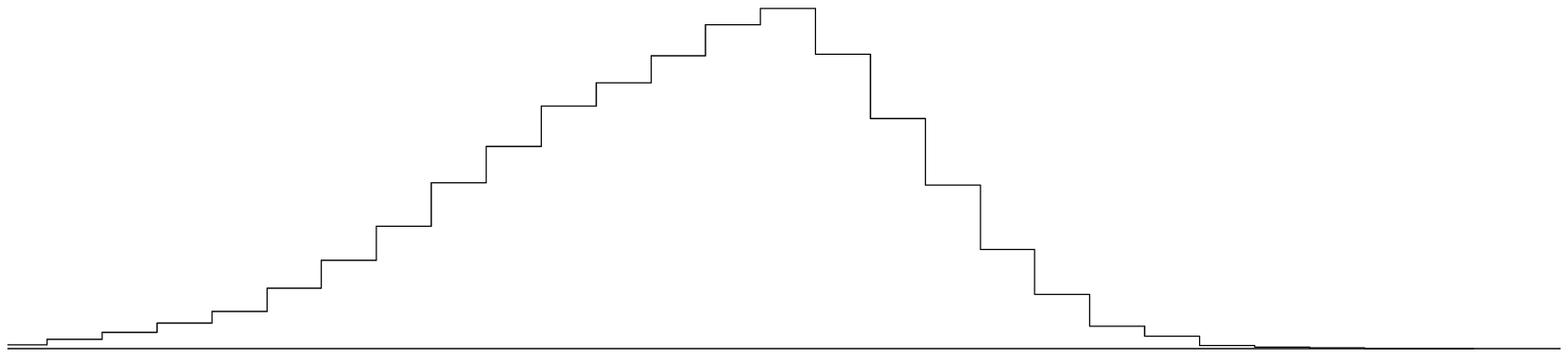}
\includegraphics[width=9cm]{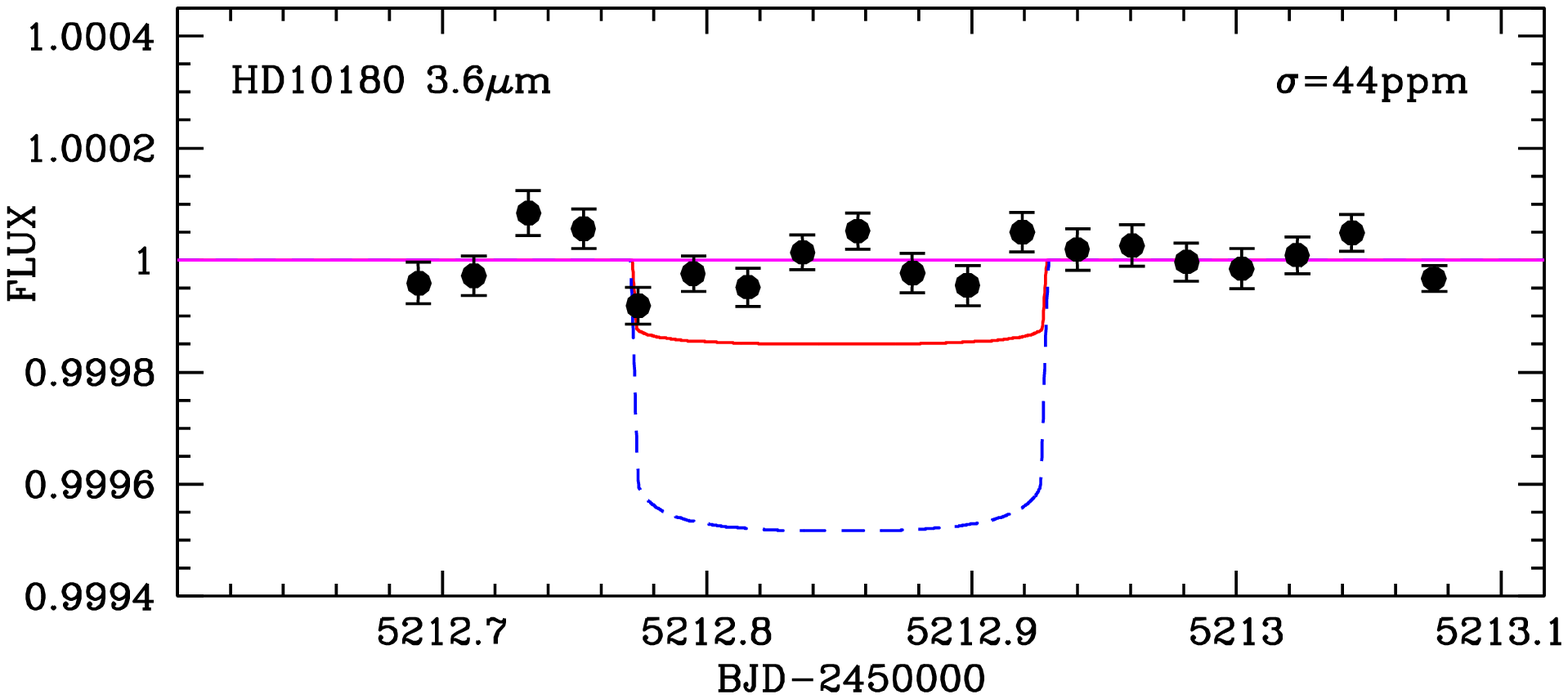}
\caption{Same as Fig. 1 for the light curve of 2010 Jan 16 for HD\,10180\,c.}
\end{figure} 

\begin{figure}
\label{fig:4}
\centering                     
\includegraphics[width=9cm]{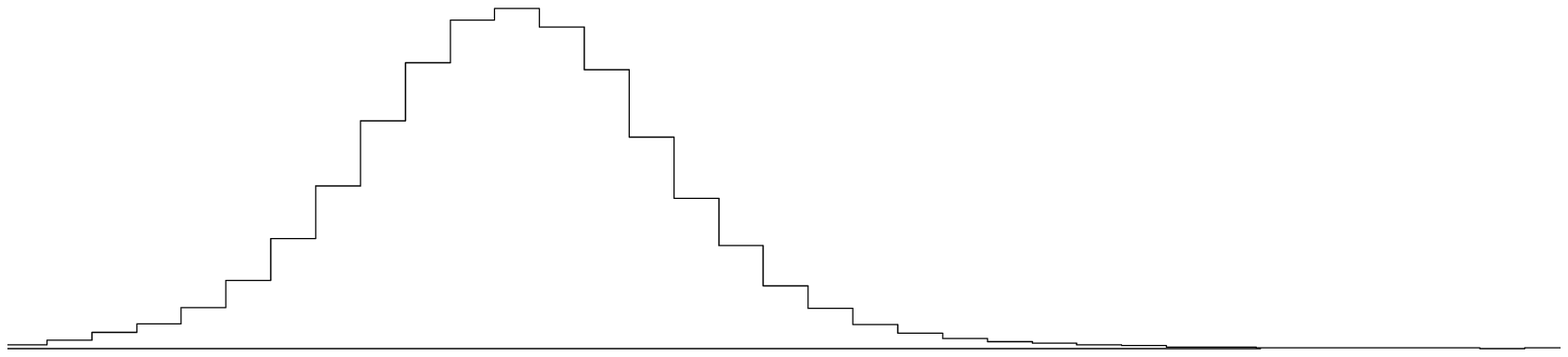}
\includegraphics[width=9cm]{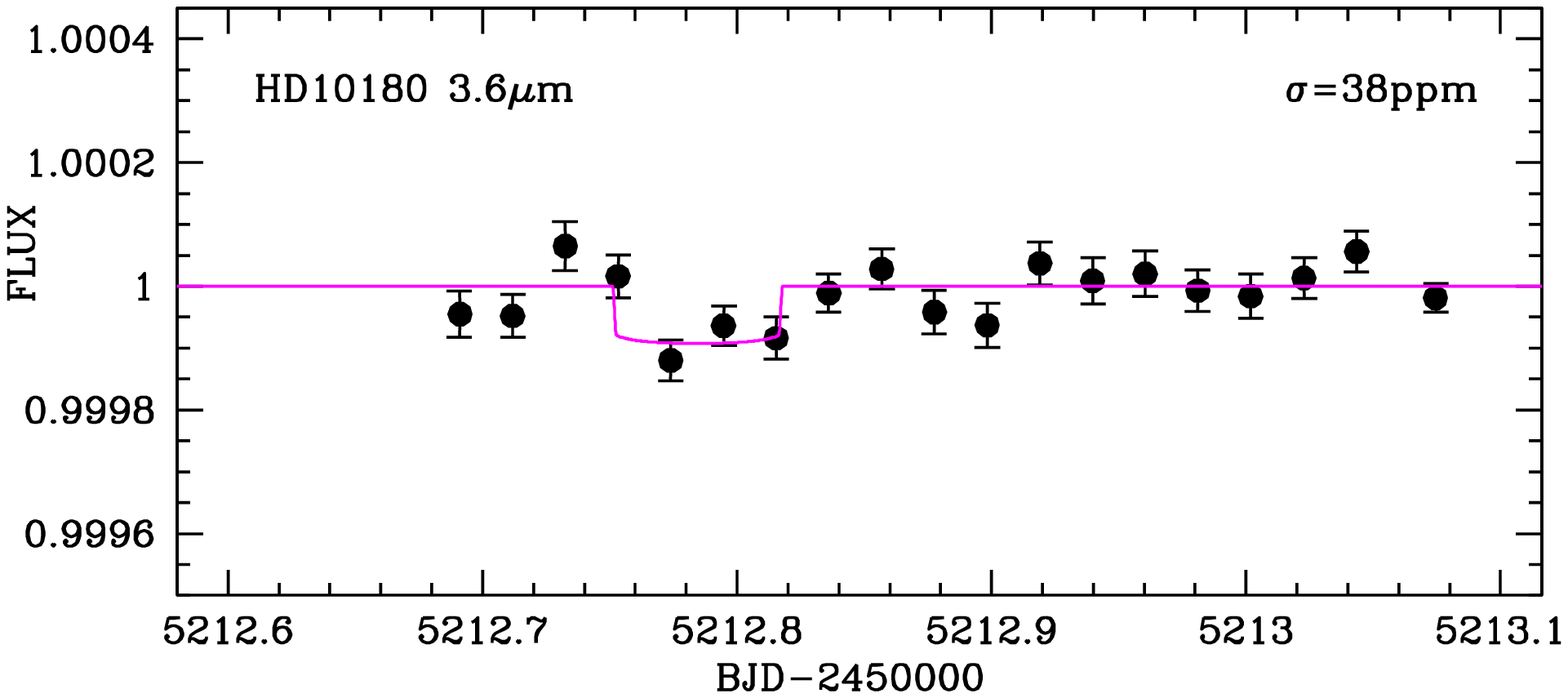}
\includegraphics[width=9cm]{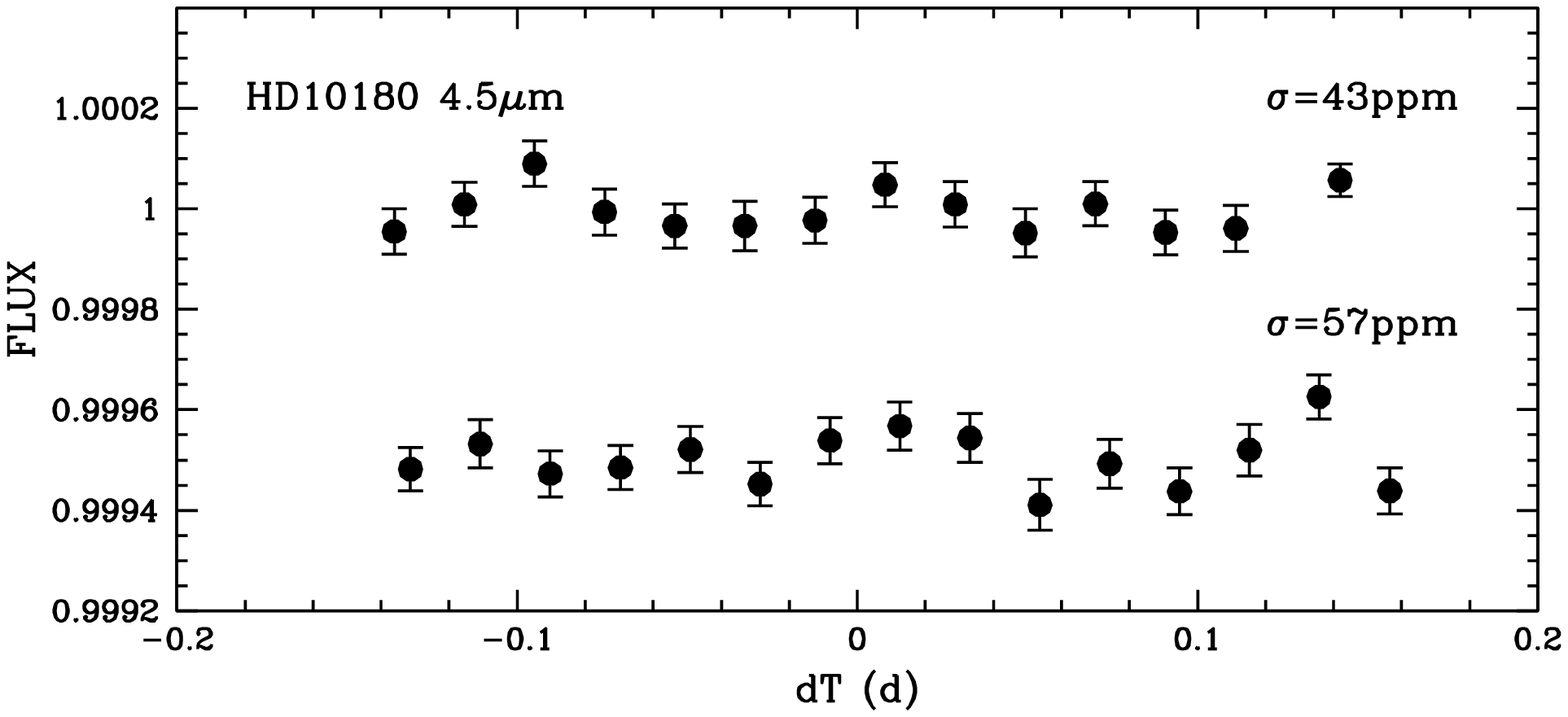}
\caption{$Top$: Same as Fig. 3, except that  a possible transit of HD\,10180\,b was included in the global model. Highest posterior probability transit model for HD\,10180\,b is shown in purple. The posterior PDF for the transit timing of HD\,10180\,b derived from an RV analysis assuming 7 planets is shown above the figure.  $Bottom$: 2013 {\it Warm Spitzer} photometry for HD\,10180 binned   per 30 min,  corrected for the instrumental effects, normalized, and folded on the transit ephemeris for HD\,10180\,b based on our tentative detection  of a transit and on the results of an RV analysis assuming 7 planets (dT = time from most likely mid-transit time). }
\end{figure} 

\subsection{HD\,13808}

In 2012, we analyzed the HARPS dataset for the $V$=8.4 early K-dwarf HD\,13808, confirming the existence  of the two planets around it with periods of  14.2~d and 
53.7~d announced by Mayor et al. (2011, hereafter M11). Furthermore, our analysis revealed the existence of (1) a low-frequency signal that was well-modeled with a quadratic trend, whose origins is the magnetic cycle of the star (Queloz et al. in prep.), and (2) a low-amplitude Doppler signal with a period of 1.091~d which corresponds to a planet of $M_p\sin{i} = 1.5 \pm 0.3$ $M_{\oplus}$ 
with an interestingly high transit probability of $\sim$20\%. The false-alarm probability (FAP) derived by {\it Systemic} for this short-period planet candidate was close to 1\%.
Based on this small FAP and the scientific importance of this putative planet if transiting, we decided to monitor two of its transit windows with {\it Spitzer}. 
Unfortunately, the resulting light curves did not show any convincing transit-like structure (Fig. 5). Our {\it Spitzer} data were not acquired during a transit window of the planet at 14.2~d, thus keeping its transiting nature unconstrained. 

In 2014, we analyzed the updated HARPS dataset. This analysis confirmed the existence of the low-frequency signal  of magnetic cycle origins, still well-modeled by a quadratic trend, but it led to a much lower significance (FAP $\sim$ 10\%) for the 1.091~d signal. A stronger signal at 18.98~d and at its alias period of 1.056~d emerged with FAP close to 1\%. Still, including a polynomial function of the CCF parameters (bisector, contrast, width) in our  MCMC modeling, these signals vanished from the residuals, indicating that their origin is stellar (activity), not planetary. The period 18.98~d could then be interpreted as the rotation period of the star, resulting in an equatorial rotation speed $\sim$2.2 km.s$^{-1}$ which is consistent with the measured $v\sin{i}$ of
 2 km.s$^{-1}$ (Glebocki \& Gnacinski 2005). This stellar signal, combined with aliasing effects, would thus be responsible for creating spurious signals with periods slightly larger than 1~d. 

\begin{figure}
\label{fig:5}
\centering                     
\includegraphics[width=8cm]{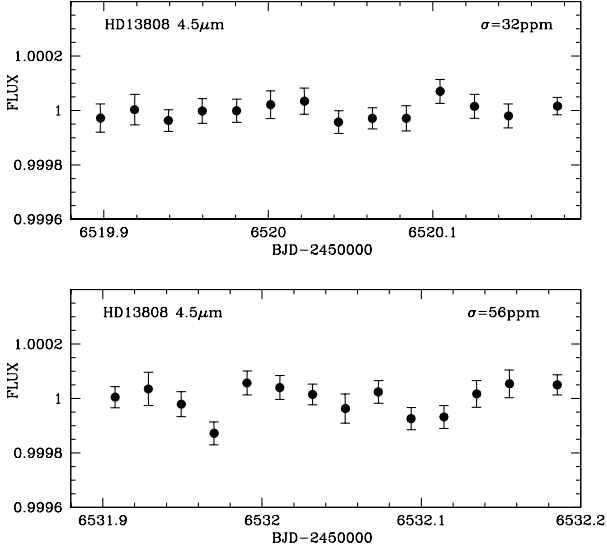}
\caption{Same as Fig. 1 for HD\,13808\,b. These data probed the 2-$\sigma$ transit window  of a planetary candidate that turned out to be a spurious signal with more RV data (see Sec. 5.4 for details). }
\end{figure} 

\subsection{HD\,20003}

HD\,20003 is a  late-G dwarf of magnitude $V$=8.4 for which M11 announced the detection by HARPS of two planets with periods of 11\,d and 33\,d.
Our analysis of the updated HARPS dataset for HD\,20003 not only confirmed these planets, but 
also revealed two longer period signals: one at $\sim$180\,d that was determined as originating from the stitching effect, and another, at $\sim$10\,yr, that originates from the magnetic cycle of the star (Udry et al. in prep.). 
Only HD\,20003\,b ($P$=11\,d) has a significantly eccentric orbit.  We used {\it Spitzer} to probe its transiting nature. We did not detect a transit (Fig. 6).  But, as can be seen in Fig. 6, we did not probe the latest part of its transit window, resulting in a small but still significant posterior full transit probability of 0.54\% (Table 1).

\begin{figure}
\label{fig:6}
\centering                     
\includegraphics[width=9cm]{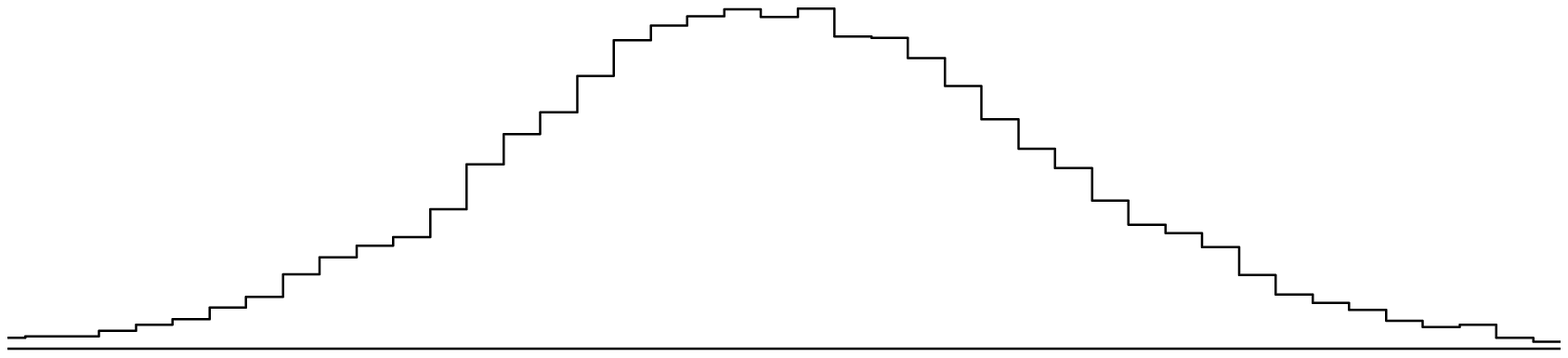}
\includegraphics[width=9cm]{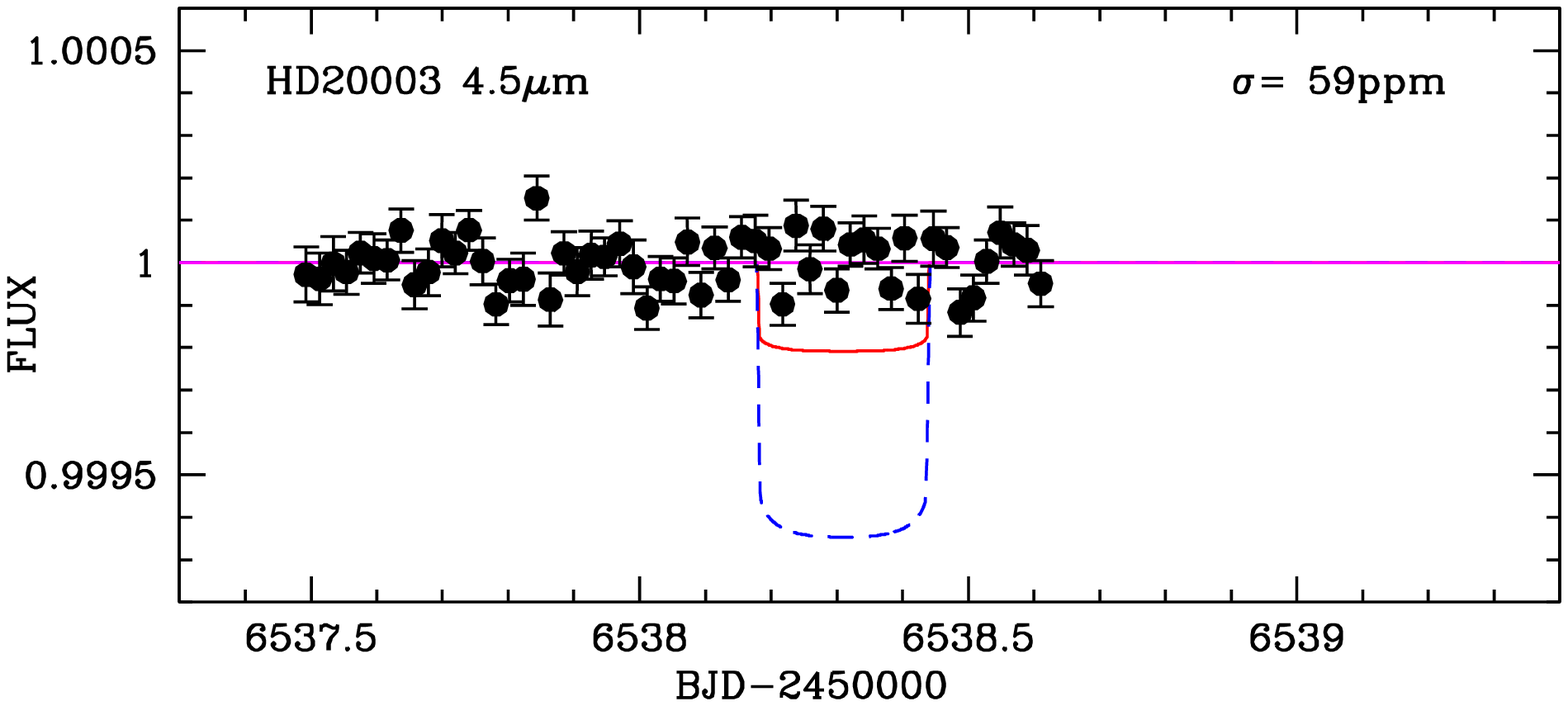}
\caption{Same as Fig. 1 for HD\,20003\,b.}
\end{figure}

\subsection{HD\,20781}

Based on 96 HARPS measurements, M11 announced the discovery of two Neptunes with orbital periods of 29.1\,d and 85.1\,d 
around this nearby $V$=8.4 K0-dwarf. The analysis of our much extended HARPS dataset (212 points) confirmed the existence of these two planets, while revealing the existence of two super-Earths in shorter orbits: a $M_p \sin{i} = 6.3 M_\oplus$ planet at 13.9\,d period, and  a $M_p \sin{i} = 2.1 M_\oplus$ planet at 5.3\,d period (Udry et al. in prep.). The low minimal mass and relatively high transit probability 
($>$7\%) of this latter planet made it an especially interesting target for a transit search, so we included it in the target list of our {\it Spitzer} program and observed one of its
transit window during Cycle 9. Because of the extreme faintness of the Doppler signal ($K \sim 90$ cm.s$^{-1}$), the orbital eccentricity of 
the planet is poorly constrained from the RVs alone, resulting in a particularly large transit window. As such, we assumed the orbit to be circular to
minimize the required {\it Spitzer} time, judging a circular orbit as  a reasonable assumption, taking into  consideration the strong tidal forces exerted by the star 
at such close distance and the compactness of the planetary system that would make  any significantly eccentric orbit unstable. 

Our {\it Spitzer} photometry did not reveal any transit-like structure (Fig. 7), the resulting posterior full transit probability being 0.15\% (Table 1). 
Still, our precision is not high enough to securely detect the transit of this very low-mass planet if its amplitude is below $\sim$150ppm, corresponding
to Mercury-like iron-rich compositions. Re-observing the planet's transit window at higher precision with, for example, CHEOPS will be mandatory to fully exclude a transiting configuration for the planet. 

\begin{figure}
\label{fig:7}
\centering                     
\includegraphics[width=9cm]{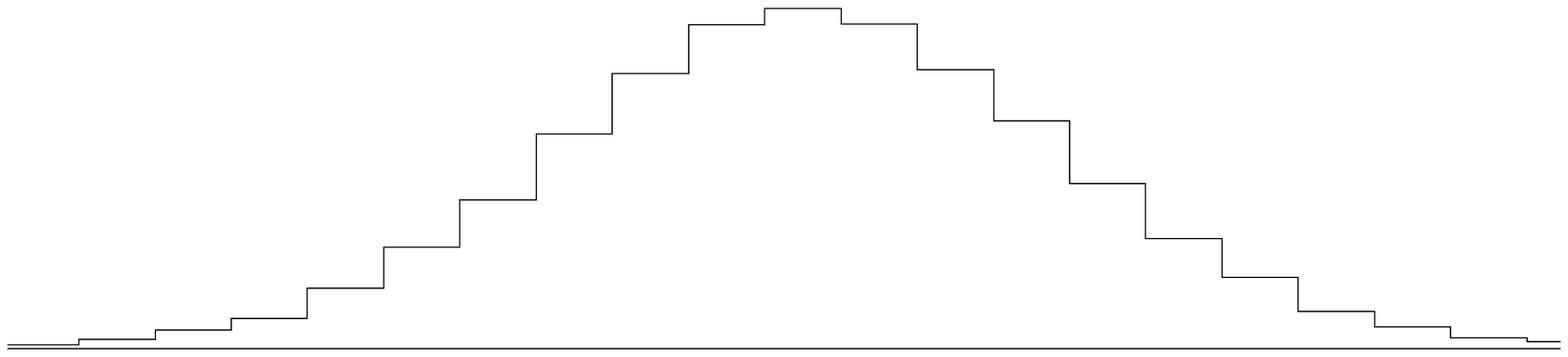}
\includegraphics[width=9cm]{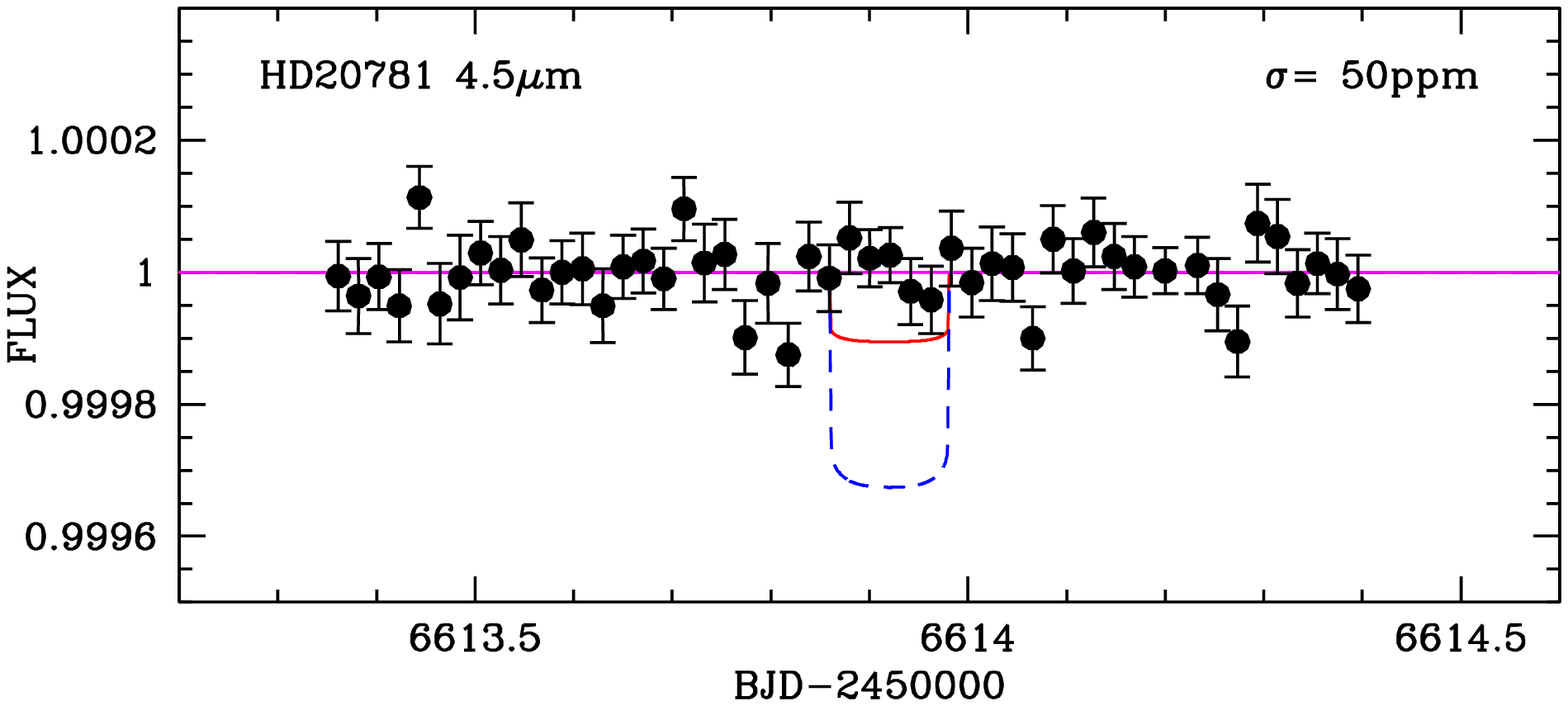}
\caption{Same as Fig. 1 for HD\,20781\,b.}
\end{figure}

\subsection{HD\,31527}

M11 announced  the existence of three Neptune-mass planets  around this nearby $V=7.5$ solar-type star, with orbital periods of 16.5\,d, 51.3\,d, and 275\,d. Our analysis of the extended HARPS dataset (242 vs 167 
measurements) confirmed the existence of these planets and improved their orbital parameters, while not revealing
any other planet (Udry et al. in prep.). We used $\sim$36 hr of continuous {\it Spitzer} observation to search for a transit of the innermost planet, HD\,31527\,b (geometric transit probability  = 4.4\%). The resulting light curve did not reveal any transit (Fig.~8), the resulting posterior full transit probability being of 0.45\% (Table 1). Compared to the orbital solution that we used to schedule our {\it Spitzer}  observations, the updated solution presented here (Table 2) results in {\it Spitzer} observations that are not well centered on the peak of the posterior PDF for the transit timing. The right wing of this PDF is thus unexplored. Its future exploration with, for example, CHEOPS would be desirable to fully exclude a transiting nature for the planet. 

\begin{figure}
\label{fig:8}
\centering                     
\includegraphics[width=9cm]{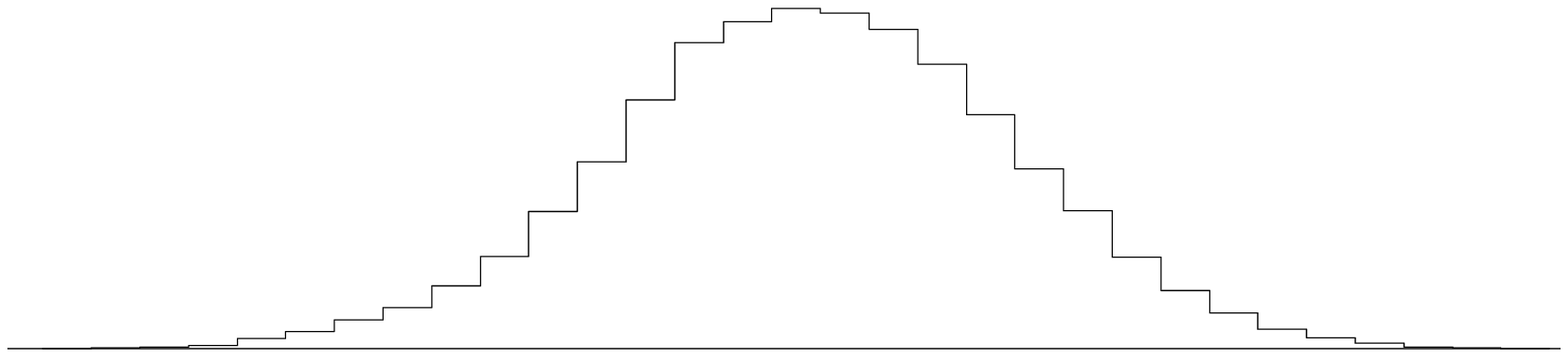}
\includegraphics[width=9cm]{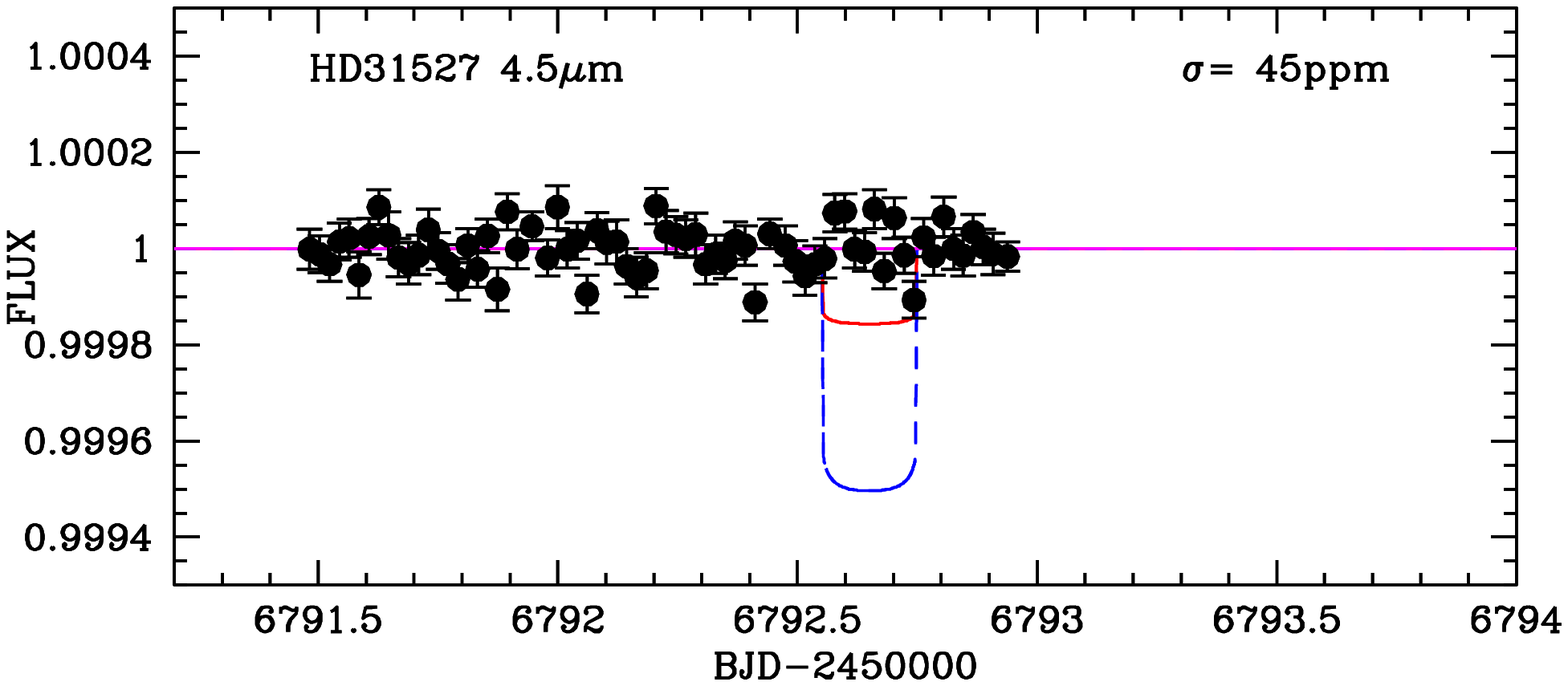}
\caption{Same as Fig. 1 for HD\,31527\,b.}
\end{figure}

\subsection{HD\,39194}

This star is a $V=8.1$ early K-dwarf around which HARPS detected three super-Earths  with orbital periods of
5.6\,d, 14.0\,d, and 33.9\,d (M11, Queloz et al. in prep.). Our analysis of the updated HARPS dataset (261 RVs vs  133
in the discovery paper) fully confirmed the existence of these planets, and revealed a low-amplitude trend. 
With a geometric transit probability of $\sim$6.5\%, theinnermost of these planets was an interesting target for 
our program, and we monitored one of its transit windows with {\it Spitzer} in Dec 2012. As shown in Fig. 9, the resulting 
light curve was flat, our deduced posterior transit probability being of 0.46\% (Table 1).

\begin{figure}
\label{fig:10}
\centering                     
\includegraphics[width=9cm]{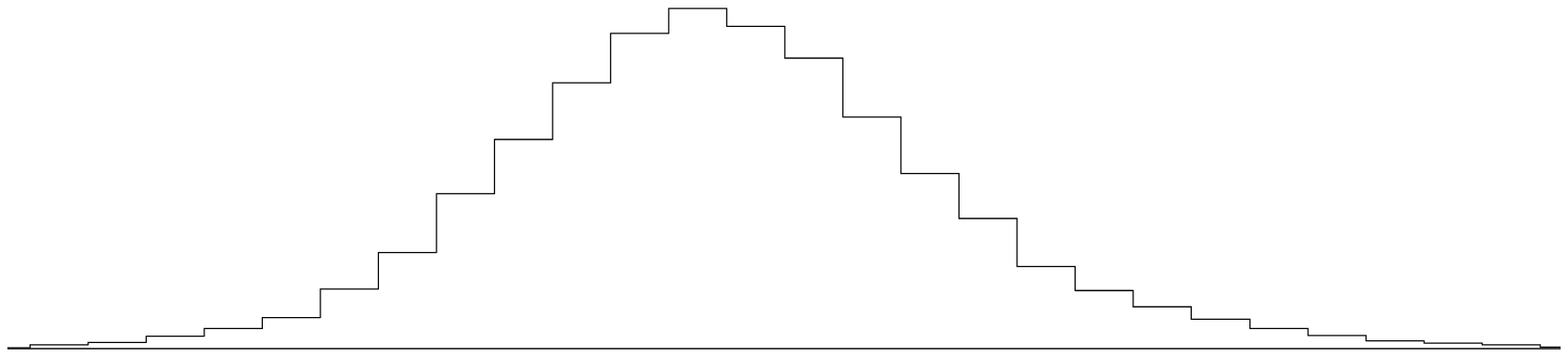}
\includegraphics[width=9cm]{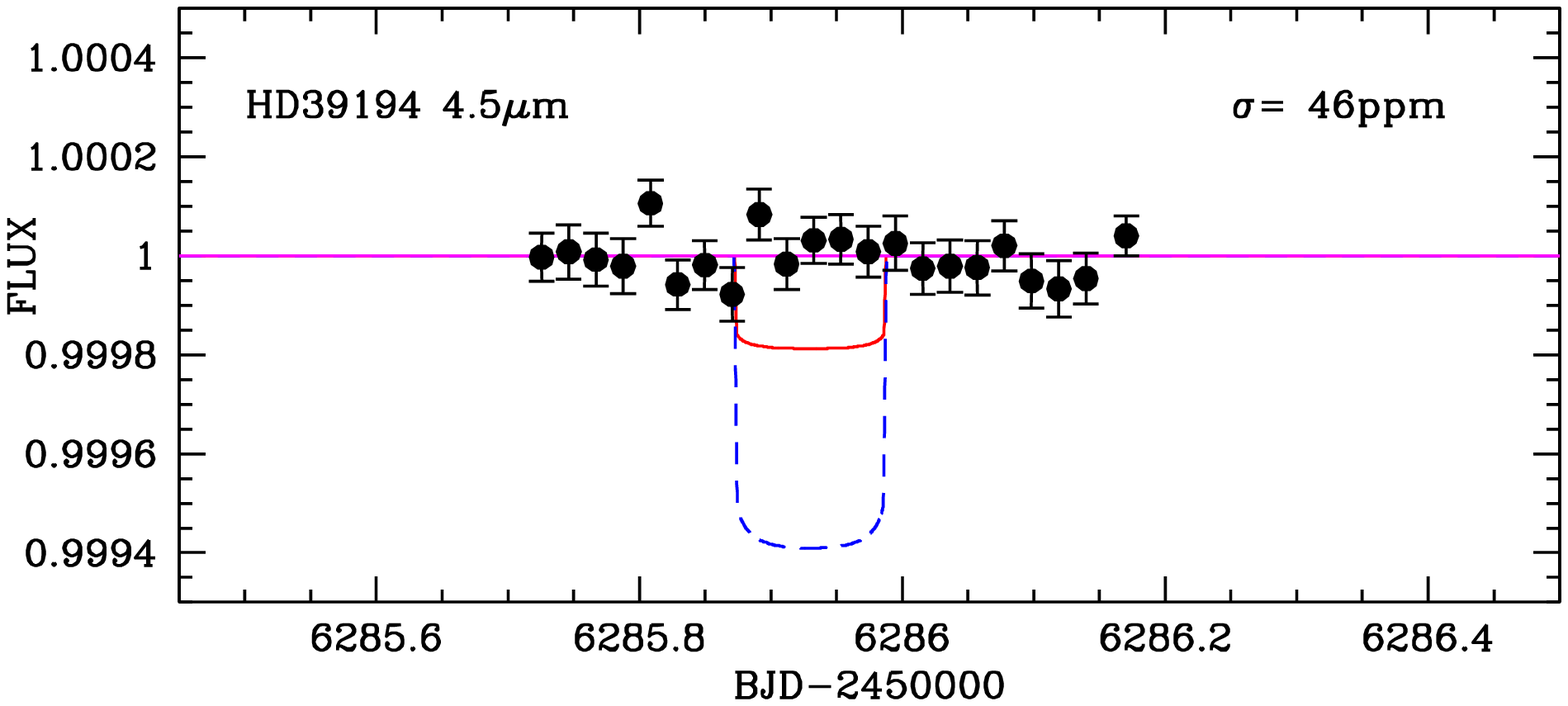}
\caption{Same as Fig. 1 for HD\,39194\,b.}
\end{figure}

\subsection{HD\,45184}

M11 reported the detection of a Neptune-mass planet on a 5.9~d period around this bright ($V=6.4$) 
solar-type star. Based on  more than double the HARPS measurements (174 vs 82), our analysis confirms 
the existence of this planet, while revealing the presence of a second planet of similar mass, $\sim$9.5 $M_\oplus$,
on a outer orbit ($P$=13.1~d), and a trend in the RVs that we could relate to the magnetic cycle of the star (Udry et al. in prep.).
 We monitored the star for  more than 11hrs with {\it Spitzer} to search for the transit of HD\,45184\,b. The resulting light
  curve is flat (Fig. 10), while its precision would have been high enough to detect the searched transit for any plausible composition. 
We did not explore the latest part of the transit window, leaving a posterior probability of 1.3\% that the planet undergoes full transit (Table 1), 
for a prior geometric probability of 7.7\%.

\begin{figure}
\label{fig:11}
\centering                     
\includegraphics[width=9cm]{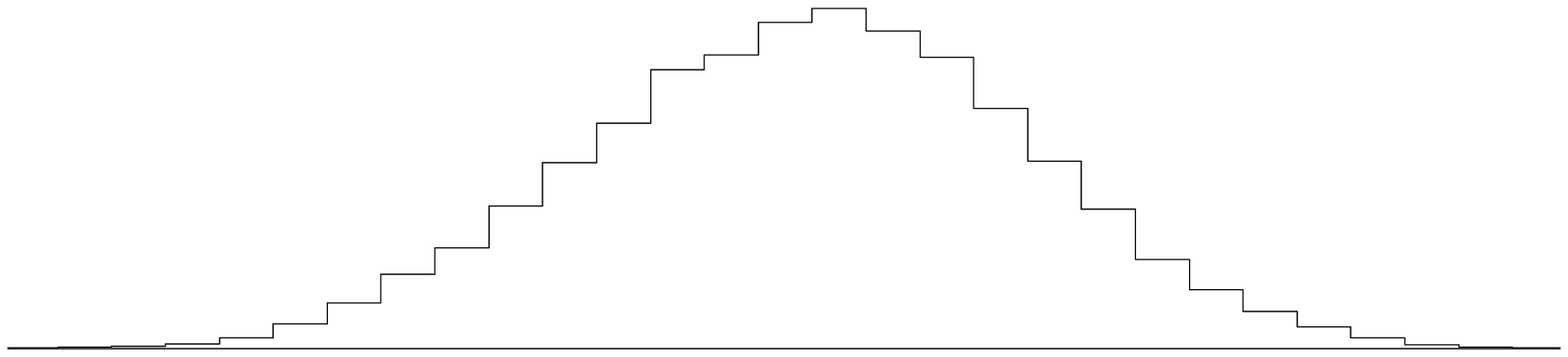}
\includegraphics[width=9cm]{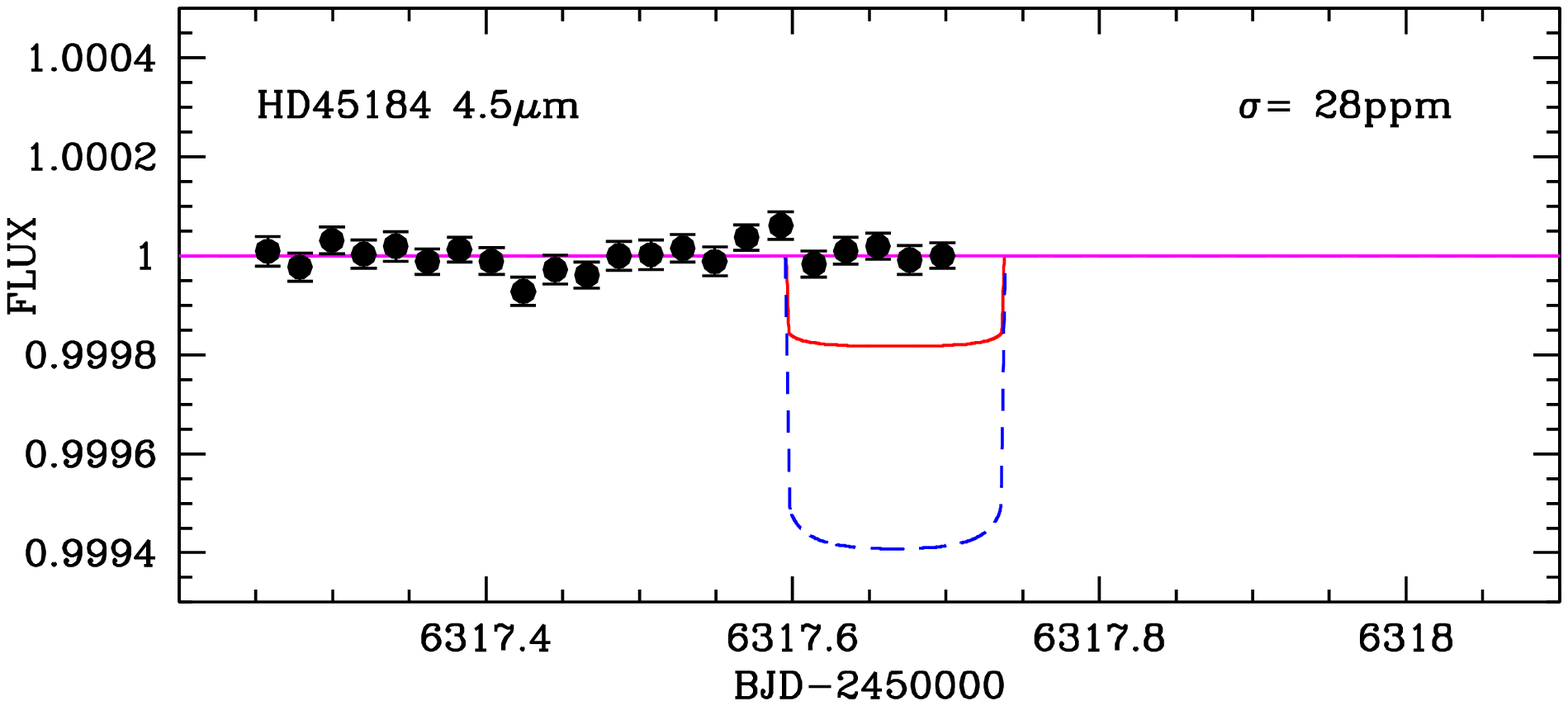}
\caption{Same as Fig. 1 for HD\,45184\,b.}
\end{figure}

\subsection{HD\,47186}

HD\,47186\,b is a short-period ($P$=4.08\,d) Neptune-mass planet discovered by HARPS in 2009 (Bouchy et al. 2009) around a $V=7.6$ 
solar-type star, which is also orbited by a giant planet at a much longer period. Our analysis of the extended HARPS dataset confirmed the existence of 
these two planets, and enabled us to derive an excellent precision on the time of inferior conjunction of the inner planet ($1\sigma$-error of 30 min for 
our {\it Spitzer} observations). We searched for its transit within our {\it Spitzer}  cycle 6 program. The resulting flat light curve (Fig. 11) enables us  to fully reject the transiting nature of the planet (Table 1). 

\begin{figure}
\label{fig:12}
\centering                     
\includegraphics[width=9cm]{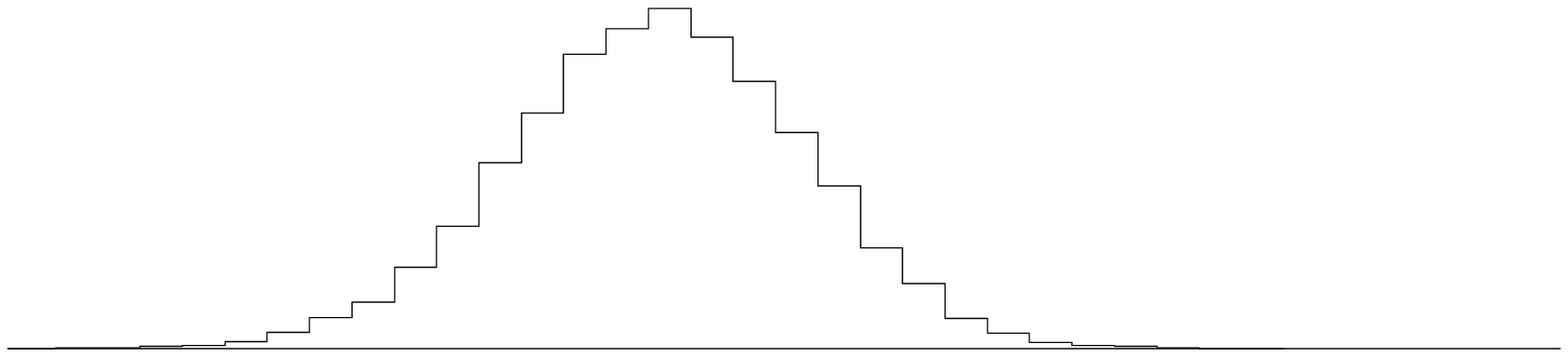}
\includegraphics[width=9cm]{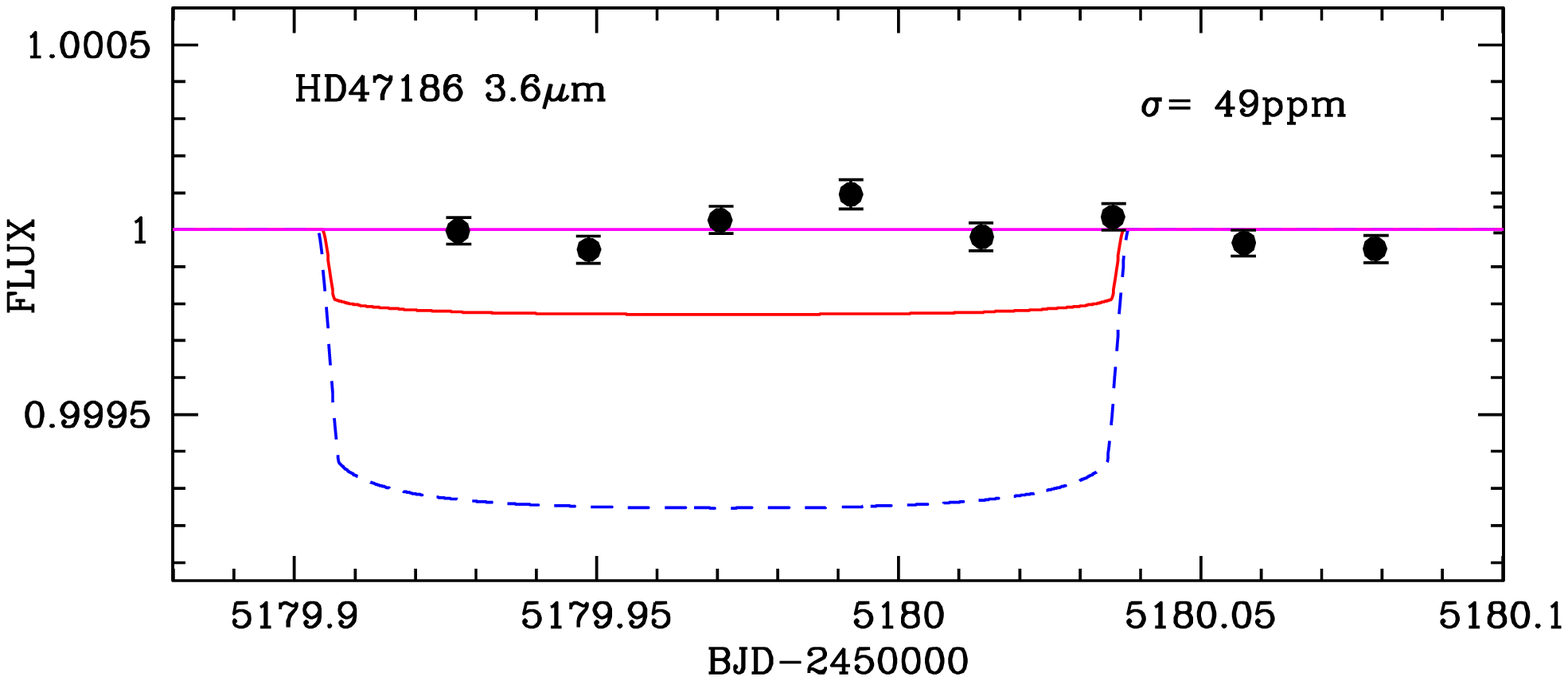}
\caption{Same as Fig. 1 for HD\,47186\,b.}
\end{figure}

\subsection{HD\,51608}

We used {\it Spitzer} to search for the transit of the $P=14.07$~d Neptune-mass planet HD\,51608\,b detected by M11
around this bright ($V=6.3$) late G-type dwarf. Our analysis of the updated HARPS dataset (210 RVs) confirmed the 
existence of the two planets ($P=14.07$d and $95$d) announced in M11, and favored a low-amplitude trend
of probable magnetic cycle origin (Udry et al. in prep.). Our {\it Spitzer} photometry (Fig. 12) enabled us to discard a 
transit of HD\,51608\, for any plausible planetary composition, our posterior full transit probability being 0.11\% (Table 1) for a
 prior geometric transit probability of $\sim$4\%.

\begin{figure}
\label{fig:13}
\centering                     
\includegraphics[width=9cm]{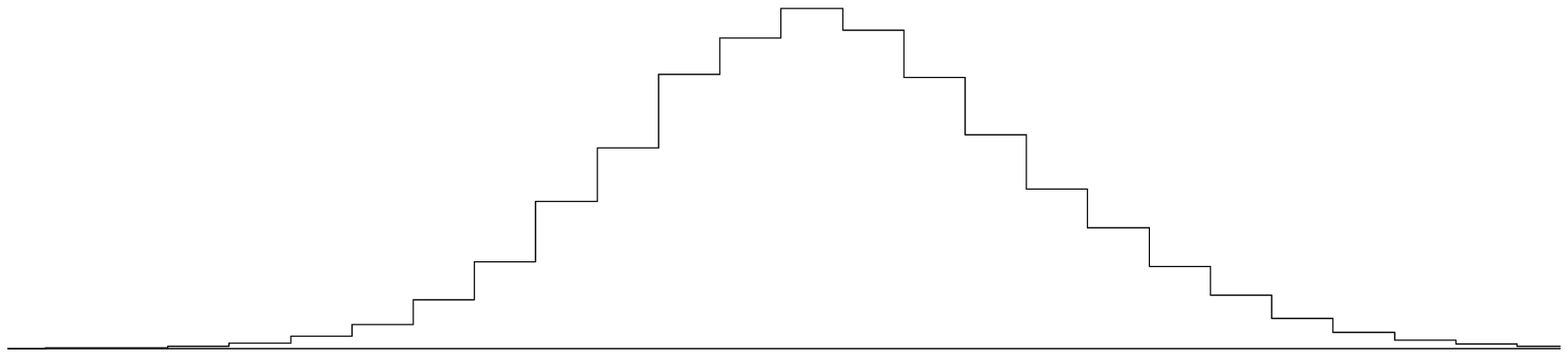}
\includegraphics[width=9cm]{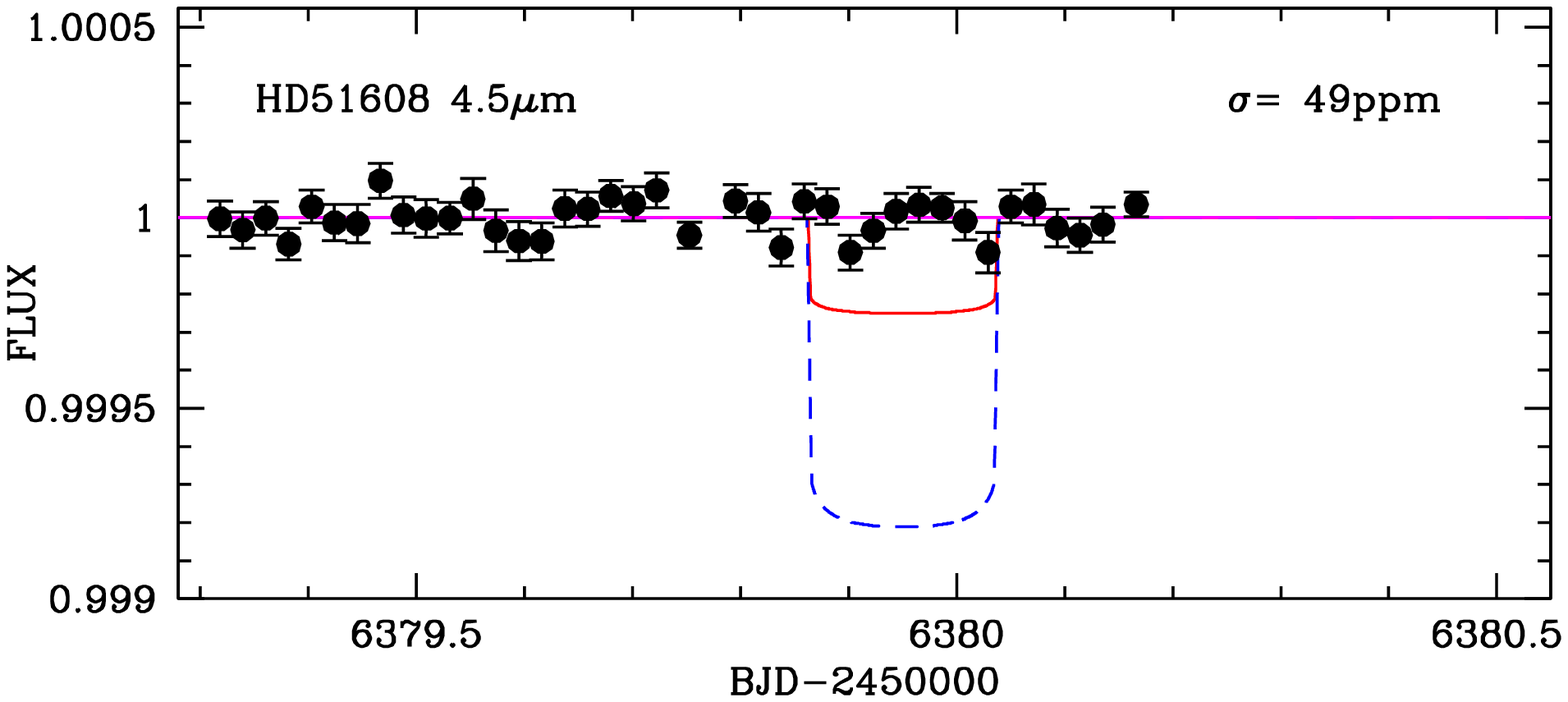}
\caption{Same as Fig. 1 for  HD\,51608\,b.}
\end{figure}

\subsection{HD\,93385}

The detection of two low-mass planets was reported by M11 for this $V=7.5$ solar-type star, with derived minimal masses of 8.4 and 10.1 $M_\oplus$ and orbital periods of 13.2~d and 46~d, respectively. Our analysis of the updated HARPS dataset (231 RVs)  revealed the existence of a third lower-mass ($Mp\sin{i} = 4.0 \pm 0.5 M_\oplus$) planet  on a 7.3~d orbit (Queloz et al. in prep., see Table A.3).  With {\it Spitzer}, we monitored a transit window ($\sim$29hr) of this new planet, HD\,93385\,d. Our resulting photometry (Fig. 13) did not reveal any clear transit-like structure, and the posterior full transit probability that we derived from its analysis is 0.23\% (Table 1), for a prior geometric transit probability of 7.9\%. By injecting transit models in this light curve and analyzing the results with our MCMC code, we concluded that its precision of $\sim$ 40 ppm per half-hour is good enough to discard any central transit of a planet with a density equal or smaller than Earth's, but cannot firmly discard the transit of denser planet. We thus recommend the re-observation of the transit window at higher photometrical precision with, for example, CHEOPS.
 
\begin{figure}
\label{fig:14}
\centering                     
\includegraphics[width=9cm]{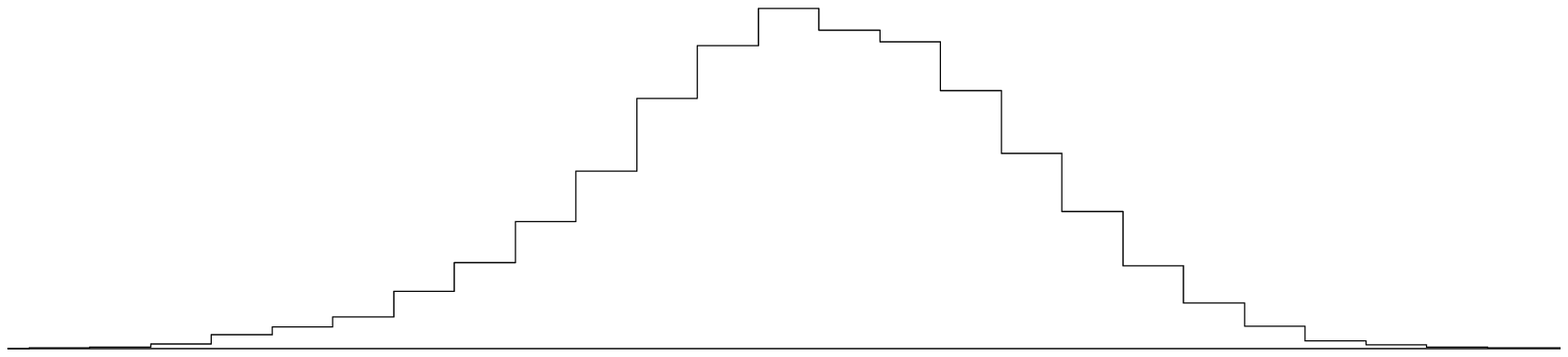}
\includegraphics[width=9cm]{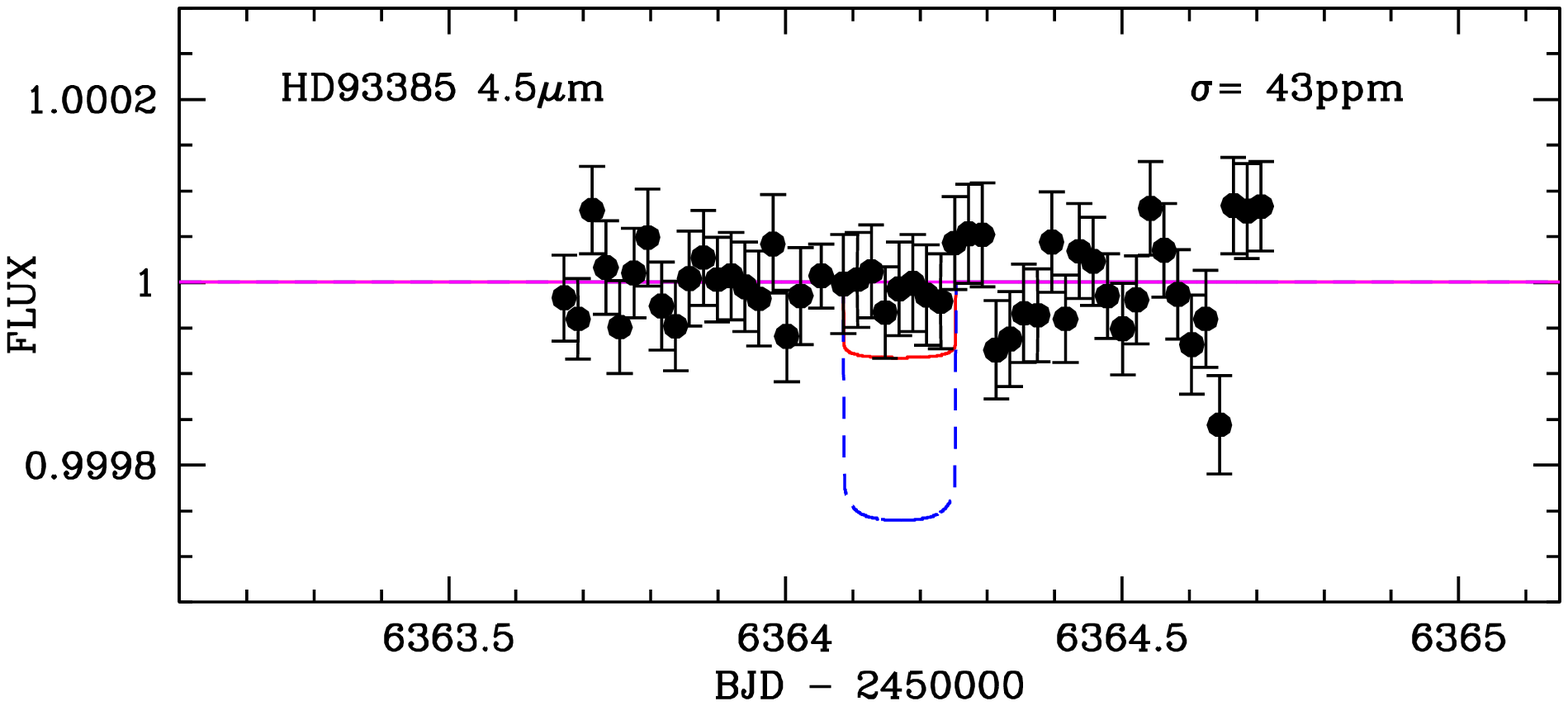}
\caption{Same as Fig. 1 for  HD\,93385\,d.}
\end{figure}

\subsection{HD\,96700}

HD\,96700\,b is a $\sim 9 M_\oplus$ planet on a 8.1~d orbit discovered by M11 around a $V=6.5$ solar-type star. 
Our analysis of the updated HARPS dataset (244 RVs) confirmed its existence and  the outer Neptune-mass planet found by M11 on a $\sim$100~d period. It also revealed a low-amplitude trend in the RVs of probable magnetic cycle origin (Queloz et al. in prep.). We searched for the transit of HD\,96700\,b with {\it Spitzer} in 2013, without success (Fig. 14).
The precision of the {\it Spitzer} light curve  is high enough to discard a transit of HD\,96700\,b in it for any possible planetary composition. The resulting posterior transit probability of the planet is 0.09\%, for a prior probability of 6.8\% (Table 1). 

\begin{figure}
\label{fig:15}
\centering                     
\includegraphics[width=9cm]{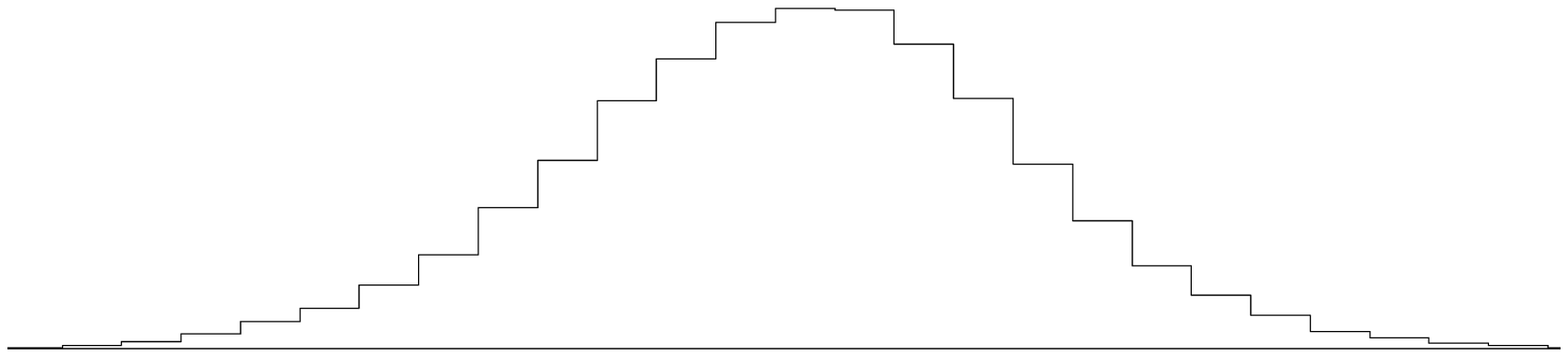}
\includegraphics[width=9cm]{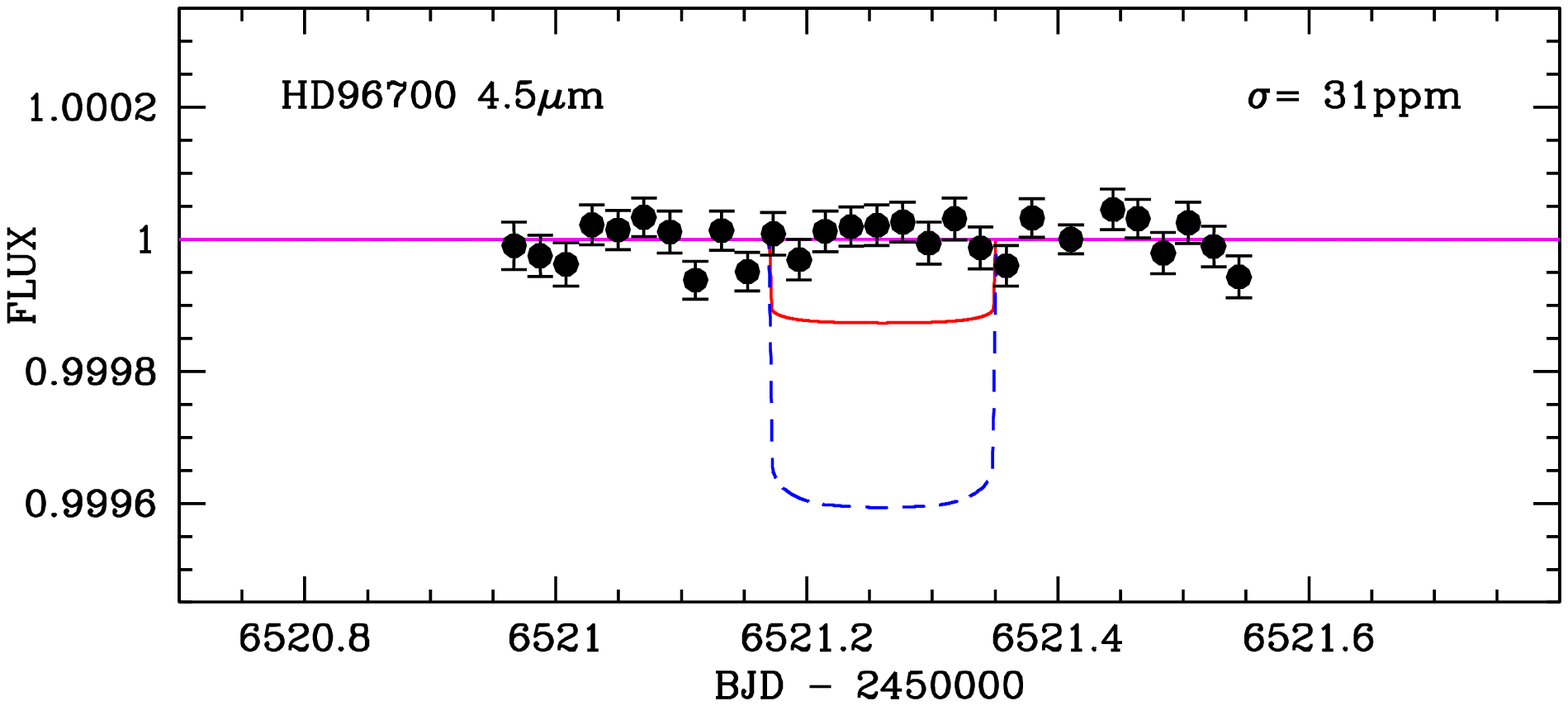}
\caption{Same as Fig. 1 for  HD\,96700\,b.}
\end{figure}

\subsection{HD\,115617}

HD\,115617 (aka 61 Vir) is a $V$=4.7 solar-type star (G5V) at only 8.5 pc from Earth. 
A close-in super-Earth ($M_p \sin i = 5 M_\oplus$, $P$ = 4.215 d) and two outer Neptunes ($M_p \sin i  = 
18$ \& $23 M_\oplus$, $P = 38$ \& $123$ d) were discovered around it by Vogt et al. (2010) using RVs obtained 
with Keck/HIRES and the Anglo-Australian Telescope (AAT). We performed a global analysis of the Keck, and AAT RVs that confirmed the existence of the three planets, without revealing any additional object orbiting the star. 
We used {\it Spitzer} to observe a transit window of HD\,115617 in March 2010.

With $K=2.96$,  HD\,115617 is an extremely bright star for {\it Spitzer}. 
At 4.5 $\mu$m, it is nevertheless faint enough to be unsaturated for the shortest available 
integration time (0.01s). However, the {\it Spitzer} Science Center (SSC) informed us that the requested long observation of HD\,15617 could not be performed at 4.5 $\mu$m for technical reasons. At 3.6 $\mu$m, the star has a flux density 
of $19000 \pm 4000$ mJy while the saturation limit is 20000 mJy for an integration time of 0.01s. 
SSC informed us the observations could still be performed without any risk of saturation if the star was 
placed at the corner of four pixels, leading to an effective flux density that was low enough to avoid saturation. We decided to test this strategy.  
Unfortunately, the  light curve corrected for known systematic effects  is corrupted by a clear variability
at the level of a few tens of percent (Fig. 15, top).  A Lomb-Scargle periodogram (Press et al. 1992) of the corrected photometry 
reveals a power excess at $\sim$0.11 days (Fig~15, bottom). We could not identify the origin of this variability. 
Nevertheless, an astrophysical origin is very unlikely, since HD\,115617 is known to be an old inactive star (Vogt et al. 2010). 
This variability probably originates from the near-saturation of the detector. Indeed, the  center of 
the stellar image was not located at the corner of four pixels as intended.
The instrumental effect at work has a typical timescale similar to the signal we are trying to detect, and its amplitude is much
 larger than the one of the searched transit. Even worse, it seems highly variable in nature and we could not find any analytical function
 of external parameters (PSF peak value, center position and width, background, etc.) able to represent it satisfactorily.  
 Without a thorough understanding of the effect, we concluded that searching for a transit of 
  a few hundreds of ppm with such data was illusory. The transiting nature of HD\,115617\,b thus remains unconstrained by our project. 
   
\begin{figure}
\label{fig:17}
\centering                
\includegraphics[width=9cm]{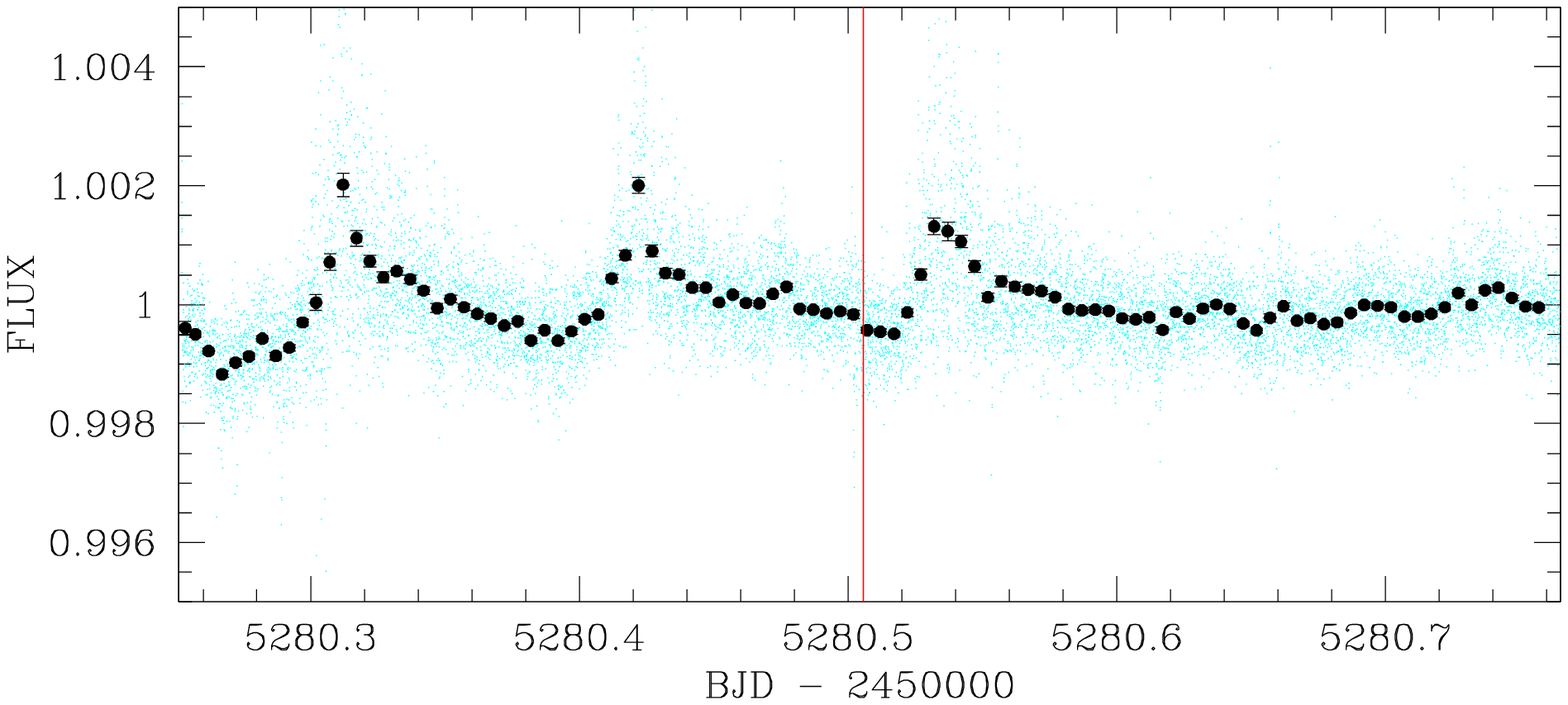}
\includegraphics[width=9cm]{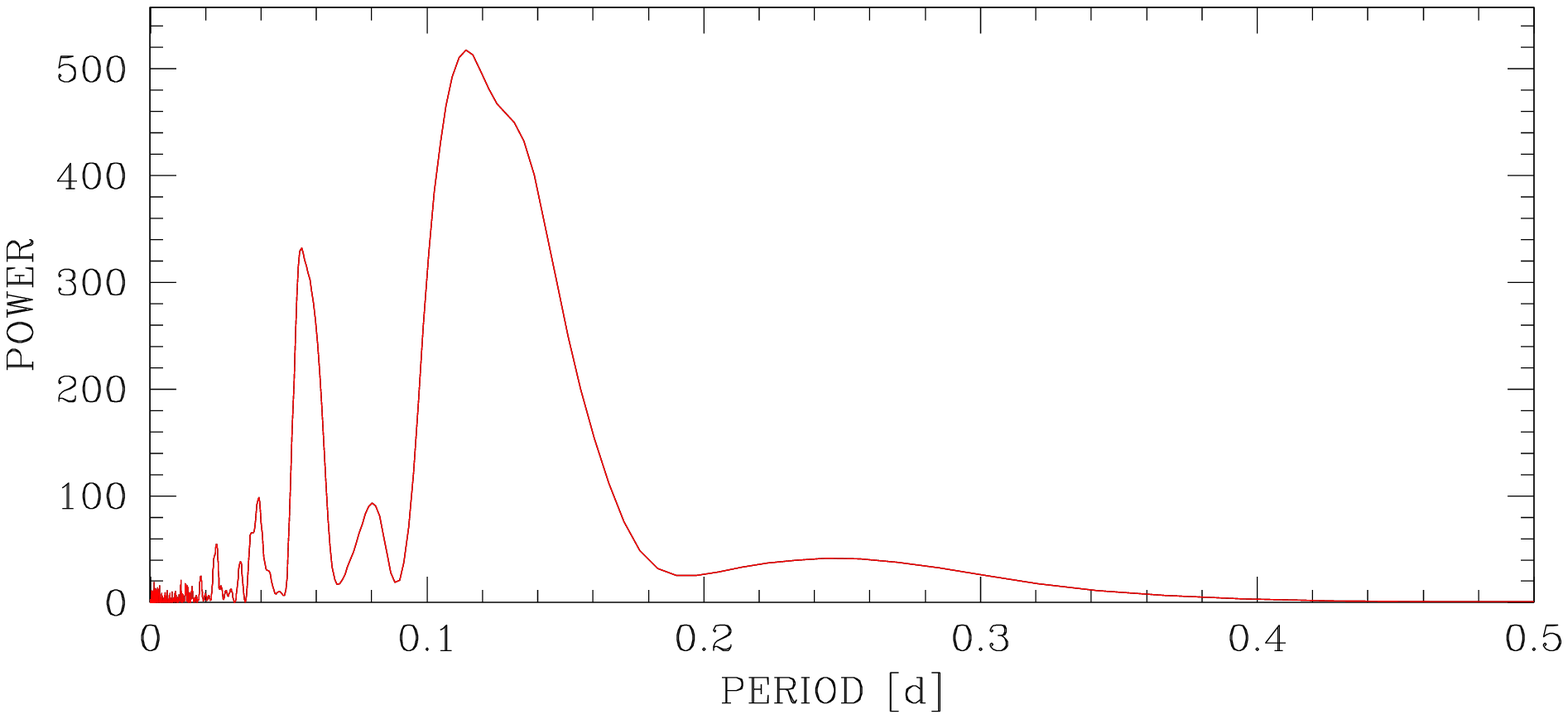}
\caption{$Top$: 61\,{\it Vir} light curve  obtained by {\it Spitzer} to search for the transit of its planet b, divided for the best-fit phase-pixel model (here a third order $x$ and $y$-position polynomial), unbinned (green dots) and binned to intervals of 0.005d (7.2 min). The start of the second AOR is represented by the red vertical line. $Bottom$: Lomb-Scargle periodogram showing a clear power excess at $\sim$0.11 days and its first harmonic.}
\end{figure}   

\subsection{HD\,125612}
HD\,1215612 is a $V=8.3$ solar-type star around which three planets have been detected so far by RVs, first a $P \sim$ 510~d gas giant  in 2007 (Fischer et al.) with Keck+HIRES data, then  a close-in ($P$=4.15~d) Neptune-mass planet and a $\sim$3000~d-period gas giant with HARPS (Lo Curto et al. 2010). In April 2010, we used {\it Spitzer} to search for the transit of the close-in planet HD\,125612\,c. Our initial light curve showed a structure consistent with a transit, so we triggered new observations of the transit window with {\it Spitzer} in September 2010 that did not confirm the transit signal. Reanalyzing the April data, we noticed that the transit signal disappeared  when we included terms in the PSF widths, revealing that the signal originated from the {\it Spitzer} PSF breathing effect (Lanotte et al. 2014). Our global analysis of the two {\it Spitzer} light curves allowed us to discard a transiting configuration for the planet for any possible planetary composition (Fig. 16), the resulting posterior probability for a full transit being of only 0.24\%, for a prior probability of 9.7\% (Table 1). 
We note that the standard deviations of our {\it Spitzer} light curves binned per 30 min intervals are the largest for HD\,125612 (92 and 102 ppm, see Fig. 16). This is also the case for the RV  jitter measured from the HARPS RVs (3.2 \ms), suggesting that HD\,125612 is significantly more active star than our other targets. 

\begin{figure}
\label{fig:18}
\centering                     
\includegraphics[width=9cm]{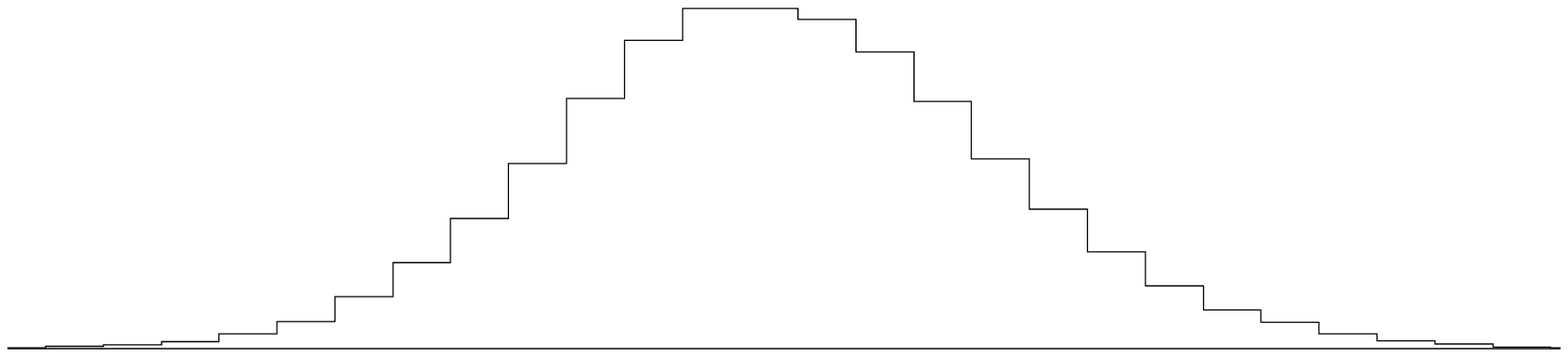}
\includegraphics[width=9cm]{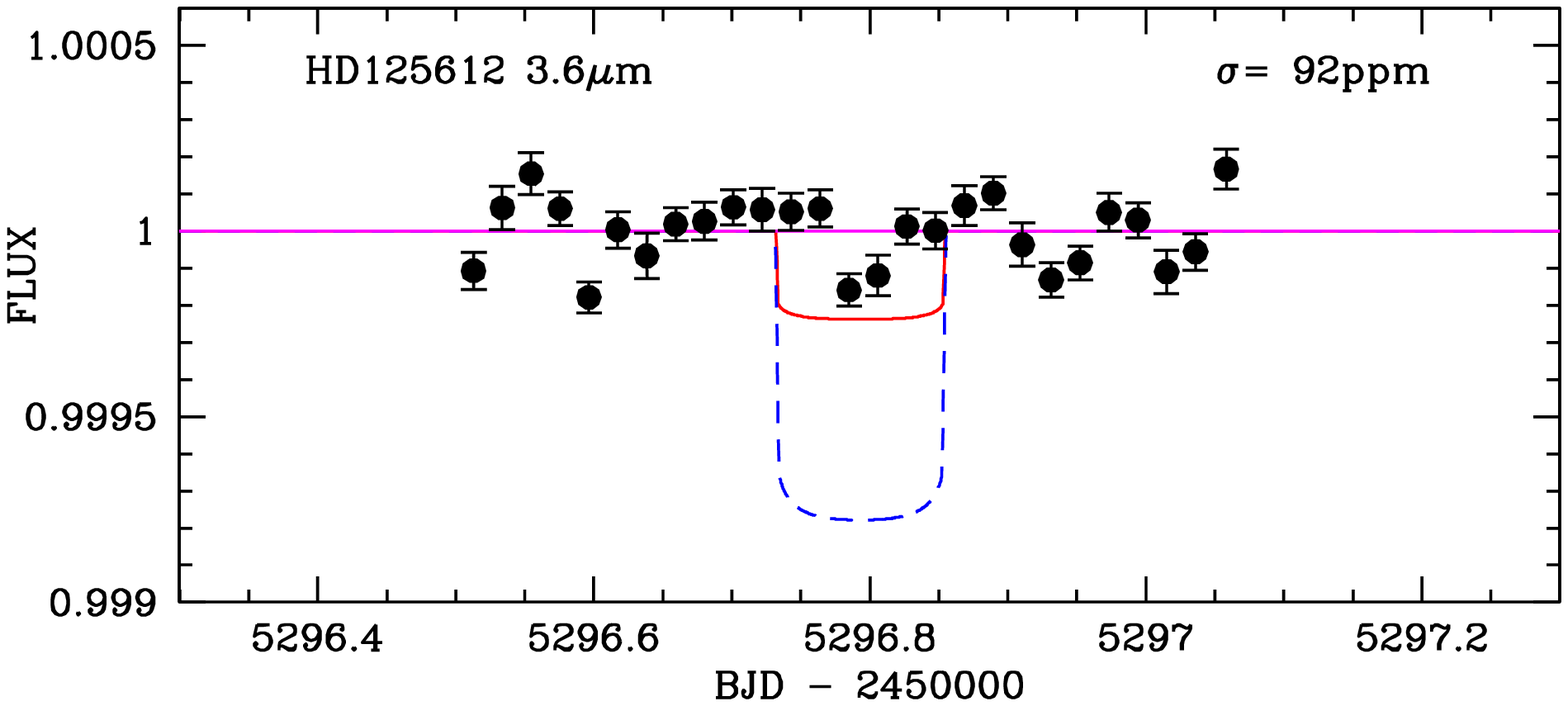}
\includegraphics[width=9cm]{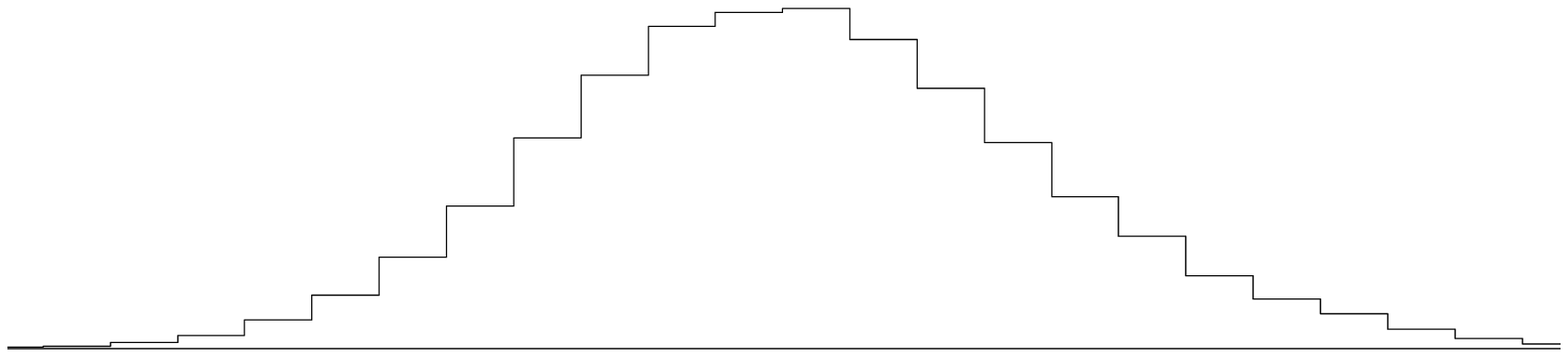}
\includegraphics[width=9cm]{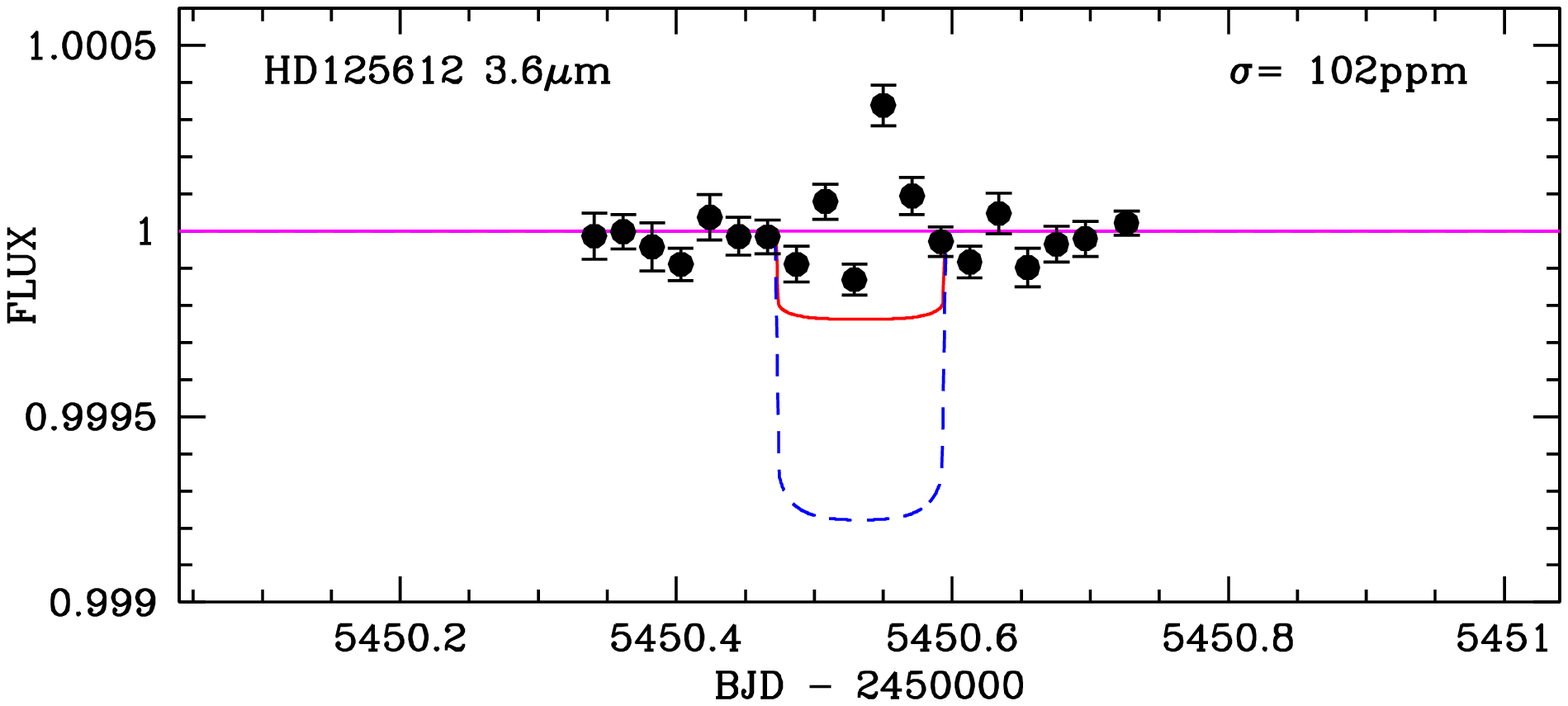}
\caption{Same as Fig. 1 for HD\,125612\,c.}
\end{figure} 

\subsection{HD\,134060}

HD\,134060 is a G0-type dwarf with a $V$-magnitude of 6.3 at 24pc from Earth, for which M11 and Udry et al. (in prep.) reported the discovery by HARPS of a Neptune (minimal mass = 11 $M_\oplus$) on a 3.3d orbit, HD\,134060\,b, in addition to a giant planet on a much wider orbit. Interestingly, the HARPS RVs showed that HD\,134060\,b has a significant eccentricity of $0.40 \pm 0.04$, despite its very short orbit, reminiscent of the still poorly understood eccentricity of the hot Neptune prototype GJ\,436\,b (Lanotte et al. 2014). Our analysis 
of the updated HARPS dataset confirmed this significant eccentricity. The orientation of the elliptic orbit of the planet made its occultation by its host star much more likely than its transit (23 vs 8\%, see Table A.4), and its timing much better constrained. We thus decided to use a few hours of {\it Spitzer} first to try to detect the occultation of the planet. We choose to perform the measurement at 3.6$\mu$m, based on the clear detection of the occultation of GJ\,436\,b at 3.6 $\mu$m and its non-detection at 4.5 $\mu$m (Stevenson et al. 2010, Lanotte et al. 2014). Assuming black-body spectra for the star and the planet, we estimated the expected occultation depth to range from a few tens to more than 200ppm, depending on the unknown planet's size, albedo, and heat-distribution efficiency. The resulting light curve (Fig. 17, bottom panel) did not reveal a single eclipse, but its precision was not high enough to discard an occultation of the planet for a large range of the plausible planetary parameters cited above. We then attempted a transit search, this time at 4.5 $\mu$m.  The resulting light curve was flat (Fig. 17, top panel) and enabled us to fully discard a full transit of the planet (Table 1).

\begin{figure}
\label{fig:19}
\centering                     
\includegraphics[width=9cm]{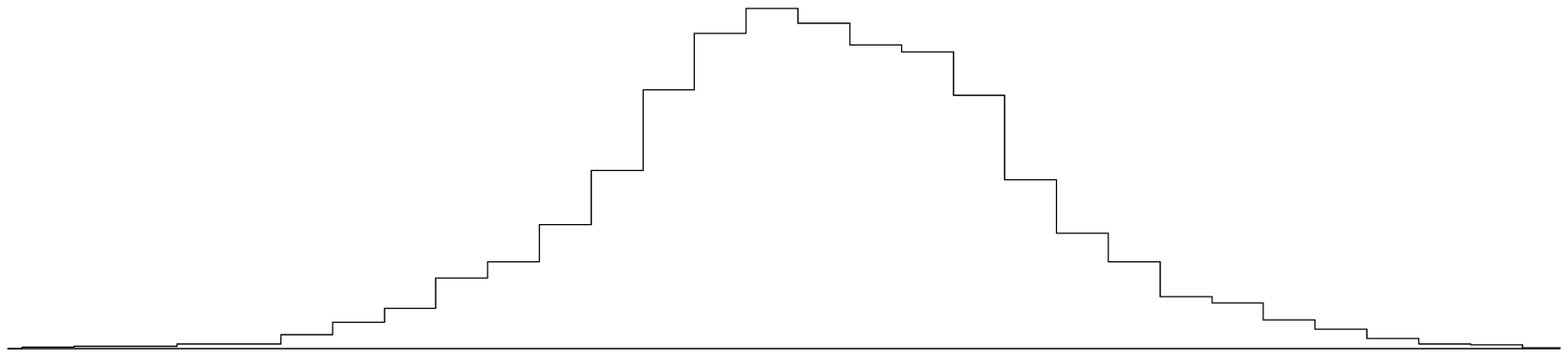}
\includegraphics[width=9cm]{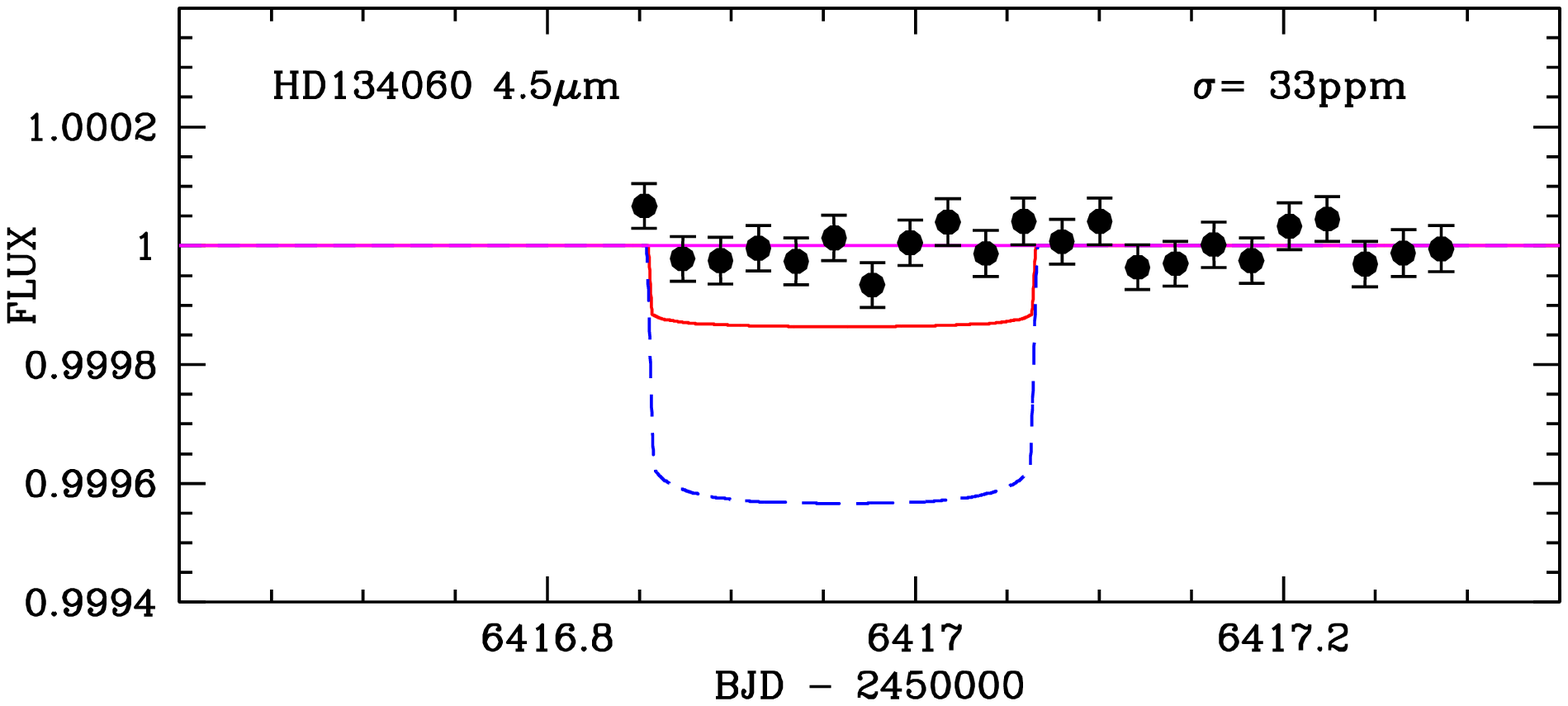}
\includegraphics[width=9cm]{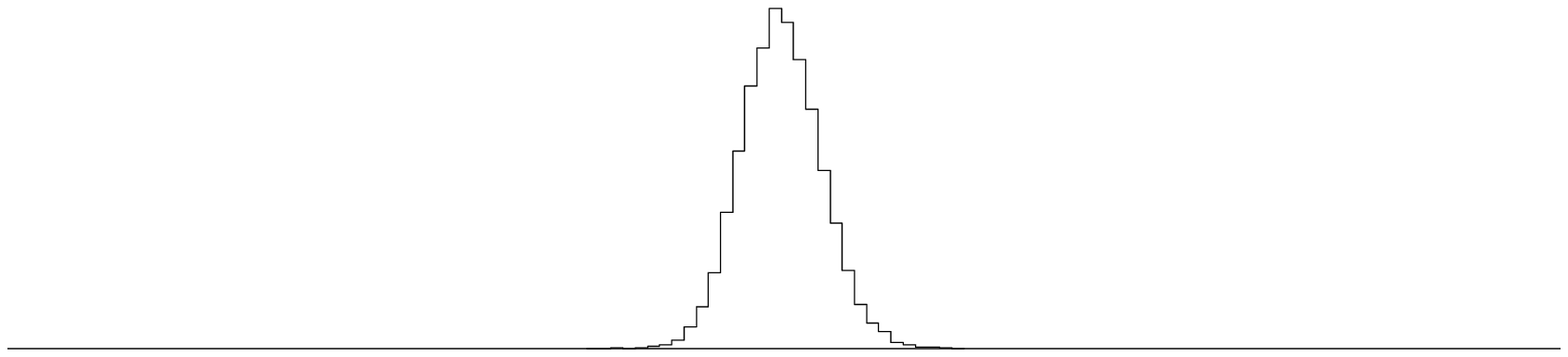}
\includegraphics[width=9cm]{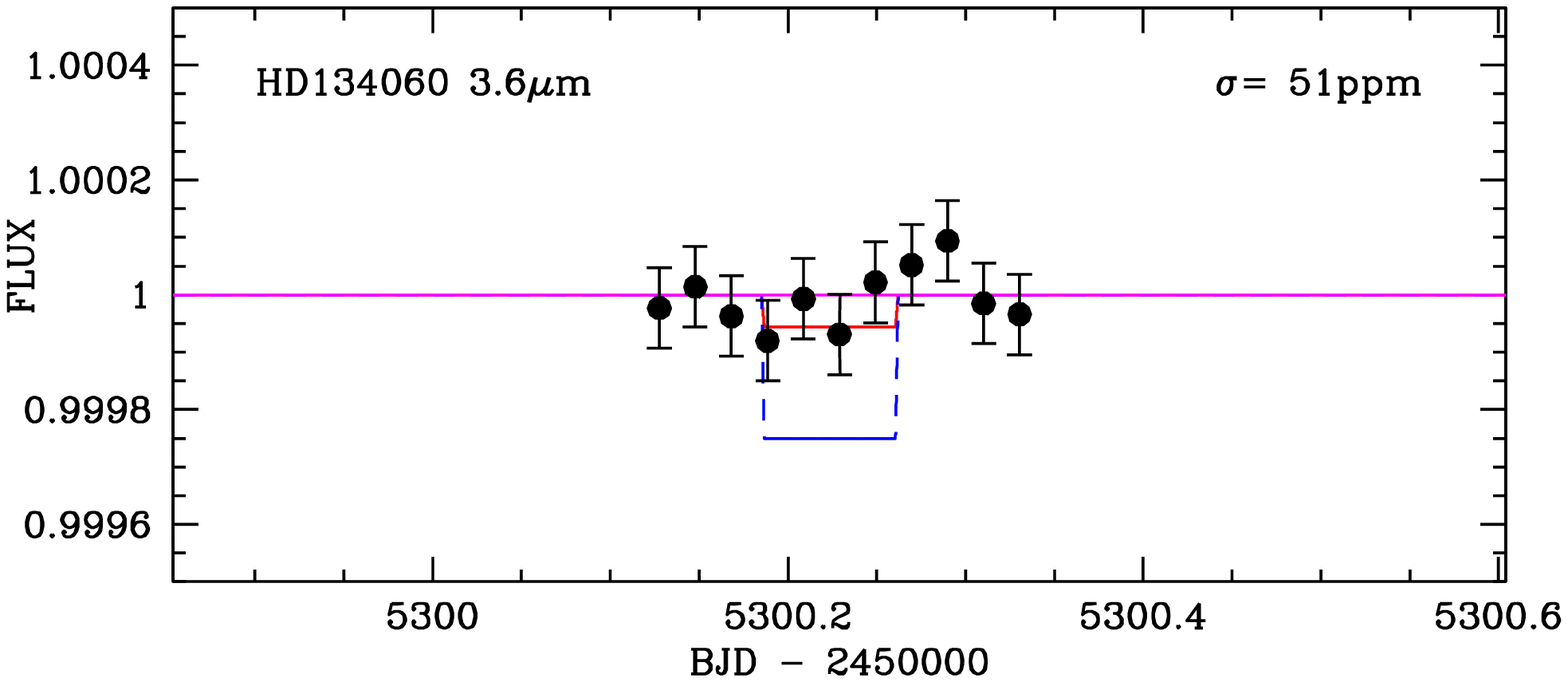}
\caption{$Top$: same as Fig. 1 for HD\,134060\,b. $Bottom$: same for the planet's occultation.  The red solid line and blue dashed line show, respectively, models for a central occultation of a 1.85 $R_\oplus$ (Earth-like composition) and 5 $R_\oplus$ (H-dominated composition) planet,  assuming  a null albedo, an inefficient heat distribution to the night side, and negligible tidal effects for the planet, and assuming black-body emissions for both the planet and its star. The time range of the $x$-axes of both panels correspond to the same duration so as to outline the fact that the transit timing was much less constrained by the RVs than the occultation timing.}
\end{figure} 

\subsection{HD\,181433}
Three planets were detected by HARPS around this $V=8.4$ K-dwarf (Bouchy et al. 2008). Our analysis of the updated HARPS dataset confirmed the existence of these three planets, and resulted in a prior transit probability of 5\% for the inner most of them, the super-Earth HD\,181433\,b ($M_p \sin i \sim 7.5 M_\oplus$, $P$=9.375d). Our {\it Spitzer} observations of one of its transit windows led to a transit light curve that did not reveal any transit signature (Fig. 18), while being precise enough to discard a transit for any possible composition of the planet. Our global MCMC analysis of the RVs and {\it Spitzer} photometry led to a posterior full transit probability of 0.52\% (Table 1), which is probably large enough to justify a future exploration of the first part of the transit window that was left unexplored by our {\it Spitzer} observations (see Fig. 18).

\begin{figure}
\label{fig:20}
\centering                     
\includegraphics[width=9cm]{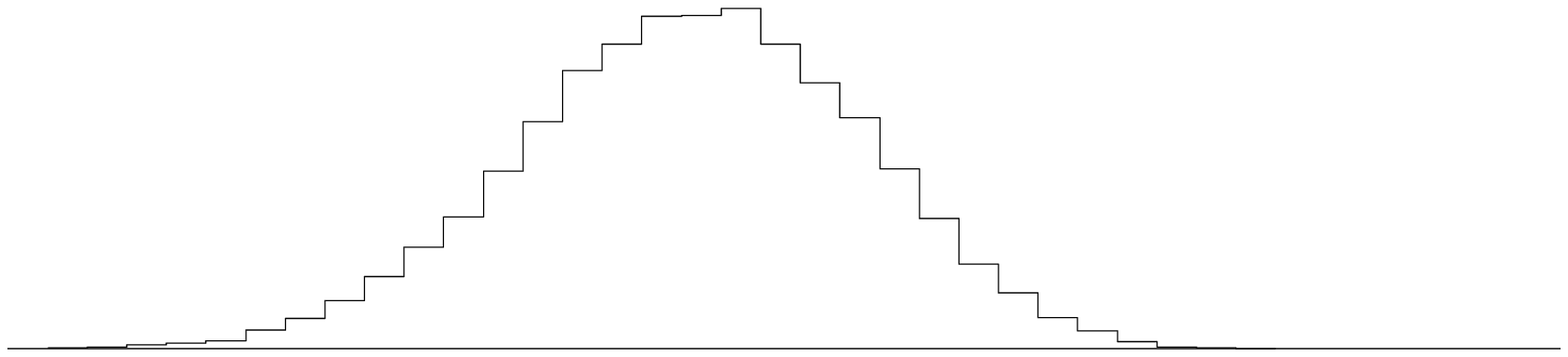}
\includegraphics[width=9cm]{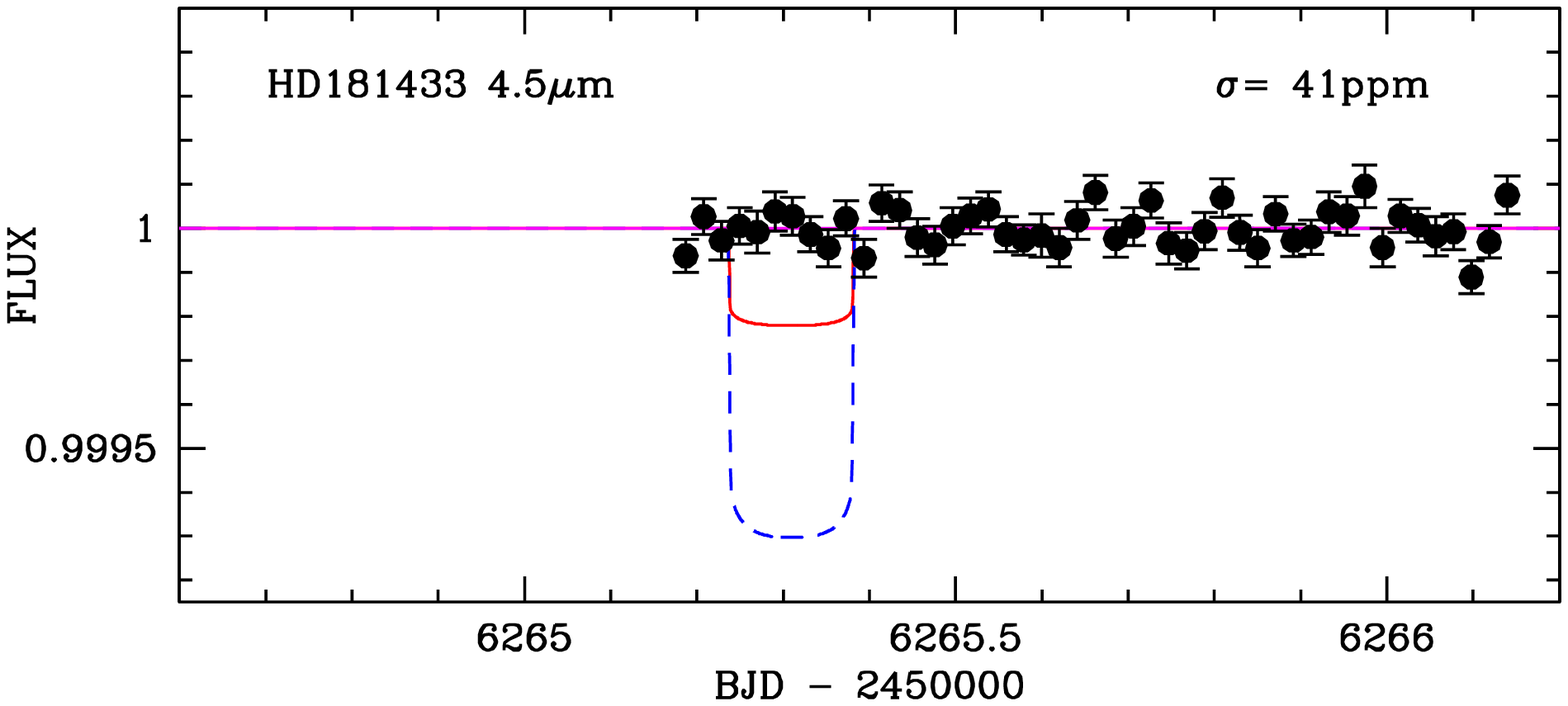}
\caption{Same as Fig. 1 for  HD\,181433\,b.}
\end{figure} 

\subsection{HD\,215497}

Two planets were detected by HARPS around this nearby ($\sim$44 pc) K-dwarf  (Lo Curto et al. 2010), one of them being a giant planet on a significantly eccentric 568d-orbit, and the other being a close-in ($P=$3.93d) super-Earth ($M_p \sin i = 6 M_\oplus$). Our analysis of the updated HARPS dataset, containing only three additional RVs, confirmed the existence of both planets. Based on its interestingly high transit probability of $\sim$ 12\% the inner super-Earth HD\,215497\,b, we targeted it in our {\it Spitzer} transit search. The resulting light curve did not reveal any transit signature and was precise enough to discard a transit for most plausible compositions. Yet, a high impact parameter transit of an iron-dominated planet is not discarded by the data (see Fig. 19), resulting in a small but significant posterior full transit probability of $\sim$ 0.31\% (Table 1). 

\begin{figure}
\label{fig:21}
\centering                     
\includegraphics[width=9cm]{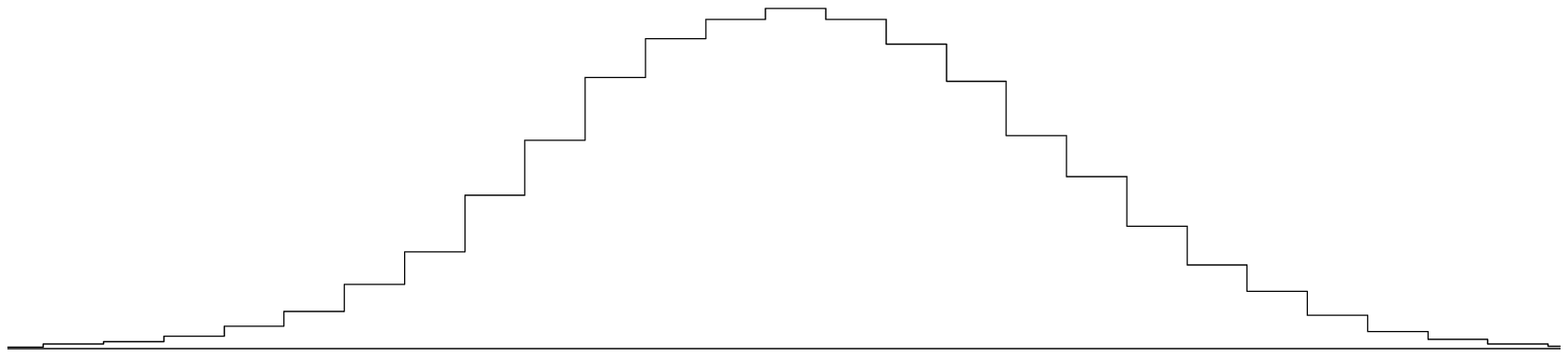}
\includegraphics[width=9cm]{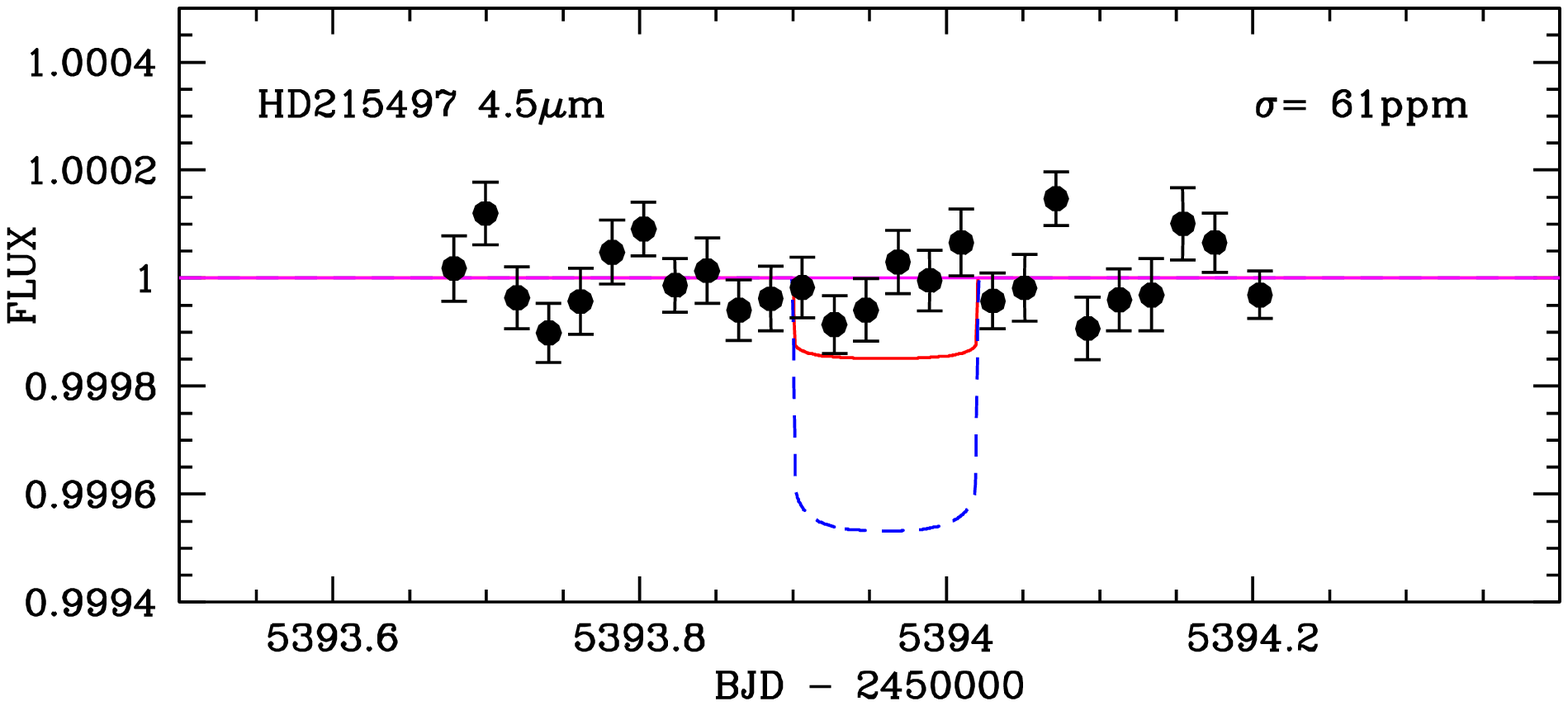}
\caption{Same as Fig. 1 for HD\,215497\,b.}
\end{figure} 

\subsection{HD\,219828}

Discovered by Melo et al. in 2007, HD\,219828\,b is a hot Neptune ($M_p \sin i = 20 M_\oplus$) with an interestingly high geometric transit probability of $\sim$14\%, thanks to its close-in orbit ($a$=0.05 au, $P$=3.83d), combined with the relatively large size of its evolved G0-type host star ($R_\ast = 1.6 R_\odot$). On the other hand, this large stellar size translated in expected transit depths as small as 100 ppm (pure iron composition), so we monitored two transit windows to reach a photometric precision that was high enough to firmly constrain the (non-)transiting nature of the planet.  The resulting light curves, both obtained at 4.5 $\mu$m, are shown in Fig. 20. They do not reveal any transit signature. Our global analysis of the RVs + photometry led to a complete rejection of a full transit configuration (Table 1-. Our analysis of the much extended HARPS dataset compared to the HD\,219828\,b discovery paper (91 vs 22 RVs) confirmed the existence of a second, more massive planet ($M_p \sin i \sim 15  M_{Jup}$) planet on a very eccentric ($e=0.81$)  long-period ($P$=13.1 years) orbit, as recently announced by Santos et al. (2016). 

\begin{figure}
\label{fig:22}
\centering                     
\includegraphics[width=9cm]{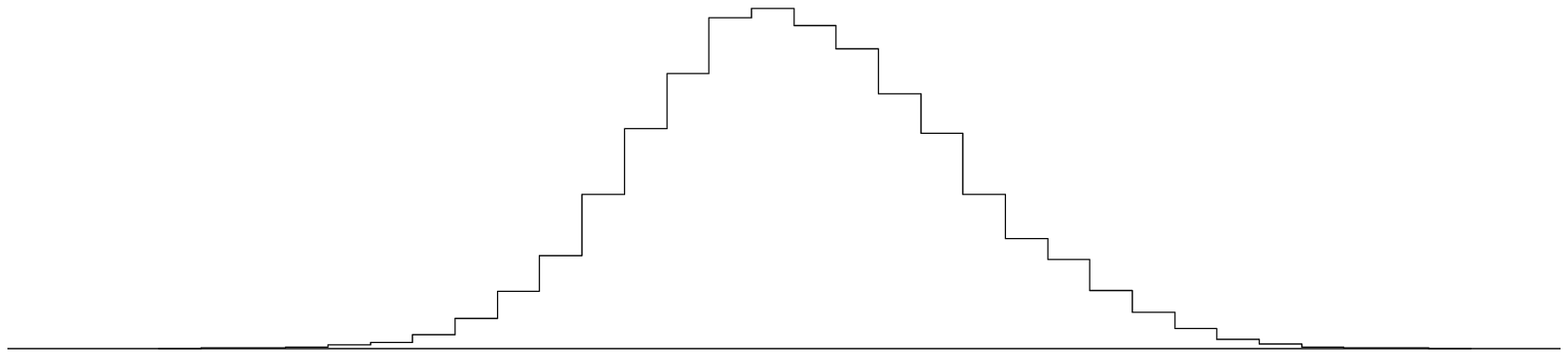}
\includegraphics[width=9cm]{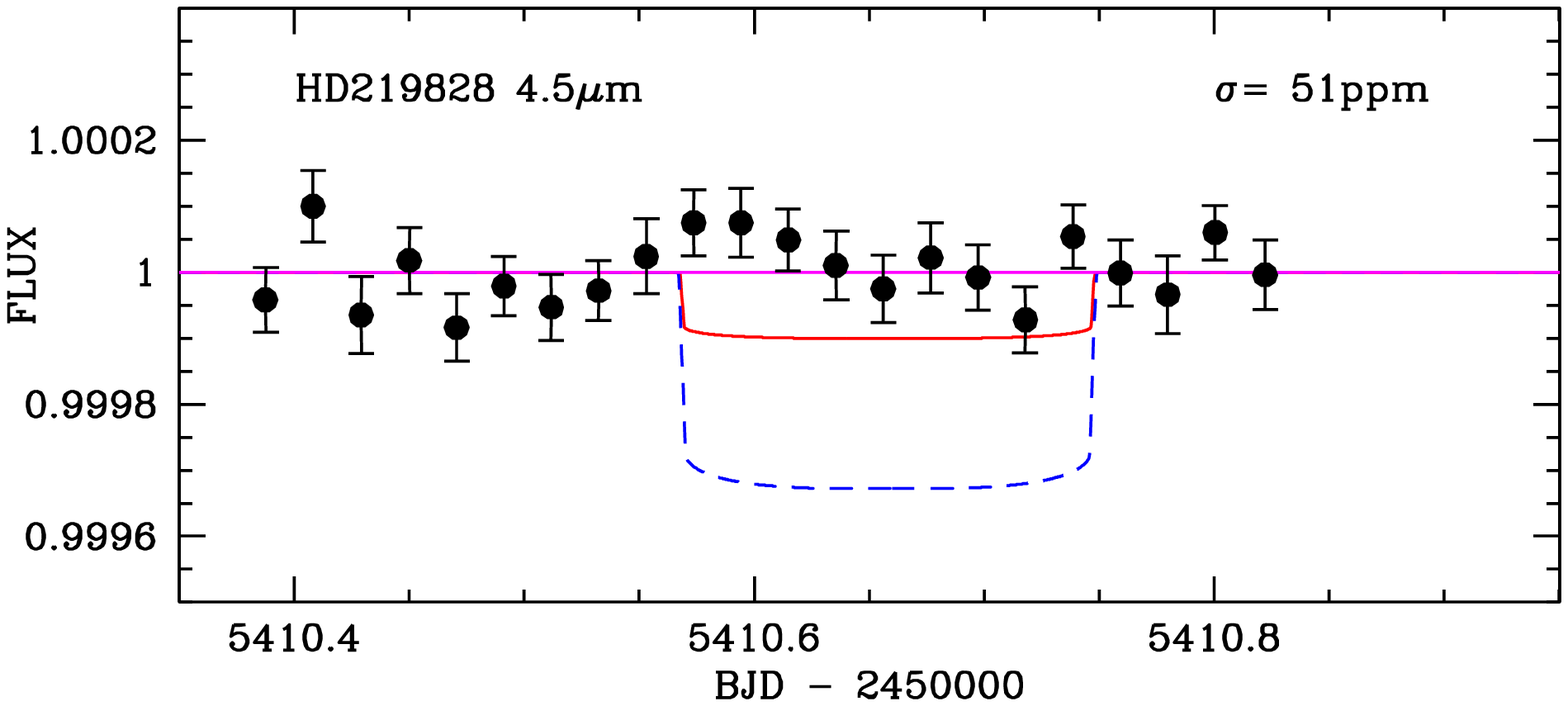}
\includegraphics[width=9cm]{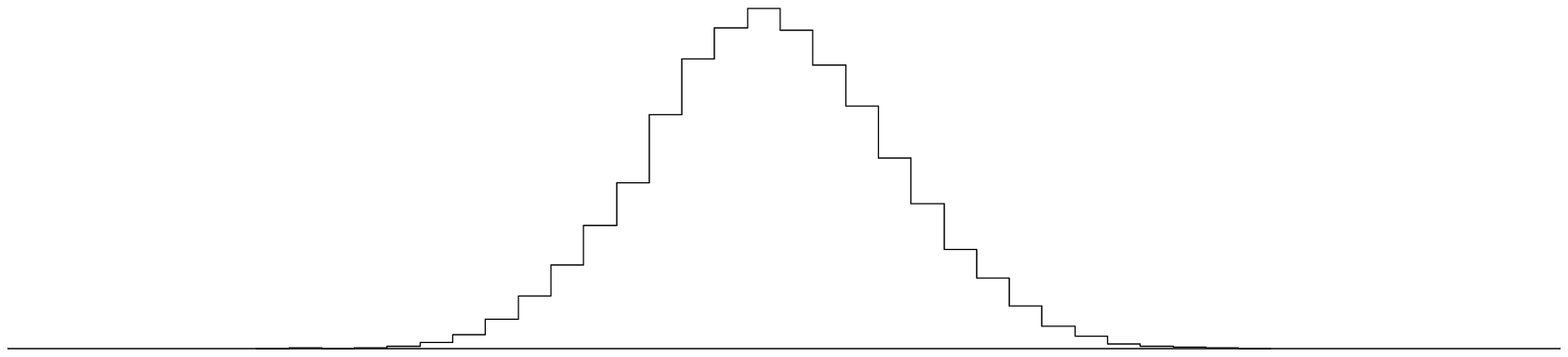}
\includegraphics[width=9cm]{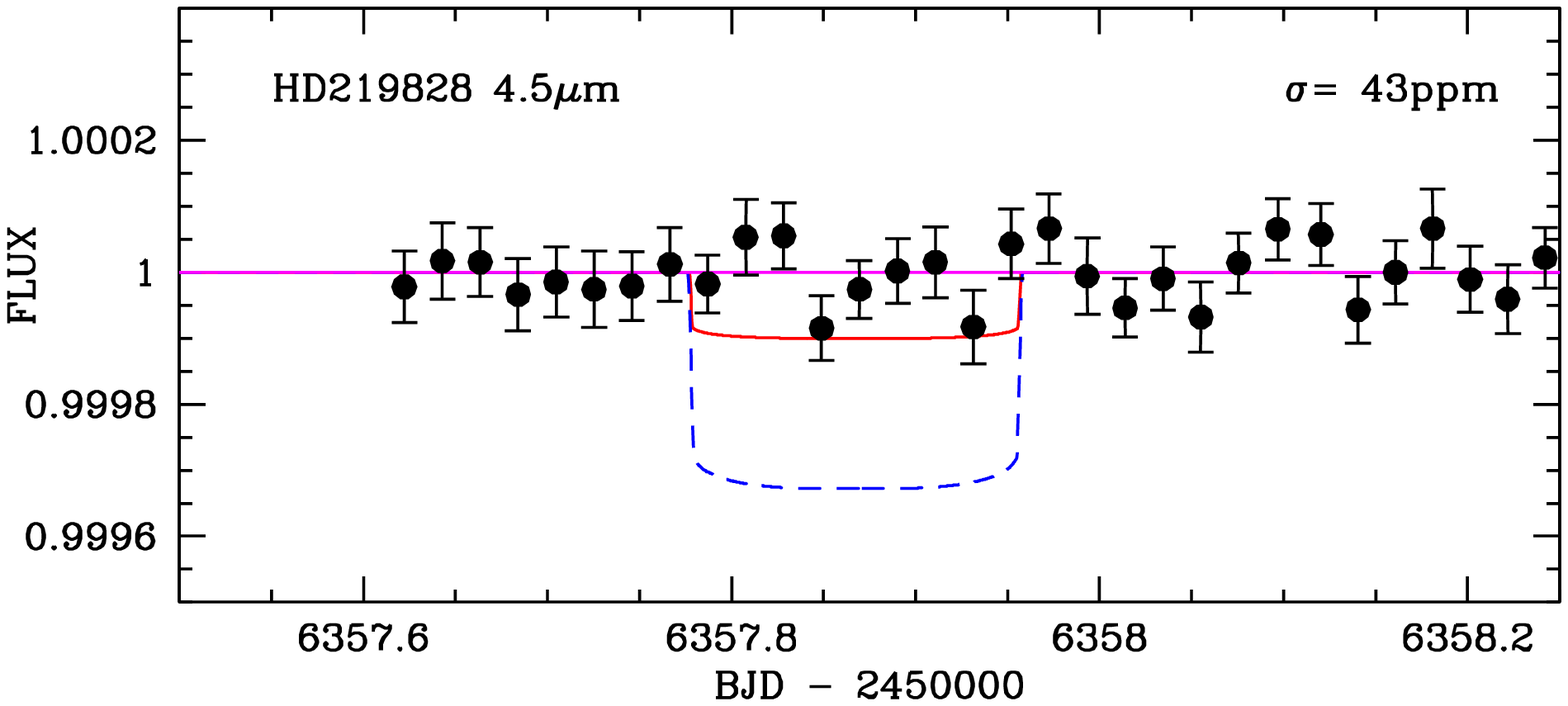}
\caption{Same as Fig. 1 for HD\,219828\,b. }
\end{figure} 

\section{Discussion and conclusion}

The constraint brought by the {\it Spitzer} photometry on the transiting nature of a given planet can be directly estimated in Table 1 by comparing the derived prior and posterior transit probabilities. For 16 out of the 19 RV planets targeted here, our {\it Spitzer} observations explored the transit window with a coverage and precision high enough to make a transiting configuration very unlikely, the posterior transit probabilities for these 16 targets all being less than 0.55\% (Table 1). For HD\,45184\,b, this posterior transit probability is still of 1.3\%, because our {\it Spitzer} observations did not explore the second part of its transit window (see Sect. 5.9 and Fig. 10). The transiting nature of HD\,13808\,b and  HD\,115617\,b is left unexplored by our observations. 

For the radius of each planet, our MCMC analysis assumed  a uniform prior PDF ranging from a pure-iron composition radius to 11 $R_{\oplus}$.  The fact that a transiting configuration was  disfavored by our MCMC analysis for all targeted planets does not preclude the possibility that one or several transits were in the data, but were just too shallow to be noticed by the Markov Chains. To estimate the actual detection threshold of our observations, we performed the following procedure for each of our targets. We created 50 fake transit light curves based on the multiplication of the actual light curve by a transit model of the targeted RV planets, each transit assuming a circular orbit for the planet, a mid-transit time drawn from the prior transit timing PDF, which was derived from the RV analysis, and an impact parameter drawn from a uniform PDF that ranged from 0 to 0.9. For each transit model, the depth was automatically tuned to have a difference in BIC of +9.2 between models neglecting and taking into account the transit. This difference in BIC corresponds to a Bayes factor of $e^{9.2/2}=100$, indicating a decisive selection of the transit model (Jeffreys 1961), i.e. a firm detection (at better than 3-3.5 sigma) of the transit. We then averaged the transit depths (and planet's radii) derived for the 50 light curves and adopted the resulting value as the detection threshold for the considered dataset. These detection thresholds are given in Table 1, expressed as transit depth (in ppm) and planetary radius. For each planet, they are compared to the planet's radius assuming pure-iron and Earth-like compositions. The detection threshold radius is smaller than the ones assuming pure-iron and Earth-like compositions for, respectively, 5 and 14 of the 17 planets for which the transiting nature was constrained by our observations. For 12 and 3 planets, we cannot thus fully reject the absence of a transit in our data, provided very metal-rich and Earth-like compositions, respectively. However, no transiting configuration is even midly favored for any planet  by our MCMC analysis (from the comparison of the last columns of Table 1), so, considering the excellent photometric precision of our {\it Spitzer} data, the hypothesis of a missed transit is clearly unlikely. Using the posterior transit probabilities shown in Table 1, the formula $1-\prod_{i=1:19} P_i(tr,D)$ indeed results  in a posterior probability that none of the probed 19 planets transits of 83\%, vs 22\% for the corresponding  prior probability ($1-\prod_{i=1:19} P_i(tr)$).  

Our multi-Cycle {\it Spitzer} transit search explored the transiting nature of 25 RV planets. It detected one or several transits for the planets HD\,75732\,e (aka 55\,Cnc\,e) (Demory et al. 2011, Gillon et al. 2012) and HD\,219234\,b (Motalebi et al 2015), confirmed the transiting nature of HD\,97658\,b (Van Grootel et al. 2015), discarded or disfavored the transiting nature of 20 planets (including one presented in S\'egransan et al., in prep.), and left the one of two planets unconstrained. By discovering the transits of two planets of a few Earth-masses  that are suitable for detailed atmospheric characterization, it brought a significant contribution to the study of super-Earths.  Statistically speaking, its final result is normal: considering  all the planets listed in Table 1 except HD\,97658\,b, which we decided to observe only because we knew that it was probably transiting (Dragomir et al. 2013),  the sum of the geometric transit probabilities amounts to 196\%, i.e. the project was expected to observe the transits of $\sim 2$ planets. 

 The photometric performances demonstrated by {\it Spitzer} in this program are illustrated in Fig. 21. This figure compares as a function of the targets' $K$-magnitude the standard deviations measured in the detrended light curves for a sampling of 30 min to the corresponding formal errors computed following the instructions of the {\it Spitzer} Observation Manual (SOM\footnote{http://ssc.spitzer.caltech.edu/warmmission/propkit/som/}). At 4.5$\mu$m, the measured standard deviations are well modeled by the linear relationship $\sigma_{30min} =  32.5 + 11.97\times(K_{mag} - 5)$ ppm, while the mean quadratic difference between the measured standard deviations and the formal errors is $35\pm5$ ppm. This quadratic difference is $64\pm9$ ppm at 3.6 $\mu$m, and $45\pm4$ ppm when neglecting HD125612, which seems to be a more active star than our other targets (Sec. 5.15).  These quadratic differences can be attributed to the low-frequency noise of instrumental and astrophysical origins that cannot be represented by our instrumental model. Figure 21 shows that this red noise dominates the photometric precision of {\it Spitzer}, especially for the brighter targets. Its values are low enough --  a few dozens of ppm -- to qualify the photometric precision of {\it Spitzer} as excellent, and to make it an optimal facility for the search for very low-amplitude transits on bright nearby stars. 

 The {\it Spitzer} mission should come to an end in early 2019. Fortunately, the CHEOPS space mission (Broeg et al. 2015)  will arrive just in time to  take over the search for the transits of super-Earths discovered by RVs around nearby stars. CHEOPS will not benefit from a targets' visibility as favorable as {\it Spitzer}, but its full dedication to transit observations  will more than compensate for its geocentric orbit. 
 
 \begin{figure}
\label{fig:21}
\centering                     
\includegraphics[width=9cm]{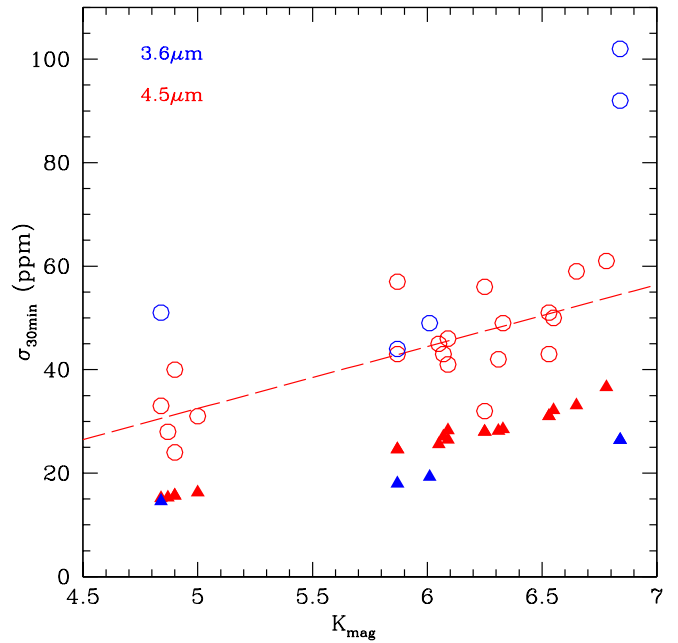}
\caption{Standard deviations of the detrended {\it Spitzer} photometry binned per 30 min (open circles) and the corresponding formal errors (triangles) as a function of the $K$-magnitude of the targets.  Blue  = 3.6 $\mu$m,  red = 4.5 $\mu$m. The dashed red line shows the best-fit linear relationship between the standard deviations measured at 4.5 $\mu$m and the $K$-magnitudes, its equation being $\sigma_{30min} =  32.5 + 11.97\times(K_{mag} - 5)$ ppm. }
\end{figure} 

\begin{table*}
\begin{center}
\label{tab:rvsmallplanets}
\begin{tabular}{cccccccccccc}
\hline\noalign {\smallskip}
Planet    &        $M_s$   &   $d$  & $K$    &  $M_p \sin{i} $  & Detection & Pure-Fe & Earth-like & Prior  & Posterior \\ \noalign {\smallskip}
               &                    &              &           &                            &   threshold & radius  & radius & $P(tr)$  & $P(tr,D)$\\ \noalign {\smallskip}
               & [$M_\odot$]   &   [pc]   &           &     [$M_\oplus$]    &  ppm/[$R_\oplus$] & [$R_\oplus$] &  [$R_\oplus$]  &   [\%]                       &     [\%]   \\ \noalign {\smallskip}
\hline \noalign {\smallskip}
HD\,40307\,b$^{1}$ & 0.78 & 12.8 & 4.79 & 4.3 & 150/0.91 & 1.13 & 1.44 &  6.6 & 0.19  \\ \noalign {\smallskip}
GJ\,3634\,b$^{2}$    & 0.45 & 19.8   & 7.47 & 7.0 & 500/1.05 & 1.31 & 1.65 & 7.0 & 0.50 \\ \noalign {\smallskip}
HD\,75732\,e$^{3, 4}$  & 0.91 & 12.3 & 4.02 &    7.8   & NA$^{a}$ & 1.34 & 1.70 &   28.9 & 100  \\ \noalign {\smallskip}
HD\,97658\,b$^{5}$ & 0.77 &  21.1 & 5.73  &  7.6 & NA$^{a}$ & 1.33 & 1.69 &  4.3  & 100 \\ \noalign {\smallskip}
HD\,219134\,b$^{6}$ & 0.78 & 6.5 & 5.57 & 4.3 & NA$^{a}$ &  1.15 & 1.44  & 9.5 & 100  \\ \noalign {\smallskip}
\hline \noalign {\smallskip}
BD-061339\,b$^{7}$ & 0.63 & 20.3  & 6.31 & 6.9 &  235/1.15 & 1.30 & 1.64 & 7.7 &  0.53 \\ \noalign {\smallskip}
HD1461\,b$^{7}$      & 1.04 & 23.2  & 4.90 & 6.7 &  170/1.56 & 1.29 & 1.63 & 8.1 &  0.14 \\ \noalign {\smallskip}
HD\,10180\,c$^{7}$  & 1.06 & 39.0  & 5.87 & 13.1 & 180/1.73 & 1.53 & 1.95 &  8.4 &  0.14   \\ \noalign {\smallskip}
HD\,13808\,b$^{7}$ & 0.77 & 28.6  & 6.78 & 11.8 &  NA$^{a}$ & 1.50 & 1.90 &  3.5 &  3.5 \\ \noalign {\smallskip}
HD\,20003\,b$^{7}$ & 0.91 & 43.8  & 6.65 & 11.8 &  280/1.79 & 1.49 & 1.90 & 3.4 &  0.54 \\ \noalign {\smallskip}
HD\,20781\,b$^{7}$ & 0.83 & 35.4  & 6.55 & 2.1 &  220/1.39 & 0.93 & 1.17 &  7.1 &  0.15  \\ \noalign {\smallskip}
HD\,31527\,b$^{7}$ & 0.96 & 38.6  & 6.05 & 10.7 & 190/1.64  & 1.46 & 1.85 & 4.4 &  0.46  \\ \noalign {\smallskip}
HD\,39194\,b$^{7}$ & 0.72 & 25.9 & 6.09 & 4.1 &  210/1.22 & 1.12 & 1.42 &  6.4 &  0.46 \\ \noalign {\smallskip}
HD\,45184\,b$^{7}$ & 1.00 & 21.9 & 4.87 & 12.1 & 150/1.39 & 1.50 & 1.91 & 7.7 &  1.29 \\ \noalign {\smallskip}
HD\,47186\,b$^{7}$ & 1.03 & 39.6  & 6.01 & 23.2 & 175/1.59 & 1.77 & 2.26 & 10.5 &  0.00  \\ \noalign {\smallskip}
HD\,51608\,b$^{7}$ & 0.86 & 34.8  & 6.33 & 13.1 & 195/1.39 & 1.54 & 1.95 &  4.0 &  0.11 \\ \noalign {\smallskip}
HD\,93385\,d$^{7}$ & 1.04 & 42.2 & 6.07 & 4.0 &  200/1.78 & 1.11 & 1.41 &  7.9 &  0.23  \\ \noalign {\smallskip}
HD\,96700\,b$^{7}$ & 0.96 & 25.7 & 5.00 & 9.1 & 155/1.58 & 1.40  & 1.77 & 6.8 &  0.09 \\ \noalign {\smallskip}
HD\,115617\,b$^{7}$& 0.94 & 8.6 & 2.96 & 6.2 &  NA$^{a}$ & 1.26  & 1.60 & 9.4 &  9.4  \\ \noalign {\smallskip}
HD\,125612\,c$^{7}$ & 1.09 & 54.2  & 6.84 & 19.3 &  290/1.92 & 1.69 & 2.16 & 9.7 &  0.24 \\ \noalign {\smallskip}
HD\,134060\,b$^{7}$ & 1.07 & 24.2 & 4.84 & 10.0 &  165/1.61 & 1.43 & 1.82 & 8.4 &  0.00  \\ \noalign {\smallskip}
HD\,181433\,b$^{7}$ & 0.86 & 26.8 & 6.09 & 7.5 & 190/1.26 & 1.33  & 1.68 & 4.9 &  0.52  \\ \noalign {\smallskip}
HD\,215497\,b$^{7}$ & 0.87 & 43.6 & 6.78 & 6.1 &  250/1.66 & 1.26 & 1.59 & 11.8 &  0.31 \\ \noalign {\smallskip}
HD\,219828\,b$^{7}$ & 1.18 & 72.3  & 6.53 & 20.2 &  155/2.18 & 1.71 & 2.19 & 14.2 &  0.00  \\ \noalign {\smallskip}
\hline \noalign {\smallskip} 
\end{tabular}
\end{center}
\caption{24 of the 25 planets targeted by our {\it Spitzer} multi-cycle  transit search. The first group of 5 planets
are the targets for which results were presented previously. The second group of 19 planets
are the targets of this work.
For each planet, column 6 provides the mean transit detection threshold (in ppm and  [$R_\oplus$]),
as inferred from injection of transit models in the data (see Sec. 6 for details). Column 7 and 8
give the theoretical radii corresponding, respectively, to a pure-iron and Earth-like compositions. Columns 9 and 10 
present the  a priori (geometric) and a posteriori (after observation) transit probabilities 
for the planet. $^{a}$Not Applicable: for HD\,13808\,b and HD\,115617\,b (61\,Vir\,b), 
our observations did not constrain the (non-)transiting configuration, while for HD\,75732\,e (55\,Cnc\,e), HD\,97658\,b, 
and HD\,219134\,b,  a transit was firmly detected in the data.  References: $^1$: Gillon et al. 2010, $^2$: Bonfils et al. 2011,
$^3$: Demory et al. 2011, $^4$: Gillon et al. 2012a, $^5$: Van Grootel et al. 2014, $^6$: Motalebi et al. 2015, $^7$: this work. }
\end{table*}

 \begin{acknowledgements} 
This work is based in part on observations made with the {\it Spitzer Space Telescope}, which is operated by the 
Jet Propulsion Laboratory, California Institute of Technology under a contract with NASA. Support for this work
 was provided by NASA. M. Gillon is Research Associate at the Belgian Scientific Research Fund (F.R.S-FNRS). 
 This publication makes use of data products from the Two Micron All Sky Survey, which is a joint project of the 
 University of Massachusetts and the Infrared Processing and Analysis Center/California Institute of Technology, 
 funded by the National Aeronautics and Space Administration and the National Science Foundation.
\end{acknowledgements} 

\bibliographystyle{aa}

\begin{appendix}

\section{Tables}
The following tables describe the targets of this work (Tables A.1 to A.4) and the {\it Spitzer} observations (Tables A.5 to A.11).

\begin{table*}
\begin{center}
{\scriptsize
\label{tab:targets}
\begin{tabular}{cccccc}
\hline\noalign {\smallskip}
Star                                      & BD-061339                             & HD\,1461                              & HD\,10180                 &      HD\,13808                & HD\,20003                         \\ \noalign {\smallskip}
\hline \noalign {\smallskip}
$d$ [parsec]                          & $20.3 \pm 0.7^{(1)}$        & $23.2 \pm 0.3^{(1)}$           & $39.0 \pm 0.6^{(1)}$         &  $28.6 \pm 0.5^{(1)}$      & $43.8 \pm 1.2^{(1)}$               \\ \noalign {\smallskip}
$V$ [mag]                                       & $9.67\pm 0.04^{(2)}$      & $6.46 \pm 0.01^{(2)}$         & $7.32 \pm 0.01^{(12)}$        &  $8.38 \pm 0.01^{(2)}$     & $8.37 \pm 0.01^{(2)}$   \\ \noalign {\smallskip}
$K$ [mag]                                       & $6.31 \pm 0.02^{(3)}$      & $4.90 \pm 0.02^{(3)}$         &$5.87 \pm 0.02^{(3)}$        &  $6.25 \pm 0.02^{(3)}$     & $6.65 \pm 0.02^{(3)}$   \\ \noalign {\smallskip} 
Spectral type                        & K7V/M0V$^{(4)}$             & G3V$^{(7)}$                        & G1V$^{(13)}$                    & K2V$^{(13)}$                    & G8V$^{(16)}$                      \\ \noalign {\smallskip} 
$T_{eff}$ [K]$^a$                  & $4040 \pm 50^{(5)}$        & $5765 \pm 50^{(8)}$           & $5910 \pm 50^{(14)}$       &  $5035 \pm 50^{(8)}$        & $5495 \pm 50^{(8)}$     \\ \noalign {\smallskip} 
[Fe/H] [dex]                          & $-0.07 \pm 0.10^{(5)}$      & $0.19 \pm 0.01^{(8)}$        & $0.08 \pm 0.01^{(14)}$      & $-0.21 \pm 0.02^{(8)}$      &  $0.04 \pm 0.02^{(8)}$      \\ \noalign {\smallskip} 
$M_\ast$   [$M_\odot$]        & $0.63 \pm 0.03^{(5)}$      & $1.04 \pm 0.07^{(8)}$         & $1.06 \pm 0.05^{(14)}$      & $0.77 \pm 0.06^{(8)}$        &  $0.91 \pm 0.07^{(8)}$        \\ \noalign {\smallskip} 
Bolometric Correction         & $-1.07 \pm 0.03^{(6)}$      &  $-0.082 \pm 0.009^{(6)}$   & $-0.058 \pm 0.009^{(6)}$  &  $-0.290 \pm 0.021^{(6)}$  &  $-0.140 \pm 0.013^{(6)}$       \\ \noalign {\smallskip} 
$R_\ast$  [$R_\odot$]$^b$ & $0.69 \pm 0.02$               & $1.10 \pm 0.02$                  & $1.18 \pm 0.02$                &  $0.81 \pm 0.02$                &   $0.98 \pm 0.02$                        \\ \noalign {\smallskip} 
$\log{g}$ [cgs]                     & $4.56 \pm 0.04$               & $4.37 \pm 0.04$                  & $4.32 \pm 0.03$                &  $4.51 \pm 0.05$                &   $4.41 \pm 0.04$                      \\ \noalign {\smallskip} 
\hline \noalign {\smallskip}
 RV      &  & & & &     \\ \noalign {\smallskip}
\hline \noalign {\smallskip}
  Data                                   & HARPS: 102$^{(4)}$+10   & HARPS: 167$^{(9)}$+82$^{(10)}$+5      & HARPS: 190$^{(14)}$+63    & HARPS: 133$^{(9)}$+89              & HARPS: 104$^{(9)}$+77 \\ \noalign {\smallskip}
                                            & PFS: 15$^{(5)}$                    &  Keck: 144$^{(11)}$              &                                              &                                                        & \\ \noalign {\smallskip} 
  Model                                & 2  Keplerians                         & 3 Keplerians                         &  6 Keplerians                        & 2 Keplerians                                   & 4 Keplerians \\ \noalign {\smallskip} 
                                            &    + linear trend                     & + quadratic trend                  &                                               & + quadratic trend                           &  \\ \noalign {\smallskip}  
                                            &                                              &  +CCF function                     &                                               & +CCF function                               & \\ \noalign {\smallskip}  
 Jitter noise [\ms]                 & HARPS: 2.7                         &  HARPS: 1.5                         & 1.5                                        &   1.7                                                &  1.4  \\ \noalign {\smallskip} 
                                            & PFS: 2.7                               & Keck: 2.4 \& 1.6$^e$            &                                              &                                                        &    \\ \noalign {\smallskip} 
\hline \noalign {\smallskip}                                           
Planet                                  & BD-061339\,b$^{(4)}$             &    HD\,1461\,b$^{(11)}$        & HD\,10180\,c$^{(14)}$          &  HD\,13808\,b$^{(9, 15)}$                  & HD\, 20003\,b$^{(9, 17)}$  \\ \noalign {\smallskip}
\hline \noalign {\smallskip}
$M_p \sin{i}$ [$M_\oplus$]    & $6.93 \pm 0.96   $               & $6.73 \pm 0.47$                 & $13.11 \pm 0.62$                 &  $11.83 \pm 0.88$                            &$11.79 \pm 0.61$ \\ \noalign {\smallskip} 
Min. $R_p$ [$R_{\oplus}$]$^c$     &  $1.30 \pm 0.05$         & $1.29 \pm 0.03$                  & $1.53 \pm 0.02$                  &   $1.50 \pm 0.03$                            & $1.49 \pm 0.02$  \\ \noalign {\smallskip} 
Min. $(R_p/R_\ast)^2$ [ppm]$^c$ & $298 \pm 29$              & $115 \pm  6$                       & $142 \pm 6$                        &  $ 286 \pm 18$                                &  $191 \pm 9$      \\ \noalign {\smallskip} 
 $T_{eq}$ [K]$^d$                  & $796 \pm 17$                     & $1154 \pm 20$                    & $1223 \pm 18$                    &   $674 \pm 14$                                 &   $836 \pm 12$         \\ \noalign  {\smallskip} 
$T_{0}$-2450000 [BJD$_{TDB}$]         & $6627.48_{-0.16}^{+0.18}$     & $6549.30 \pm 0.12$        & $5212.837_{-0.074}^{+0.059}$  & $6537.49 \pm 0.26$  &  $6538.34 \pm 0.36$   \\ \noalign {\smallskip}     
 $P$ [d]                                   & $3.87310 \pm 0.00037$     & $5.77198 \pm 0.00030$      & $5.75931 \pm 0.00021$      & $14.1853 \pm 0.0019$                    &  $11.8489 \pm 0.0015$ \\ \noalign {\smallskip} 
$W_{b=0}$ [min]                     & $139 \pm 15$                     & $218 \pm 11$                        &$238 \pm 10$                     & $246 \pm 13$                                  &  $388 \pm 23$ \\ \noalign {\smallskip} 
 $K$ [\ms]                               & $3.90 \pm 0.52$                 & $2.34 \pm 0.13$                  & $4.51 \pm 0.15$                  &  $3.73 \pm 0.20$                             &   $3.84 \pm 0.20$    \\ \noalign {\smallskip} 
$a$ [AU]                                 & $0.04136 \pm 0.00066$     & $0.0638 \pm 0.0015$          & $0.0641 \pm 0.0010$          & $0.1050 \pm 0.0028$                      &  $0.09817 \pm 0.00072$ \\ \noalign {\smallskip} 
 $e$                                        & $0.11_{-0.08}^{+0.11}$       & $0.037_{-0.026}^{+0.041}$ &$0.045_{-0.030}^{+0.037}$  & $0.042_{-0.029}^{+0.043}$             & $0.377 \pm 0.047$  \\ \noalign {\smallskip}
 $\omega$ [deg]                     & $192_{-93}^{+69}$              & $134_{-120}^{+110}$          & $320_{-41}^{+51}$              & $272_{-75}^{+85}$                          & $267.1_{-8.3}^{+7.6}$    \\ \noalign {\smallskip} 
Prior $P_{transit}$ [\%]           & $7.72_{-0.71}^{+0.77}$       & $8.06 \pm 0.38$                   & $8.38_{-0.33}^{+0.30}$     & $3.52 \pm 0.18$                              & $3.43 \pm 0.15$ \\ \noalign {\smallskip} 
Prior $P_{occultation}$ [\%]    & $7.9_{-0.7}^{+1.1}$            & $8.00 \pm 0.38$                  &  $8.78_{-0.32}^{+0.38}$      & $3.68_{-0.18}^{+0.21}$                   & $7.50_{-0.56}^{+0.62}$   \\ \noalign {\smallskip} 
\hline \noalign {\smallskip} 
\end{tabular}
}
\end{center}
\caption{Targets 1-5 of our {\it  Warm Spitzer} program. For each target, the table presents the assumed stellar parameters, the RV data and details of their 
analysis, and the parameters that we derived for the planet searched for transit from our RV analysis. $^a$For the sake of realism, a minimal error of 50K was assumed here. $^b$From 
luminosity and $T_{eff}$. $^c$Assuming $M_p \sin{i} = M_p$. These minimum values correspond to a pure iron planet (Seager et
 al. 2007). $^d$Assuming a null albedo and a heat distribution factor $f'$ = 1/4 (Seager 2010). $^e$For the first and second Keck datasets presented 
 in Rivera et al. (2010), respectively. References: $^{(1)}$Van Leeuwen (2007),  $^{(2)}$Kharchenko (2001), $^{(3)}$Skrutskie et al. (2006), $^{(4)}$Lo Curto et al. (2013), $^{(5)}$Arriagada et al. (2013), $^{(6)}$Flower (1996), $^{(7)}$Gray et al. (2003), $^{(8)}$Santos et al. (2013),  $^{(9)}$Mayor et al. (2011), $^{(10)}$D\'iaz et al. (2015), $^{(11)}$Rivera et al. (2010), $^{(12)}$Hog et al. (2000), $^{(13)}$Gray et al. (2006), 
 $^{(14)}$Lovis et al. (2011), $^{(15)}$Queloz et al. in prep., $^{(16)}$Houk \& Cowley (1975), $^{(17)}$Udry et al. in prep.}
\end{table*}

\begin{table*}
\begin{center}
{\scriptsize
\label{tab:targets}
\begin{tabular}{cccccc}
\hline\noalign {\smallskip}
Star                                      & HD\,20781                            & HD\,31527                            &      HD\,39194                & HD\,45184                         \\ \noalign {\smallskip}
\hline \noalign {\smallskip}
$d$ [parsec]                          & $35.4 \pm 1.3^{(1)}$        & $38.6 \pm 0.9^{(1)}$             & $21.9 \pm 0.2^{(1)}$               \\ \noalign {\smallskip}
$V$ [mag]                                       & $8.44 \pm 0.01^{(2)}$      & $7.48 \pm 0.01^{(2)}$            &  $8.08 \pm 0.01^{(2)}$     & $6.37 \pm 0.01^{(4)}$   \\ \noalign {\smallskip}
$K$ [mag]                                       & $6.55 \pm 0.02^{(3)}$      & $6.05 \pm 0.02^{(3)}$            &  $6.09 \pm 0.02^{(3)}$     & $4.87 \pm 0.02^{(3)}$   \\ \noalign {\smallskip} 
Spectral type                        & K0V$^{(4)}$                      & G0V$^{(9)}$                          & K0V$^{(4)}$                    & G2V$^{(4)}$                      \\ \noalign {\smallskip} 
$T_{eff}$ [K]$^a$                  & $5255 \pm 50^{(5)}$        & $5900 \pm 50^{(5)}$             &  $5205 \pm 50^{(5)}$        & $5870 \pm 50^{(5)}$     \\ \noalign {\smallskip} 
[Fe/H] [dex]                          & $-0.11 \pm 0.02^{(5)}$      & -$0.17\pm 0.01^{(5)}$          & $-0.61 \pm 0.02^{(5)}$      &  $0.04 \pm 0.01^{(5)}$      \\ \noalign {\smallskip} 
$M_\ast$   [$M_\odot$]        & $0.83 \pm 0.06^{(5)}$      & $0.96 \pm 0.07^{(5)}$           & $0.72 \pm 0.05^{(2)}$        &  $1.00 \pm 0.07^{(5)}$        \\ \noalign {\smallskip} 
Bolometric Correction         & $-0.208 \pm 0.016^{(6)}$  &  $-0.059 \pm 0.009^{(6)}$   &  $-0.225 \pm 0.017^{(6)}$  &  $-0.058 \pm 0.008^{(6)}$       \\ \noalign {\smallskip} 
$R_\ast$  [$R_\odot$]$^b$ & $0.86 \pm 0.02$               & $1.09 \pm 0.02$                   &  $0.77 \pm 0.02$                &   $1.04 \pm 0.02$                        \\ \noalign {\smallskip} 
$\log{g}$ [cgs]                     & $4.49 \pm 0.04$               & $4.35 \pm 0.04$                   &  $4.52 \pm 0.04$                &   $4.40 \pm 0.04$                      \\ \noalign {\smallskip} 
\hline \noalign {\smallskip}
 RV      &  & & & &     \\ \noalign {\smallskip}
\hline \noalign {\smallskip}
  Data                                   & HARPS: 96$^{(7)}$+117      & HARPS: 167$^{(7)}$+75                      & HARPS: 133$^{(7)}$+128             & HARPS: 82$^{(7)}$+92 \\ \noalign {\smallskip}
  Model                                & 4 Keplerians                         & 3 Keplerians                                          &   3 Keplerians                                 & 2 Keplerians    \\ \noalign {\smallskip} 
                                            &                                               &                                                              & + quadratic trend                           & + quartic trend \\ \noalign {\smallskip}  
                                            &                                              &                                                               &  + CCF function                            & + CCF function \\ \noalign {\smallskip}  
 Jitter noise [\ms]                 &  1.2                                      &   1.2                                                        &   1.1                                                &  1.95  \\ \noalign {\smallskip} 
\hline \noalign {\smallskip}                                           
Planet                                  & HD\,20781\,b$^{(8)}$             &    HD\,31527\,b$^{(7, 8)}$              &  HD\,39194\,b$^{(7, 9)}$                  & HD\, 45184\,b$^{(7, 8)}$  \\ \noalign {\smallskip}
\hline \noalign {\smallskip}
$M_p \sin{i}$ [$M_\oplus$]    & $2.12 \pm 0.35   $               & $10.68 \pm 0.71$                      &  $4.08 \pm 0.32$                            &$12.08 \pm 0.86$ \\ \noalign {\smallskip} 
Min. $R_p$ [$R_{\oplus}$]$^c$     &  $0.93 \pm 0.04$         & $1.46 \pm 0.03$                       &   $1.12 \pm 0.03$                            & $1.50 \pm 0.03$  \\ \noalign {\smallskip} 
Min. $(R_p/R_\ast)^2$ [ppm]$^c$ & $99 \pm 11$              & $150 \pm  8$                              &  $ 179 \pm 13$                                &  $175 \pm 10$      \\ \noalign {\smallskip} 
 $T_{eq}$ [K]$^d$                  & $993 \pm 20$                     & $839 \pm 15$                           &   $935 \pm 19$                                 &   $1143 \pm 20$         \\ \noalign  {\smallskip} 
$T_{0}$-2450000 [BJD$_{TDB}$]         & $6613.92 \pm 0.23$     & $6792.65 \pm 0.36$        & $6285.93 \pm 0.15$  &  $6317.67 \pm 0.13$   \\ \noalign {\smallskip}     
 $P$ [d]                                   & $5.3144 \pm 0.0011$     & $16.5547 \pm 0.0033$                & $5.63675 \pm 0.00044$                    &  $5.88607 \pm 0.00032$ \\ \noalign {\smallskip} 
$W_{b=0}$ [min]                     & $180 \pm 6$                     & $293 \pm 19$                          & $174 \pm 10$                                  &  $209 \pm 13$ \\ \noalign {\smallskip} 
 $K$ [\ms]                               & $0.88 \pm 0.14$                 & $2.78 \pm 0.13$                     &  $1.83 \pm 0.12$                             &   $4.33 \pm 0.23$    \\ \noalign {\smallskip} 
$a$ [AU]                                 & $0.0560 \pm 0.0014$         & $0.1253 \pm 0.0031$               & $0.0555\pm 0.0013$                      &  $0.0638 \pm 0.0015$ \\ \noalign {\smallskip} 
 $e$                                        & 0 (fixed)                              & $0.117 \pm 0.052$ &$0.033_{-0.023}^{+0.035}$             &  $0.122_{-0.057}^{+0.053}$  \\ \noalign {\smallskip}
 $\omega$ [deg]                     & -                                       & $42_{-26}^{+23}$                      & $224_{-90}^{+97}$                          & $178_{-27}^{+30}$    \\ \noalign {\smallskip} 
Prior $P_{transit}$ [\%]           & $7.14 \pm 0.25 $       & $4.40\pm 0.29$                   & $6.38 \pm 0.36$                              & $7.72_{-0.44}^{+0.51}$ \\ \noalign {\smallskip} 
Prior $P_{occultation}$ [\%]    & $7.14 \pm 0.25$            & $3.81 \pm 0.22$                  & $6.57_{-0.33}^{+0.42}$                   & $7.67_{-0.43}^{+0.48}$   \\ \noalign {\smallskip} 
\hline \noalign {\smallskip} 
\end{tabular}
}
\end{center}
\caption{Same as Table A.1 for targets 6-9. References: $^{(1)}$Van Leeuwen (2007), $^{(2)}$Hog et al. (2000), 
 $^{(3)}$Skrutskie et al. (2006), $^{(4)}$Kharchenko (2001), $^{(5)}$Santos et al. (2013), $^{(6)}$Flower (1996), 
 $^{(7)}$Mayor et al. (2011), $^{(8)}$Udry et al. in prep., $^{(9)}$Queloz et al. in prep.}
\end{table*}

\begin{table*}
\begin{center}
{\scriptsize
\label{tab:targets}
\begin{tabular}{cccccc}
\hline\noalign {\smallskip}
Star                                      & HD\,47186                            & HD\,51608                              & HD\,93385                 &      HD\,96700                & HD\,115617                        \\ \noalign {\smallskip}
\hline \noalign {\smallskip}
$d$ [parsec]                          & $39.6 \pm 1.0^{(1)}$        & $34.8 \pm 0.7^{(1)}$           & $42.2 \pm 1.3^{(1)}$          &  $25.7 \pm 0.4^{(1)}$      & $8.56 \pm 0.02^{(1)}$               \\ \noalign {\smallskip}
$V$ [mag]                                     & $7.63 \pm 0.01^{(2)}$      & $8.17 \pm 0.01^{(2)}$         & $7.49 \pm 0.01^{(2)}$       &  $6.50 \pm 0.01^{(2)}$     & $4.73 \pm 0.01^{(4)}$   \\ \noalign {\smallskip}
$K$ [mag]                                       & $6.01\pm 0.03^{(3)}$      & $6.33 \pm 0.02^{(3)}$         &$6.07 \pm 0.02^{(3)}$        &  $5.00 \pm 0.02^{(3)}$     & $2.96 \pm 0.24^{(3)}$   \\ \noalign {\smallskip} 
Spectral type                        & G5V$^{(4)}$                      & G7V$^{(4)}$                        & G2/G3V$^{(4)}$                    & G1/G2V$^{(4)}$                    & G5V$^{(4)}$                      \\ \noalign {\smallskip} 
$T_{eff}$ [K]$^a$                  & $5675 \pm 50^{(5)}$        & $5360 \pm 50^{(5)}$           & $5975 \pm 50^{(5)}$       &  $5845 \pm 50^{(5)}$        & $5575 \pm 50^{(5)}$     \\ \noalign {\smallskip} 
[Fe/H] [dex]                          & $0.23 \pm 0.02^{(5)}$      & -$0.07 \pm 0.01^{(5)}$        & $0.02 \pm 0.01^{(5)}$      & $-0.18 \pm 0.01^{(5)}$      &  $0.01 \pm 0.05^{(5)}$      \\ \noalign {\smallskip} 
$M_\ast$   [$M_\odot$]        & $1.03 \pm 0.07^{(5)}$      & $0.86 \pm 0.06^{(5)}$         & $1.04 \pm 0.07^{(5)}$      & $0.96 \pm 0.07^{(5)}$        &  $0.94 \pm 0.08^{(5)}$        \\ \noalign {\smallskip} 
Bolometric Correction         & $-0.100 \pm 0.010^{(6)}$  &  $-0.176 \pm 0.016^{(6)}$   & $-0.049 \pm 0.007^{(6)}$  &  $-0.069 \pm 0.008^{(6)}$  &  $-0.120 \pm 0.012^{(6)}$       \\ \noalign {\smallskip} 
$R_\ast$  [$R_\odot$]$^a$ & $1.10 \pm 0.02$               & $0.91 \pm 0.02$                  & $1.15  \pm 0.02$                &  $1.16 \pm 0.02$                &   $0.98 \pm 0.02$                        \\ \noalign {\smallskip} 
$\log{g}$ [cgs]                     & $4.37 \pm 0.04$               & $4.45 \pm 0.04$                  & $4.33 \pm 0.04$                &  $4.29 \pm 0.04$                &   $4.43 \pm 0.05$                      \\ \noalign {\smallskip} 
\hline \noalign {\smallskip}
 RV      &  & & & &     \\ \noalign {\smallskip}
\hline \noalign {\smallskip}
  Data                                   & HARPS: 66$^{(7)}$+67      & HARPS: 118$^{(8)}$+92       & HARPS: 127$^{(8)}$+106    & HARPS: 146$^{(8)}$+98             & AAT: 126$^{(10)}$ \\ \noalign {\smallskip}
 & & & & &  + Keck: 80$^{(10)}$ \\ \noalign {\smallskip}
  Model                                & 2  Keplerians                         & 2 Keplerians                         &  3 Keplerians                        & 2 Keplerians                                   & 3 Keplerians \\ \noalign {\smallskip} 
                                            &                                              & + quadratic trend                  &                                               & + quadratic trend                           & \\ \noalign {\smallskip}  
                                            &                                              &  + CCF function                    &                                               & + CCF function                               & \\ \noalign {\smallskip}  
 Jitter noise [\ms]                 &  0.9                                       &   1.2                                     & 1.3                                       &   1.6                                                &  Keck: 2.3 \\ \noalign {\smallskip} 
 & & & & & AAT: 2.2  \\ \noalign {\smallskip} 
\hline \noalign {\smallskip}                                           
Planet                                  & HD\,47186\,b$^{(7)}$             &    HD\,51608\,b$^{(8, 9)}$        & HD\,93385\,d$^{(10)}$          &  HD\,96700\,b$^{(7, 10)}$                  & HD\, 115617\,b$^{(11)}$  \\ \noalign {\smallskip}
\hline \noalign {\smallskip}
$M_p \sin{i}$ [$M_\oplus$]    & $23.2 \pm 1.1   $               & $13.12 \pm 0.77$                 & $3.97 \pm 0.48$                 &  $9.05 \pm 0.64$                            &$5.86 \pm 0.66$ \\ \noalign {\smallskip} 
Min. $R_p$ [$R_{\oplus}$]$^c$     &  $1.77 \pm 0.02$         & $1.54 \pm 0.02$                  & $1.11 \pm 0.03$                  &   $1.40 \pm 0.02$                            & $1.24 \pm 0.04$  \\ \noalign {\smallskip} 
Min. $(R_p/R_\ast)^2$ [ppm]$^c$ & $217 \pm 9$              & $239 \pm  13$                       & $79 \pm 7$                        &  $ 121 \pm 6$                                &  $135 \pm 9$      \\ \noalign {\smallskip} 
 $T_{eq}$ [K]$^d$                  & $1277 \pm 22$                     & $749 \pm 14$                    & $1129 \pm 19$                    &   $1087 \pm 19$                                 &   $1190 \pm 24$         \\ \noalign  {\smallskip} 
$T_{0}$-2450000 [BJD$_{TDB}$]         & $5179.972 \pm 0.021$    & $6379.95_{-0.16}^{+0.18}$        & $6364.17 \pm 0.27$  & $6521.26 \pm 0.16$  &  $5280.71 \pm 0.17$   \\ \noalign {\smallskip}     
 $P$ [d]                                   & $4.084575 \pm 0.000043$     & $14.0702 \pm 0.0015$      & $7.3422 \pm 0.0014$      & $8.12541 \pm 0.00068$                    &  $4.21504 \pm 0.00061$ \\ \noalign {\smallskip} 
$W_{b=0}$ [min]                     & $190 \pm 6$                     & $255 \pm 11$                        &$227_{-24}^{+20}$                     & $273_{-13}^{+16}$                 &  $174 \pm 9$ \\ \noalign {\smallskip} 
 $K$ [\ms]                               & $9.1 \pm 1.3$                 & $3.85 \pm 0.14$                  & $4.51 \pm 0.15$                  &  $2.97 \pm 0.15$                             &   $2.44 \pm 0.24$    \\ \noalign {\smallskip} 
$a$ [AU]                                 & $0.0505 \pm 0.0012$     & $0.1084 \pm 0.0025$          & $0.0749 \pm 0.0017$          & $0.0780 \pm 0.0019$                      &  $0.0500 \pm 0.0014$ \\ \noalign {\smallskip} 
 $e$                                        & $0.039\pm 0.014$       & $0.033_{-0.023}^{+0.033}$ &$0.13_{-0.09}^{+0.10}$     & $0.049_{-0.034}^{+0.049}$             & $0.078_{-0.055}^{+0.083}$   \\ \noalign {\smallskip}
 $\omega$ [deg]                     & $57 \pm 23$              & $130_{-66}^{+74}$          & $87 \pm 60$                       & $293 \pm 74$                                     &   $58_{-160}^{+73}$   \\ \noalign {\smallskip} 
Prior $P_{transit}$ [\%]           & $10.45 \pm 0.34$       &  $3.97_{-0.16}^{+0.19}$           & $7.9_{-0.7}^{+1.1}$     &   $6.76_{-0.36}^{+0.33}$                & $9.45_{-0.63}^{+0.99}$ \\ \noalign {\smallskip} 
Prior $P_{occultation}$ [\%]    & $9.84 \pm 0.31$            & $3.84 \pm 0.17$              &  $6.68_{-0.61}^{+0.54}$      & $7.12_{-0.34}^{+0.45}$                   &   $8.89_{-0.72}^{+0.63}$ \\ \noalign {\smallskip} 
\hline \noalign {\smallskip} 
\end{tabular}
}
\end{center}
\caption{Same as Table A.1 for targets 10-14. References: $^{(1)}$Van Leeuwen (2007), $^{(2)}$Hog et al. (2000), $^{(3)}$Skrutskie et al. (2006),   $^{(4)}$Kharchenko (2001),$^{(5)}$Santos et al. (2013), $^{(6)}$Flower (1996),  $^{(7)}$Bouchy et al. (2009), $^{(8)}$Mayor et al. (2011), $^{(9)}$Udry et al. in prep.,  $^{(10)}$Queloz et al. in prep.,  $^{(11)}$Vogt et al. (2010)} 
\end{table*}

\begin{table*}
\begin{center}
{\scriptsize
\label{tab:targets}
\begin{tabular}{cccccc}
\hline\noalign {\smallskip}
Star                                      & HD\,125612                            & HD\,134060                             & HD\,181433                &      HD\,215497                & HD\,219828                        \\ \noalign {\smallskip}
\hline \noalign {\smallskip}
$d$ [parsec]                          & $54.2 \pm 3.0^{(1)}$        & $24.2 \pm 0.3^{(1)}$           & $26.8 \pm 0.8^{(1)}$          &  $43.6 \pm 2.0^{(1)}$      & $72.3 \pm 3.9^{(1)}$               \\ \noalign {\smallskip}
$V$ [mag]                                       & $8.32 \pm 0.02^{(2)}$      & $6.29 \pm 0.01^{(2)}$         & $8.38 \pm 0.01^{(4)}$       &  $8.95 \pm 0.02^{(2)}$     & $8.01 \pm 0.01^{(2)}$   \\ \noalign {\smallskip}
$K$ [mag]                                      & $6.84\pm 0.03^{(3)}$      & $4.84 \pm 0.03^{(3)}$         &$6.09 \pm 0.02^{(3)}$        &  $6.78 \pm 0.02^{(3)}$     & $6.53\pm 0.02^{(3)}$   \\ \noalign {\smallskip} 
Spectral type                        & G3V$^{(4)}$                      & G3IV$^{(4)}$                        & K5V$^{(4)}$                    & K3V$^{(4)}$                    & G0IV$^{(4)}$                      \\ \noalign {\smallskip} 
$T_{eff}$ [K]$^a$                  & $5915 \pm 50^{(5)}$        & $5965 \pm 50^{(5)}$           & $4880 \pm 50^{(5)}$       &  $5000 \pm 100^{(5)}$        & $5890 \pm 50^{(5)}$     \\ \noalign {\smallskip} 
[Fe/H] [dex]                          & $0.24 \pm 0.01^{(5)}$      & $0.14 \pm 0.01^{(5)}$        & $0.36 \pm 0.18^{(5)}$      & $0.25 \pm 0.05^{(5)}$      &  $0.19 \pm 0.03^{(5)}$      \\ \noalign {\smallskip} 
$M_\ast$   [$M_\odot$]        & $1.09 \pm 0.07^{(5)}$      & $1.07 \pm 0.07^{(5)}$         & $0.86 \pm 0.17^{(5)}$      & $0.87 \pm 0.11^{(5)}$        &  $1.18 \pm 0.08^{(5)}$        \\ \noalign {\smallskip} 
Bolometric Correction         & $-0.057\pm 0.008^{(6)}$  &  $-0.050 \pm 0.007^{(6)}$   & $-0.360 \pm 0.024^{(6)}$  &  $-0.304 \pm 0.048^{(6)}$  &  $-0.061 \pm 0.008^{(6)}$       \\ \noalign {\smallskip} 
$R_\ast$  [$R_\odot$]$^a$ & $1.03 \pm 0.02$               & $1.15 \pm 0.02$                  & $0.84  \pm 0.02$                &  $0.97 \pm 0.04$                &   $1.60 \pm 0.03$                        \\ \noalign {\smallskip} 
$\log{g}$ [cgs]                     & $4.43 \pm 0.03$               & $4.35 \pm 0.04$                  & $4.51 \pm 0.11$                &  $4.40 \pm 0.08$                &   $4.10 \pm 0.04$                      \\ \noalign {\smallskip} 
\hline \noalign {\smallskip}
 RV      &  & & & &     \\ \noalign {\smallskip}
\hline \noalign {\smallskip}
  Data                                   & HARPS: 58$^{(7)}$+10      & HARPS: 100$^{(9)}$+50      & HARPS: 107$^{(11)}$+78    & HARPS: 105$^{(7)}$ + 3             & HARPS: 22$^{(21)}$+69 \\ \noalign {\smallskip}
  Model                                & Keck: 19$^{(8)}$                  & 2 Keplerians                         &  3 Keplerians                        & 2 Keplerians                                   & 2 Keplerians \\ \noalign {\smallskip} 
                                            &  3 Keplerians                        &  + CCF \& logR'(HK)              &                                               &  + CCF function                          &  \\ \noalign {\smallskip}  
                                            &                                              & function                                &                                              &                                                       & \\ \noalign {\smallskip} 
 Jitter noise [\ms]                 &  HARPS: 3.2                          &   1.3                                    & 1.0                                        &   1.3                                                 &  1.5  \\ \noalign {\smallskip} 
                                             & Keck: 4.7                               &                                            &                                              &                                                       &            \\ \noalign {\smallskip} 
\hline \noalign {\smallskip}                                           
Planet                                  & HD\,125612\,c$^{(7)}$             &    HD\,134060\,b$^{(9, 10)}$        & HD\,181433\,b$^{(11)}$          &  HD\,215497\,b$^{(7)}$                  & HD\,219828\,b$^{(12)}$  \\ \noalign {\smallskip}
\hline \noalign {\smallskip}
$M_p \sin{i}$ [$M_\oplus$]    & $19.3 \pm 2.1   $               & $9.97 \pm 0.60$                 & $7.5 \pm 1.1$                 &  $6.11 \pm 0.78$                            &$20.2 \pm 1.2$ \\ \noalign {\smallskip} 
Min. $R_p$ [$R_{\oplus}$]$^c$     &  $1.69 \pm 0.04$         & $1.43 \pm 0.03$                  & $1.33 \pm 0.06$                  &   $1.26 \pm 0.04$                            & $1.71 \pm 0.02$  \\ \noalign {\smallskip} 
Min. $(R_p/R_\ast)^2$ [ppm]$^c$ & $226 \pm 14$              & $130 \pm  8$                       & $210 \pm 18$                        &  $ 141 \pm 15$                                &  $96 \pm 5$      \\ \noalign {\smallskip} 
 $T_{eq}$ [K]$^d$                  & $1269 \pm 21$                     & $1469 \pm 24$                    & $753_{-23}^{+29}$                    &   $1101 \pm 39$                                 &   $1596 \pm 27$         \\ \noalign  {\smallskip} 
$T_{0}$-2450000 [BJD$_{TDB}$]         & $5296.79 \pm 0.14$    & $6416.96_{-0.09}^{+0.10}$        & $6265.33 \pm 0.17$  & $5393.96 \pm 0.14$  &  $5410.658 \pm 0.050$   \\ \noalign {\smallskip}     
 $P$ [d]                                   & $4.15514 \pm 0.00044$     & $3.269555_{-0.000080}^{+0.000092}$       & $9.37518 \pm 0.00056$      & $3.93394 \pm 0.00065$                    &  $3.834863 \pm 0.000094$ \\ \noalign {\smallskip} 
$W_{b=0}$ [min]                     & $172_{-16}^{+12}$             & $309\pm 16$                        &$227 \pm 19$                     & $152 \pm 19$                                  &  $271 \pm 13$ \\ \noalign {\smallskip} 
 $K$ [\ms]                               & $7.33 \pm 0.73$                 & $4.69 \pm 0.19$                  & $2.72 \pm 0.13$                  &  $2.81 \pm 0.27$                             &   $7.43 \pm 0.27$    \\ \noalign {\smallskip} 
$a$ [AU]                                 & $0.0520 \pm 0.0011$     & $0.0441 \pm 0.0010$          & $0.0822_{-0.0058}^{+0.0050}$          & $0.0465 \pm 0.0020$                      &  $0.0507 \pm 0.0012$ \\ \noalign {\smallskip} 
 $e$                                        & $0.093_{-0.064}^{+0.090}$       & $0.480 \pm 0.034$      &   $0.380 \pm 0.041$  & $0.215 \pm 0.096$             & $0.063 \pm 0.036$  \\ \noalign {\smallskip}
 $\omega$ [deg]                     & $120_{-71}^{+66}$              & $258.5 \pm 5.2$          & $198.2 \pm 7.1 $              & $122 \pm 30 $                         & $228 \pm 39$    \\ \noalign {\smallskip} 
Prior $P_{transit}$ [\%]           & $9.7_{-0.7}^{+1.1}$       & $8.39 \pm 0.32$                   & $4.93_{-0.39}^{+0.49}$     & $11.8 _{-1.4}^{+1.8}$                              & $14.18 \pm 0.64$ \\ \noalign {\smallskip} 
Prior $P_{occultation}$ [\%]    & $8.86_{-0.69}^{+0.59}$            & $23.2_{-1.5}^{+1.7}$                  &  $6.25_{-0.51}^{+0.64}$      & $8.57_{-0.87}^{+0.97}$                   & $15.32_{-0.67}^{+0.76}$   \\ \noalign {\smallskip} 
\hline \noalign {\smallskip} 
\end{tabular}
}
\end{center}
\caption{Same as Table A.1 for targets 15-19. 
 References: $^{(1)}$Van Leeuwen (2007), $^{(2)}$Hog et al. (2000), $^{(3)}$Skrutskie et al. (2006), $^{(4)}$Kharchenko (2001),$^{(5)}$Santos et al. (2013), $^{(6)}$Flower (1996), $^{(7)}$Lo Curto et al. (2010), $^{(8)}$Fischer et al. (2007), $^{(9)}$Mayor et al. (2011), $^{(10)}$Udry et al. in prep., $^{(11)}$Bouchy et al. (2009), ${(12)}$Melo et al. (2007).}
\end{table*}

\begin{table*}
\begin{center}
{\scriptsize
\label{tab:targets}
\begin{tabular}{cccccc}
\hline\noalign {\smallskip}
Star &  BD\,-061339 & HD\,1461 & HD\,10180& &    \\ \noalign {\smallskip}
\hline \noalign {\smallskip}  
Program ID                                                          & 90072                                & (1): 80220$^{(1)}$        &  (1): 60027                               & & \\ \noalign {\smallskip} 
                                                                             &                                           & (2): 90072                    &  (2) \& (3): 90072                                & & \\ \noalign {\smallskip} 
Observation date                                                 & 2013-11-30                        & (1): 2011-08-31             &  (1): 2010-01-16                  & & \\ \noalign {\smallskip} 
                                                                            &                                            & (2): 2013-09-13            & (2) : 2013-08-10                   & & \\ \noalign {\smallskip} 
                                                                            &                                            &                                     & (3): 2013-09-09             & & \\ \noalign {\smallskip} 
Channel       [$\mu$m]                                         & 4.5                                      & 4.5                              & (1): 3.6                                  & & \\ \noalign {\smallskip} 
                                                                             &                                           &                                     & (2) \& (3): 4.5                            & & \\ \noalign {\smallskip} 
AOR(s)$^a$                                                         & (S1): 48815616                    & (1): 42790656            & (1): 38139392                      & &  \\ \noalign {\smallskip}                                                                                      
                                                                           & (1): 48815360                      &  (S2): 48816128           & (S2): 48596224                     & &   \\ \noalign {\smallskip} 
                                                                           & (2): 48815104                      &  (2): 48815872            & (2): 48595968                     & &   \\ \noalign {\smallskip} 
                                                                           &                                              &                                   & (S3): 48595712                      & &   \\ \noalign {\smallskip} 
                                                                           &                                              &                                   & (3): 48595456                     & &   \\ \noalign {\smallskip} 
{\it Spitzer} pipeline version                               & S19.1.0                                 & S19.1.0                       & (1): S18.18.0                              & &   \\ \noalign {\smallskip}     
                                                                           &                                              &                                     & (2) \& (3): S19.1.0                               & &   \\ \noalign {\smallskip}     
Exposure time [s]                                               & 0.4                                       & 0.1                                & 0.1                               & &   \\ \noalign {\smallskip} 
$N_{BCD}$$^b$                                                & (1):1038                         & (1): 5548                       & (1): 4400                             & &  \\ \noalign {\smallskip} 
                                                                           &  (2): 970                                           & (2): 3600                       & (2) \& (3): 3175                      & &  \\ \noalign {\smallskip} 
Duration  [hr]                                                      & (1): 7.9                                      & (1):  12.9                      & (1): 10.4                               & &   \\ \noalign {\smallskip}  
                                                                           &  (2): 7.4                                           & (2):  8.4                        & (2) \& (3): 7.4                             & &  \\ \noalign {\smallskip} 
Photometric aperture [pixels]                             & (1) \& (2): 3.5                       & (1) \& (2): 3.5                         & (1): 3                               & &   \\ \noalign {\smallskip}  
                                                                          &                                               &                                     & (2): 2.5                            & &   \\ \noalign {\smallskip}  
                                                                          &                                              &                                     & (3): 3                              & &   \\ \noalign {\smallskip}  
Baseline  model$^c$                                        & (1): $p(w_x + [xy]^2) $ +BM       &  (1): $p(w_x^2 + w_y3 + [xy]^3 + l^2) $ +BM  & (1): $p(t + w_x + w_y^2 + [xy]^2 + l)$ + BM &    \\ \noalign {\smallskip} 
                                                                         & (2): $p(w_x + w_y + [xy]^2)$   + BM    & (2): $p(w_x + [xy]^2) $ + BM & (2): $p(w_x + [xy]^2 + l^2)$ + BM &    \\ \noalign {\smallskip} 
                                                                          &   & & (3): $p(w_x + w_y + [xy]^2 + l)$ + BM &    \\ \noalign {\smallskip}                                                       
Error correction factors$^d$                             & (1): $\beta_w$ = 0.94, $\beta_r$=1.21     &  (1):  $\beta_w$ = 0.97, $\beta_r$=1.91  & (1):  $\beta_w$ = 0.95, $\beta_r$=1.62  &    \\ \noalign {\smallskip} 
                                                                          & (2):  $\beta_w$ = 0.91, $\beta_r$=1.24    & (2):  $\beta_w$ = 0.94, $\beta_r$=1.34   &  (2):  $\beta_w$ = 0.97, $\beta_r$=1.26 &     \\ \noalign {\smallskip} 
                                                                          &                                                                   &                                                                  &  (3):  $\beta_w$ = 0.99, $\beta_r$=1.42 &     \\ \noalign {\smallskip} 
\hline \noalign {\smallskip}                                    
\end{tabular}
}
\end{center}
\caption{Details of our {\it  Warm Spitzer} data and their analysis for targets 1-3. 
$^a$AOR = astronomical observation fequest = {\it Spitzer} observing sequence. (S) designates a short pre-run AOR performed to 
stabilize the pointing and instrument.  $^b$BCD = Basic Calibrated Data = block of 64 {\it Spitzer}/IRAC subarray exposures.  
$^c$For the baseline model, $p(\epsilon^N)$ denotes, respectively, an $N$-order polynomial function
 of the logarithm of time ($\epsilon=l$), of the PSF $x$- and $y$-positions ($\epsilon=[xy]$), and widths ($\epsilon=w_x$ \& $w_y$). 
 BM=BLISS mapping. $^d$see Sec. 4. References: $^{(1)}$Kammer et al. (2014).}
\end{table*}

\begin{table*}
\begin{center}
{\scriptsize
\label{tab:targets}
\begin{tabular}{cccccc}
\hline\noalign {\smallskip}
Star &  HD\,13808 & HD\,20003 & HD\,20781& &    \\ \noalign {\smallskip}
\hline \noalign {\smallskip}  
Program ID                                                          & 90072                                & 90072                          &  90072                       \\ \noalign {\smallskip} 
Observation date                                                 & (1) 2013-08-15                  & 2013-09-01                 &  2013-11-16                \\ \noalign {\smallskip} 
                                                                            &  (2) 2013-08-27                 &                                     &                                    \\ \noalign {\smallskip}                                          
Channel       [$\mu$m]                                         & 4.5                                      & 4.5                              & 4.5                              \\ \noalign {\smallskip} 
AOR(s)$^a$                                                        & (S1): 48814848                   & (S): 48408576            & (S): 48817664             \\ \noalign {\smallskip}                                                                                      
                                                                           & (1): 48814592                      & (1): 48408320            & (1): 48817408              \\ \noalign {\smallskip} 
                                                                           & (S2): 48818176                    & (2): 48408064            & (2): 488817152            \\ \noalign {\smallskip} 
                                                                           & (2): 48817920                      & (3): 48407808            & (3): 48816896               \\ \noalign {\smallskip}                                                                
{\it Spitzer} pipeline version                               & S19.1.0                                 & S19.1.0                       & (1): S19.1.0                  \\ \noalign {\smallskip}     
Exposure time [s]                                               & 0.4                                       & 0.4                                & 0.4                               \\ \noalign {\smallskip} 
$N_{BCD}$$^b$                                                & (1) \& (2): 970                      & (1) \& (2): 1498              & (1) \& (2): 1367                     \\ \noalign {\smallskip} 
                                                                           &                                           &   (3): 580                    & (3): 580                 \\ \noalign {\smallskip} 
Duration  [hr]                                                      & (1) \& (2): 7.4                                & (1) \& (2):  11.5                         & (1) \& (2): 10.4                         \\ \noalign {\smallskip}  
                                                                           &                                   & (3):  4.4                        & (3): 4.4                       \\ \noalign {\smallskip} 
Photometric aperture [pixels]                             & (1): 2.75                              & (1) \& (2) \& (3): 3.5        & (1) \& (2) \& (3): 2.75                         \\ \noalign {\smallskip}  
                                                                          & (2): 3.5                                 &                                      &                                     \\ \noalign {\smallskip}  
Baseline  model$^c$   & (1): $p(w_x +w_y + [xy]^2) $ +BM         &  (1): $p(w_x^2 + w_y^2 + [xy]^2) $ +BM  & (1): $p(w_x + w_y + [xy]^2 )$ + BM     \\ \noalign {\smallskip} 
                                    & (2): $p(w_x^2 + w_y^2 + [xy])$   + BM   & (2): $p(w_x + w_y + [xy]) $ + BM                      & (2): $p(w_x^2 + w_y + [xy]^2 )$ + BM     \\ \noalign {\smallskip} 
                                    &                                                                 &  (3): $p(w_x + w_y + [xy])$ + BM                     & (3): $p(w_x + [xy]^2 )$ + BM \\ \noalign {\smallskip}                                                       
Error correction factors$^d$         & (1): $\beta_w$ = 0.93, $\beta_r$=1.00     &  (1):  $\beta_w$ = 0.93, $\beta_r$=1.30  & (1):  $\beta_w$ = 0.97, $\beta_r$=1.06     \\ \noalign {\smallskip} 
                                                     & (2):  $\beta_w$ = 0.96, $\beta_r$=1.55    & (2):  $\beta_w$ = 0.95, $\beta_r$=1.12  &  (2):  $\beta_w$ = 0.91, $\beta_r$=1.14     \\ \noalign {\smallskip} 
                                                    &                                                                   & (3): $\beta_w$ = 0.98, $\beta_r$=2.03    &  (3):  $\beta_w$ = 0.97 $\beta_r$=1.31     \\ \noalign {\smallskip} 
\hline \noalign {\smallskip}                                        
\end{tabular}
}
\end{center}
\caption{Same as Table A.5 for targets 4-6.}
\end{table*}

\begin{table*}
\begin{center}
{\scriptsize
\label{tab:targets}
\begin{tabular}{cccccc}
\hline\noalign {\smallskip}
Star &  HD\,31527  & HD\,39194& &    \\ \noalign {\smallskip}
\hline \noalign {\smallskip}  
Program ID                                                          & 90072                                                  &  90072                       \\ \noalign {\smallskip} 
Observation date                                                 & 2014-05-13                                  &  2012-12-24                \\ \noalign {\smallskip} 
Channel       [$\mu$m]                                         & 4.5                                                       & 4.5                              \\ \noalign {\smallskip} 
AOR(s)$^a$                                                        & (S): 50091776                            & (S): 46914816             \\ \noalign {\smallskip}                                                                                      
                                                                           & (1): 50091520                         &  (1): 46914560             \\ \noalign {\smallskip} 
                                                                           & (2): 50091264                                          &                                      \\ \noalign {\smallskip} 
                                                                           & (3): 50091008                                      &                                        \\ \noalign {\smallskip}                                                                
{\it Spitzer} pipeline version                               & S19.1.0                                             & (1): S19.1.0                  \\ \noalign {\smallskip}     
Exposure time [s]                                               & 0.4                                                          & 0.1                               \\ \noalign {\smallskip} 
$N_{BCD}$$^b$                                                & (1) \& (2) \& (3): 1558                                           & (1): 4890                     \\ \noalign {\smallskip} 
Duration  [hr]                                                      & (1) \& (2) \& (3): 11.9                                             & (1): 11.4                         \\ \noalign {\smallskip}  
   Photometric aperture [pixels]                             & (1) \& (2) \& (3) : 3.25                                 & (1): 3.0                          \\ \noalign {\smallskip}  
 Baseline  model$^c$   & (1): $p(w_x^2 + [xy]^2) $ +BM        & (1): $p(w_x + w_y + [xy]^2 + l )$ + BM     \\ \noalign {\smallskip} 
                                    & (2): $p(w_x + w_y + [xy]^2)$   + BM                                                  &                                       \\ \noalign {\smallskip} 
                                    & (3): $p(w_x + w_y^3 + [xy]^2)$   + BM                                   &                                      \\ \noalign {\smallskip}                                                       
Error correction factors$^d$         & (1): $\beta_w$ = 0.93, $\beta_r$=1.19       & (1):  $\beta_w$ = 0.97, $\beta_r$=1.15      \\ \noalign {\smallskip} 
                                                     & (2):  $\beta_w$ = 0.95, $\beta_r$=1.20                                                    &              \\ \noalign {\smallskip} 
                                                     & (3):  $\beta_w$ = 0.93, $\beta_r$=1.38                                                             &              \\ \noalign {\smallskip} 

\hline \noalign {\smallskip}                          
\end{tabular}
}
\end{center}
\caption{Same as Table A.5 for targets 7-8.}
\end{table*}

\begin{table*}
\begin{center}
{\scriptsize
\label{tab:targets}
\begin{tabular}{cccccc}
\hline\noalign {\smallskip}
Star &  HD\,45184 & HD\,47186 & HD\,51608      \\ \noalign {\smallskip}
\hline \noalign {\smallskip}  
Program ID                                                          & 90072                                & 60027                          &  90072                       \\ \noalign {\smallskip} 
Observation date                                                 & 2013-01-24                       & 2009-12-14                 &  2013-03-27                \\ \noalign {\smallskip} 
Channel       [$\mu$m]                                         & 4.5                                    & 3.6                              & 4.5                              \\ \noalign {\smallskip} 
AOR(s)$^a$                                                        & (S): 46917376                   & 38065664                   & (S): 48411392             \\ \noalign {\smallskip}                                                                                      
                                                                           & (1): 46917120                   &                                      & (1): 48411136             \\ \noalign {\smallskip} 
                                                                           &                                            &                                    & (2): 48410880                \\ \noalign {\smallskip} 
{\it Spitzer} pipeline version                               & S19.1.0                                 & S19.1.0                       & (1): S19.1.0                  \\ \noalign {\smallskip}     
Exposure time [s]                                               & 0.1                                       & 0.1                               & 0.4                               \\ \noalign {\smallskip} 
$N_{BCD}$$^b$                                                & (1): 4880                               & 2115                      & (1): 1498                    \\ \noalign {\smallskip} 
                                                                           &                                            &                                   & (2): 1300                            \\ \noalign {\smallskip} 
Duration  [hr]                                                      & (1): 11.4                               & 5                                   & (1): 11.4                        \\ \noalign {\smallskip}  
                                                                           &                                             &                                      &  (2): 9.9                                    \\ \noalign {\smallskip} 
Photometric aperture [pixels]                             & (1): 3.25                              &  2.75                        & (1) \& (2): 2.25                         \\ \noalign {\smallskip}  
Baseline  model$^c$   & (1): $p(w_x^2 +  w_y + [xy]^2 + l) $ +BM         &  $p(w_x^2 + [xy]^2 + l) $ +BM  & (1): $p(w_x + w_y + [xy]^2 )$ + BM     \\ \noalign {\smallskip} 
                                    &                                                           &                                                               &  (2):   $p(w_x + w_y + [xy]^2 )$ + BM     \\ \noalign {\smallskip} 
Error correction factors$^d$         & (1): $\beta_w$ = 0.94, $\beta_r$=1.54     &  (1):  $\beta_w$ = 1.00, $\beta_r$=1.61  & (1):  $\beta_w$ = 0.95, $\beta_r$=1.14      \\ \noalign {\smallskip} 
                                                     &                                                &                                                                  &   (2):  $\beta_w$ = 0.94, $\beta_r$=1.12           \\ \noalign {\smallskip} 

\hline \noalign {\smallskip}                            
\end{tabular}
}
\end{center}
\caption{Same as Table A.5  for targets 9-11. }
\end{table*}

\begin{table*}
\begin{center}
{\scriptsize
\label{tab:targets}
\begin{tabular}{cccccc}
\hline\noalign {\smallskip}
Star &  HD\,93385 & HD\,96700 & HD\,115617      \\ \noalign {\smallskip}
\hline \noalign {\smallskip}  
Program ID                                                          & 90072                                & 90072                          &  60027                      \\ \noalign {\smallskip} 
Observation date                                                 & 2013-03-12                       & 2013-08-16                 &  2010-03-25                \\ \noalign {\smallskip} 
Channel       [$\mu$m]                                         & 4.5                                    & 4.5                              & 3.6                              \\ \noalign {\smallskip} 
AOR(s)$^a$                                                        & (S): 48407552                   & (S): 46916864            & (1): 39138816            \\ \noalign {\smallskip}                                                                                      
                                                                           & (1): 48407296                   &  (1): 46916608              & (2): 39139072             \\ \noalign {\smallskip} 
                                                                           & (2):  48407040                    & (2): 46916352            &                                    \\ \noalign {\smallskip} 
                                                                           & (3): 48406784                    &                                     &                                        \\ \noalign {\smallskip}                                                                
{\it Spitzer} pipeline version                               & S19.1.0                                 & S19.1.0                       & S19.1.0                         \\ \noalign {\smallskip}     
Exposure time [s]                                               & 0.1                                       & 0.1                               & 0.01                               \\ \noalign {\smallskip} 
$N_{BCD}$$^b$                                                & (1) \& (2): 4245                   & (1): 4880                      & (1): 6600                   \\ \noalign {\smallskip} 
                                                                           & (3): 2325                               &  (2): 1240                   & (2): 6640                            \\ \noalign {\smallskip} 
Duration  [hr]                                                      & (1) \& (2): 11.9                               & (1): 11.4                      & (1): 6.2                      \\ \noalign {\smallskip}  
                                                                           & (3): 5.4                           &  (2): 2.9                       &   (2): 6.3                                    \\ \noalign {\smallskip} 
Photometric aperture [pixels]                             & (1) \& (2) \& (3) : 2.5                              &  (1) \& (2): 2.75                       & (1)  \& (2): 2.25                         \\ \noalign {\smallskip}  
Baseline  model$^c$   & (1): $p(w_x +  w_y^2 + [xy] + l) $ +BM         & (1): $p(w_x + w_y + [xy]^2) $ +BM  & (1): -     \\ \noalign {\smallskip} 
                                    & (2): $p(w_x +  w_y + [xy]) $ +BM          &  (2):  $p(w_x + [xy]^2) $ +BM        &  (2):  -    \\ \noalign {\smallskip} 
                                    & (3): $p(w_x  + [xy]^2) $ +BM         &                                                       &                                      \\ \noalign {\smallskip}                                                       
Error correction factors$^d$         & (1): $\beta_w$ = 0.96, $\beta_r$=1.00     &  (1):  $\beta_w$ = 0.99, $\beta_r$=1.36 & (1):  -     \\ \noalign {\smallskip} 
                                                     & (2): $\beta_w$ = 1.00, $\beta_r$=1.10    & (1):  $\beta_w$ = 0.97, $\beta_r$=1.65    &   (2):  -            \\ \noalign {\smallskip} 
                                                     & (3): $\beta_w$ = 1.00, $\beta_r$=1.85   &                                                                   &            \\ \noalign {\smallskip} 

\hline \noalign {\smallskip}                                                
\end{tabular}
}
\end{center}
\caption{Same as Table A.5 for targets 12-14.
The HD\,115617 photometry is affected by a strong systematic error that prevented any searching for the transit
of HD\,115617\,b. }
\end{table*}

\begin{table*}
\begin{center}
{\scriptsize
\label{tab:targets}
\begin{tabular}{cccccc}
\hline\noalign {\smallskip}
Star &  HD\,125612 & HD\,134060 & HD\,181433      \\ \noalign {\smallskip}
\hline \noalign {\smallskip}  
Program ID                                                          & 60027                                & (1): 60027                    &  90072                      \\ \noalign {\smallskip} 
                                                                           &                                             & (2): 90072                         &                                \\ \noalign {\smallskip} 
Observation date                                                 & (1): 2010-04-09                      & (1): 2010-04-13                &  2012-12-03                \\ \noalign {\smallskip} 
                                                                            & (2): 2010-09-10                      & (2): 2013-05-04           &                                    \\ \noalign {\smallskip}                                          
Channel       [$\mu$m]                                         & 3.6                                    & (1): 3.6                              & 4.5                              \\ \noalign {\smallskip} 
                                                                            &                                           & (2): 4.5                              &                          \\ \noalign {\smallskip} 
AOR(s)$^a$                                                        & (1): 38110464                  & (1): 38110720                    & (S): 46915584            \\ \noalign {\smallskip}                                                                                      
                                                                           & (2): 40313600                   &  (S2): 46916096              & (1): 46915328             \\ \noalign {\smallskip} 
                                                                           &                                          & (2): 46915840                  & (2): 46915072            \\ \noalign {\smallskip} 
{\it Spitzer} pipeline version                               & S18.18.0                                 & (1): S18.18.0                 & S19.1.0                         \\ \noalign {\smallskip}     
                                                                           &                                            &   (2): S19.1.0                      &                                      \\ \noalign {\smallskip}     
Exposure time [s]                                               & 0.4                                       & (1): 0.02                               & 0.4                               \\ \noalign {\smallskip} 
                                                                           &                                             & (2): 0.1                              &                                    \\ \noalign {\smallskip} 
$N_{BCD}$$^b$                                                & (1): 1790                               & (1): 6200                      & (1): 1563                  \\ \noalign {\smallskip} 
                                                                           & (2): 1310                               &  (2): 4480                   & (2): 1498                           \\ \noalign {\smallskip} 
Duration  [hr]                                                      & (1): 13.8                               & (1): 6                          & (1): 11.9                      \\ \noalign {\smallskip}  
                                                                           & (2): 10.2                              &  (2): 11.5                       &(2): 11.4                                    \\ \noalign {\smallskip} 
Photometric aperture [pixels]                             & (1): 1.75                             &  (1): 1.75                       & (1): 3.25                         \\ \noalign {\smallskip}  
                                                                          &  (2): 1.9                                & (2):  2.25                     &  (2): 3.25                                   \\ \noalign {\smallskip}  
Baseline  model$^c$   & (1): $p(w_x^2 +  w_y^2 + [xy]^3 + l) $ +BM         & (1): $p(w_x + [xy]^2 + l) $ +BM  & (1): $p(w_x + w_y + [xy]^2 )$ + BM     \\ \noalign {\smallskip} 
                                    & (2): $p(t + w_x +  w_y^2 + [xy] + l^2) $ +BM          &  (2):  $p(w_x^2 + w_y + [xy]^4) $ +BM        &  (2):   $p(w_x + w_y + [xy]^2 )$ + BM     \\ \noalign {\smallskip} 
Error correction factors$^d$         & (1): $\beta_w$ = 0.81, $\beta_r$=2.02     &  (1):  $\beta_w$ = 1.00, $\beta_r$=1.62 & (1):  $\beta_w$ = 0.95, $\beta_r$=1.27     \\ \noalign {\smallskip} 
                                                     & (2): $\beta_w$ = 0.72, $\beta_r$=2.34    & (1):  $\beta_w$ = 0.95, $\beta_r$=1.30    &   (2):  $\beta_w$ = 0.93, $\beta_r$=1.11          \\ \noalign {\smallskip} 
\hline \noalign {\smallskip}                           
\end{tabular}
}
\end{center}
\caption{Same as  Table A.5 for targets 15-17. }
\end{table*}

\begin{table*}
\begin{center}
{\scriptsize
\label{tab:targets}
\begin{tabular}{ccc}
\hline\noalign {\smallskip}
Star &  HD\,215497 & HD\,219828       \\ \noalign {\smallskip}
\hline \noalign {\smallskip}  
Program ID                                                          & 90072                                & 60027 \& 90072                                      \\ \noalign {\smallskip} 
Observation date                                                 & 2010-07-16                       & (1): 2010-08-01                             \\ \noalign {\smallskip}       
                                                                             &                                           & (2): 2013-03-06                               \\ \noalign {\smallskip}       
Channel       [$\mu$m]                                         & 4.5                                    & 4.5                                            \\ \noalign {\smallskip} 
AOR(s)$^{a}$                                                      & 38701568                           & (1): 38702336                          \\ \noalign {\smallskip}    
                                                                            &                                            & (2S): 46914304                                \\ \noalign {\smallskip}          
                                                                           &                                            & (2A): 46914048                                \\ \noalign {\smallskip}          
                                                                           &                                            & (2B): 46913792                                \\ \noalign {\smallskip}          
{\it Spitzer} pipeline version                               & S18.18.0                                 & S19.1.0                                     \\ \noalign {\smallskip}     
Exposure time [s]                                               & 0.4                                       & 0.4                                             \\ \noalign {\smallskip} 
$N_{BCD}$$^{b}$                                                &   1750                                 & (1): 1440                                     \\ \noalign {\smallskip} 
                                                                           &                                            & (2A):  1564                                     \\ \noalign {\smallskip} 
                                                                            &                                          &  (2B):   448                                          \\ \noalign {\smallskip} 
 Duration  [hr]                                                      & 13.4                                   & (1): 11                                          \\ \noalign {\smallskip}  
                                                                            &                                          & (2A): 11.9                                    \\ \noalign {\smallskip}  
                                                                           &                                          & (2B): 3.4                                       \\ \noalign {\smallskip}  
 Photometric aperture [pixels]                             & 3                                      &  (1): 2.5                                         \\ \noalign {\smallskip}  
                                                                          &                                          & (2A): 3                                          \\ \noalign {\smallskip}  
                                                                           &                                          & (2B): 3                                          \\ \noalign {\smallskip}  
Baseline  model$^{c}$     & $p(w_x +  w_y + [xy] + l) $ +BM         & (1): $p(w_x + w_y + [xy]^2 + l^2) $ +BM            \\ \noalign {\smallskip}     
                                    &                                                                & (2A): $p(w_x + [xy]^2 + l) $ +BM            \\ \noalign {\smallskip}  
                                    &                                                                 & (2B): $p(w_x +  [xy]^2 + l^2) $ +BM            \\ \noalign {\smallskip}                        
Error correction factors$^{d}$           & $\beta_w$ = 0.95, $\beta_r$=1.20     &  (1):  $\beta_w$ = 0.93, $\beta_r$=1.53    \\ \noalign {\smallskip} 
                                                     &                                                            &  (2A):  $\beta_w$ = 0.96, $\beta_r$=1.01    \\ \noalign {\smallskip} 
                                                    &                                                            &  (2B):  $\beta_w$ = 0.94, $\beta_r$=1.14   \\ \noalign {\smallskip} 
\hline \noalign {\smallskip}                                            
\end{tabular}
}
\end{center}
\caption{Same as Table A.5 for targets 18-19. }
\end{table*}

\end{appendix}

\end{document}